\documentclass[12pt]{oxarticle}

\usepackage[usenames, dvipsnames]{xcolor}
\usepackage{graphicx, float, array, xspace, amscd, amsmath, amsthm, amssymb, latexsym, longtable,bbm}
\usepackage[letterpaper, left=1.5in, right=0.5in, top=0.8in, bottom=1.2in]{geometry}
\usepackage[bbgreekl]{mathbbol}
\usepackage{amsfonts}
\usepackage{url}
\usepackage{tikz}
\usepackage{lscape}

\DeclareSymbolFontAlphabet{\mathbb}{AMSb}
\DeclareSymbolFontAlphabet{\mathbbl}{bbold}

\usepackage[sort&compress, comma, square, numbers]{natbib}

\usetikzlibrary{arrows.meta}
\usetikzlibrary{shapes.geometric}
\usetikzlibrary{positioning}

\let\SS=\S 
\renewcommand{\a}{\alpha}
\renewcommand{\b}{\beta}
\newcommand{\g}{\gamma}\newcommand{\G}{\Gamma}
\renewcommand{\d}{\delta}\newcommand{\D}{\Delta}
\newcommand{\e}{\epsilon}
\newcommand{\z}{\zeta}

\renewcommand{\th}{\theta}\newcommand{\vth}{\vartheta}

\renewcommand{\k}{\kappa}
\renewcommand{\l}{\lambda}\renewcommand{\L}{\Lambda}
\newcommand{\m}{\mu}
\newcommand{\n}{\nu}
\newcommand{\x}{\xi}

\newcommand{\p}{\pi}\renewcommand{\P}{\Pi}\newcommand{\vp}{\varpi}
\renewcommand{\r}{\rho}
\newcommand{\s}{\sigma}\renewcommand{\S}{\Sigma}
\renewcommand{\t}{\tau}

\newcommand{\ph}{\phi}\newcommand{\vph}{\varphi}

\renewcommand{\O}{\Omega}

\DeclareFontFamily{OT1}{pzc}{}
\DeclareFontShape{OT1}{pzc}{m}{it}{<-> s * [1.200] pzcmi7t}{}
\DeclareMathAlphabet{\mathpzc}{OT1}{pzc}{m}{it}

\newcommand{\ccA}{\mathpzc A}
\newcommand{\ccB}{\mathpzc B}

\newcommand{\cE}{\mathcal{E}}
\newcommand{\cF}{\mathcal{F}}

\newcommand{\cI}{\mathcal{I}}

\newcommand{\cL}{\mathcal{L}}

\newcommand{\cN}{\mathcal{N}}
\newcommand{\cO}{\mathcal{O}}
\newcommand{\cP}{\mathcal{P}}\newcommand{\ccP}{\mathpzc P}

\newcommand{\cS}{\mathcal{S}}

\DeclareFontFamily{U}{bbold}{}
\DeclareFontShape{U}{bbold}{m}{n}
 {  <-5.5> s*[1.05] bbold5
    <5.5-6.5> s*[1.05] bbold6
    <6.5-7.5> s*[1.05] bbold7
    <7.5-8.5> s*[1.05] bbold8
    <8.5-9.5> s*[1.05] bbold9
    <9.5-11.5> s*[1.05] bbold10
    <11.5-16> s*[1.05] bbold12
    <16-> s*[1.05] bbold17
 }{}

\newcommand{\IC}{\mathbbl{C}}

\newcommand{\IF}{\mathbbl{F}}

\newcommand{\IH}{\mathbbl{H}}

\newcommand{\IK}{\mathbbl{K}}

\newcommand{\IP}{\mathbbl{P}}
\newcommand{\IQ}{\mathbbl{Q}}
\newcommand{\IR}{\mathbbl{R}}

\newcommand{\IT}{\mathbbl{T}}

\newcommand{\IZ}{\mathbbl{Z}}

\font\eightrm=cmr8 at 8pt
\font\eightbf=cmb10 at 8pt
\font\csc=cmcsc10

\newcommand{\beq}{\begin{equation}}
\newcommand{\eeq}{\end{equation}}
\newcommand{\beqnn}{\begin{equation*}}
\newcommand{\eeqnn}{\end{equation*}}
\newcommand{\bea}{\begin{eqnarray}}
\newcommand{\eea}{\end{eqnarray}}
\newcommand{\bean}{\begin{eqnarray*}}
\newcommand{\eean}{\end{eqnarray*}}

\newcommand{\fref}[1]{Figure~\ref{#1}}
\newcommand{\tref}[1]{Table~\ref{#1}}
\newcommand{\sref}[1]{\SS\ref{#1}}

\newcommand{\pd}[2]{\frac{\partial #1}{\partial #2}}

\newcommand{\ee}{\text{e}}
\newcommand{\ii}{\text{i}}
\newcommand{\dd}{\text{d}}

\newcommand{\place}[3]{\vbox to0pt{\kern-\parskip\kern-7pt
                             \kern-#2truein\hbox{\kern#1truein #3}
                             \vss}\nointerlineskip}
\newcommand{\smallfrac}[2]{\frac{\scriptstyle #1}{\scriptstyle #2}}

\DeclareFontFamily{U}{wncy}{}
\DeclareFontShape{U}{wncy}{m}{n}{<->wncyr10}{}
\DeclareSymbolFont{mcy}{U}{wncy}{m}{n}
\DeclareMathSymbol{\sha}{\mathord}{mcy}{"58}

\newcommand{\capt}[3]{\parbox{#1}{\renewcommand{\baselinestretch}{1.0}
                                                           \caption{\label{#2}\small\it #3}}}

\newcommand{\del}{\partial}

\newcommand{\cy}{Calabi-Yau\xspace}
\newcommand{\cym}{Calabi-Yau manifold\xspace}
\newcommand{\cys}{Calabi-Yau manifolds\xspace}
\newcommand{\K}{K\"ahler\xspace}

\renewcommand{\Re}{\text{Re~}}

\newcommand{\+}{\phantom{-}}

\newcommand{\lra}{\longrightarrow}
\newcommand{\Frob}{\text{Frob}}

\renewcommand{\=}{\;=\;}

\makeatletter
\g@addto@macro\bfseries{\boldmath}

\def\blindfootnote{\xdef\@thefnmark{}\@footnotetext}
\makeatother

\renewcommand{\baselinestretch}{1.1}
\numberwithin{equation}{section}
\proofmodefalse
\begin{document}
\pagestyle{empty}      

\begin{center}
{\Huge A One Parameter Family of \cy Manifolds with Attractor Points of Rank Two\\[0.3in]}
{\csc Philip Candelas$^{*\,1}$, Xenia de la Ossa$^{*\,2}$, Mohamed Elmi$^{* \,3}$\\
and\\
Duco van Straten$^{\dagger \,4}$\\[0.2in]}

{\it $^*$Mathematical Institute\hphantom{$^1$}\\
University of Oxford\\
Andrew Wiles Building\\
Woodstock Road, Radcliffe Observatory Quarter\\
Oxford, OX2 6GG, UK\\[2ex]

 $^\dagger$Fachbereich 08\hphantom{$^1$}\\ 
AG Algebraische Geometrie\\
Johannes Gutenberg-Universit\"at\\
D-55099 Mainz, Germany\\
}

\footnotetext[1]{{\tt candelas@maths.ox.ac.uk} \hfil
$^2\,${\tt delaossa@maths.ox.ac.uk} \hfil
$^3\,${\tt elmi@maths.ox.ac.uk}\\[2pt]
\hspace*{5.35cm}$^4\,${\tt straten@mathematik.uni-mainz.de}}
\vfill
{\bf Abstract}
\end{center}
\vskip-7pt
\begin{minipage}{\textwidth}
\baselineskip=14.5pt
In the process of studying the $\z$-function for one parameter families of \cys we have been led to a manifold, first studied by Verrill, for which the quartic numerator of the $\z$-function factorises into two quadrics remarkably often. Among these factorisations, we find \emph{persistent factorisations}; these are determined by a parameter that satisfies an algebraic equation with coefficients in $\IQ$, so independent of any particular prime. Such factorisations are expected to be modular with each quadratic factor associated to a modular form. If the parameter is defined over $\IQ$ this modularity is assured by the proof of the Serre Conjecture. We identify three values of the parameter that give rise to persistent factorisations, one of which is defined over $\IQ$, and identify, for all three cases, the associated modular groups. We~note that these factorisations are due a splitting of Hodge structure and that these special values of the parameter are rank two attractor points in the sense of IIB supergravity. To~our knowledge, these points provide the first explicit examples of non-singular, non-rigid rank two attractor points for \cys of full SU(3) holonomy. The values of the periods and their covariant derivatives, at the attractor points, are identified in terms of critical values of the $L$-functions of the modular groups. Thus the critical $L$-values enter into the calculation of physical quantities such as the area of the black hole in the 4D spacetime. In our search for additional rank two attractor points, we perform a statistical analysis of the numerator of the $\z$-function and are led to conjecture that the coefficients in this polynomial are distributed according to the statistics of random $\text{USp}(4)$~matrices.
\end{minipage}
%
\newpage
{\baselineskip=12pt
\tableofcontents}

%
%
\newpage
\setcounter{page}{1}
\pagestyle{plain}     
\section{Introduction}\label{sec:Intro}
\vskip-10pt
\subsection{Preamble}\label{sec:preamble}
\vskip-10pt
The attractor mechanism, first described in \cite{Ferrara:1995ih} in the context of
$N=2$ supergravity, remains a fascinating topic that links 4D black holes to string theory and 
has led to an understanding of black hole entropy in term of the counting of microstates. 
We refer to \cite{Pioline:2006ni} and \cite{Sen:2007qy} for overviews.  
In \cite{Moore:1998pn} G.~Moore posed many questions pertaining to the arithmetic nature 
of attractor points, which are divided into being of {\em rank one} or {\em rank two}.
We report here on a specific one parameter family of \cys $X_{\vph}$ determined 
by the equation
\beq
1-\vph\, (X_1+X_2+X_3+X_4+X_5) \left(\frac{1}{X_1}+\frac{1}{X_2}+\frac{1}{X_3}+\frac{1}{X_4}+\frac{1}{X_5}\right) \= 0
\label{eq:PolynomialP}\eeq
first considered by H. Verrill in \cite{verrill1996, verrill2004sums}  and by Hulek and Verrill in \cite{hulek_verrill_2005} which has at least three attractor points of rank two, occurring at a rational value 
\[
\vph{\;=}-1/7
\]
and a pair of values correspond to the roots of the quadratic equation $\vph^2{\;-\;}66\vph{\;+\;}1{\=}0$,
\[
\vph \= \vph_{\pm} \= 33 \pm 8\sqrt{17}~.
\]
To our knowledge, these are the first nontrivial such attractor points to be identified explicitly for a \cym of holonomy $\text{SU}(3)$. Some Fermat-type points were explicitly identified as rank two attractor points in~\cite{Moore:1998pn}.
 
While attractor points of rank one are expected to be dense in the moduli space, those of rank two
are expected to be rare, as the underlying \cym has to satisfy very stringent conditions. As we will summarise the attractor mechanism in the following section, it may suffice here to recall that the condition for a rank two attractor point is that the two-dimensional vector space $V{\=}H^{3,0}{\;\oplus\;}H^{0,3}$ 
is the complexification of a rank two lattice in $H^3(X,\IZ)$.
The space $V^\perp{\=}H^{2,1}{\;\oplus\;}H^{1,2}$ is orthogonal to $V$ under the natural symplectic product on three forms and is also the complexification of a rank two sublattice of $H^3(X,\IZ)$. This results in a remarkable splitting of the Hodge structure of $H^3(X,\IQ)$. The Hodge Conjecture predicts that such a splitting must have a geometrical origin, which in turn makes this splitting visible in the arithmetic structure of $X$. In particular, this leads  to the factorisation, for infinitely many primes $p$, of the part $R(T)$ of the $\z$-function for the manifold coming from the third cohomology. By reversing the logic, the study of such persistent factorisations leads to an effective strategy for finding rank two attractor points, and it was in this way that the above attractor points were obtained for the one-parameter family of manifolds considered here. 

It follows from arithmetic considerations that, the splitting at a rank two attractor point gives rise to modular forms of weight two and four that are determined by the way that the two factors of $R(T)$ vary with $p$. The modular groups that arise in this way have pervasive consequences. For example, the periods of the attractor variety and further quantities like the central charge and so the area of the black hole horizon  can be expressed in terms of critical $L$-values of these modular forms. 

As a simple example of the identities that arise, we mention the Ramanujan-like formula
\beq 
\sum_{n=0}^{\infty} a_n (33-8\sqrt{17})^n\= \frac{119+29\sqrt{17}}{16\pi^2}\,\l_4(2)~,
\label{eq:FundamentalPeriodIdentity1}\eeq
where 
\[a_0\=1,\;\;a_1\=5,\;\;a_2\=45,\;\;a_3\=545,\;\;a_4\=7889,\;\;a_5\=127905,\ldots\]
and generally
\[
a_n\= \sum_{p+q+r+s+t=n} \left( \frac{n!}{p! q! r! s! t!}\right)^2~,
\]
and
\[ 
\l_4(2)\= \Re L(f,2)~,
\]
where 
\[
f\= q-2q^2+2iq^3+4q^4+8iq^5-4iq^7-8q^8+23q^9+\ldots
\]
is a modular form for $\G_1(34)$, that appears as $f_{\bf 34.4.b.a}$ in the LMFDB~\cite{lmfdb}.

The left hand side of the above formula is the value of the fundamental period $\vp_0{\=}\sum_n a_n\vph^n$, which will be defined in \sref{sec:periods}, evaluated at the attractor point $\vph_{-}$. It is intriguing that this should evaluate to an algebraic multiple of a critical $L$ value. Now the fundamental period is defined as a solution to the Picard-Fuchs equation and is very far from being an algebraic function, so perhaps equally intriguing is the fact that
\beq
\frac{1+\sqrt{17}}{28}\,\vp_0(33+8\sqrt{17}) \= \frac{119-29\sqrt{17}}{16\pi^2}\,\l_4(2)~.
\label{eq:FundamentalPeriodIdentity2}\eeq
Apart from the prefactor $\frac{1+\sqrt{17}}{28}$, the only change between \eqref{eq:FundamentalPeriodIdentity1} and \eqref{eq:FundamentalPeriodIdentity2} is the change of sign of $\sqrt{17}$. 

We make here a single disclaimer in relation to these identities and others that arise in the following.
     What is meant by saying that we have established an identity such as \eqref{eq:FundamentalPeriodIdentity1} or 
     \eqref{eq:FundamentalPeriodIdentity2} is that we have evaluated both sides of the identity to at least 1000 figures and found the quantities agree to this accuracy. We do not have proofs of the identities, in the classical sense.

It is interesting that the Hulek-Verrill manifold with five complex structure parameters, so before taking the quotient, appears also in other contexts. One of these is in the study of field theory amplitudes, principally in relation to the  banana or sunrise graphs. An example, with four loops, is shown in \fref{fig:bananagraph}.
\begin{figure}[H]
\begin{center}
\includegraphics[width=3.5in]{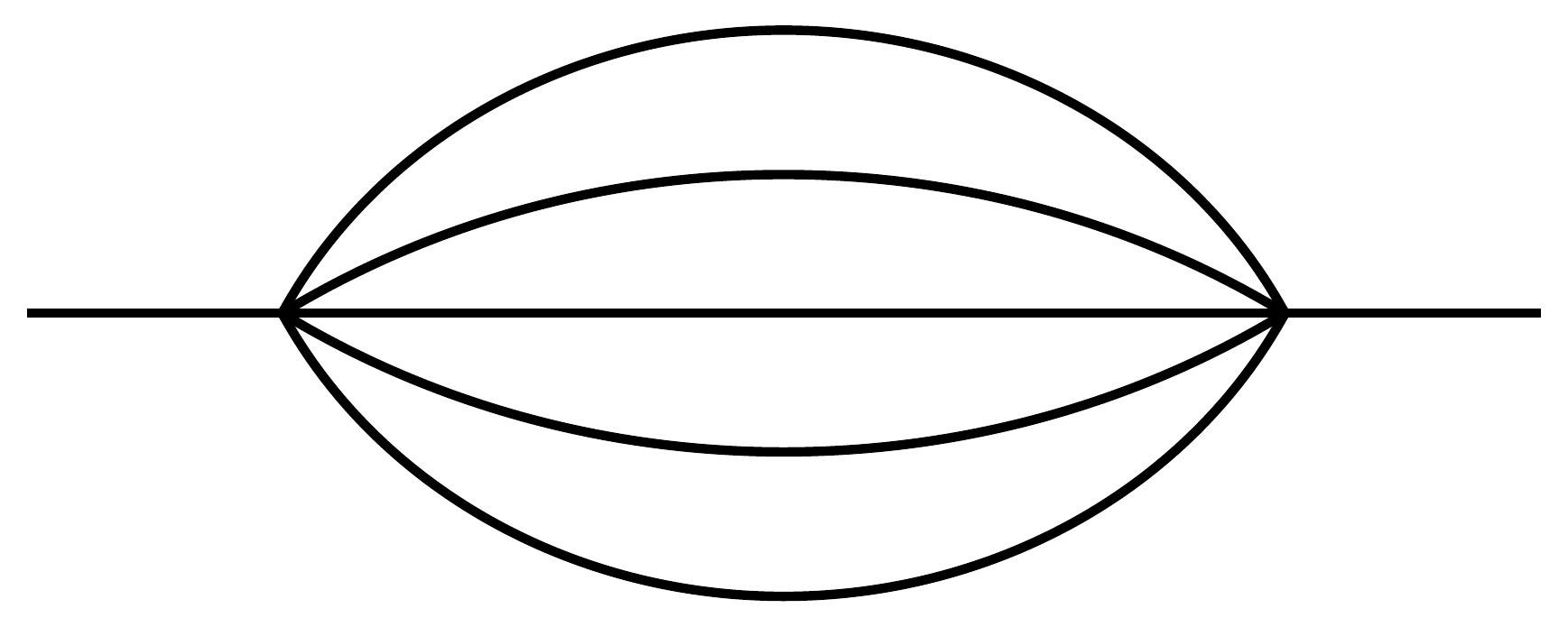}
\capt{6in}{fig:bananagraph}{The four-loop banana graph that is related to the Hulek-Verrill manifold.}
\end{center}
\end{figure}
This is a Feynman diagram for a scalar field with momentum $p$ flowing through the diagram and the internal lines refer to particles of mass $m_i$, $i=1,\ldots,5$. Denoting the maximally cut diagram in two dimensions by $F(p^2)$ and with the identifications
\beq
\m_i \= m_i^2\quad\text{and}\quad p^2 \= \frac{1}{\vph}~,
\notag\eeq
it has been observed that $p^2 F(p^2)$ is a period for the five parameter Hulek-Verrill manifold defined by the $n{\=}5$ case of the equation 
\beq
\left( \sum_{i=1}^n X_i \right) \left( \sum_{i=1}^n\frac{\m_i}{X_i} \right) \= \frac{1}{\vph}~.
\label{eq:Verrilln}\eeq
Note however that this equation is often written with coordinates related to those here by the transformation $X_i\to 1/X_i$.
In the case that all the masses are equal, the quantity $p^2 F(p^2)$ is a period for the quotient manifold.
      There is a considerable literature on this subject, to which we cannot do justice. The expository article of Vanhove~\cite{Vanhove:2018mto} and references cited therein can serve as an introduction. 
      
The fundamental periods of many \cys have an interpretation as generating functions for the numbers of lattice walks, with the n$^{\text{th}}$ coefficient $a_n$ being the number of lattice walks that return to the origin after $n$ steps. The lattice in question being the lattice generated by the monomials of the defining equation. For the Hulek-Verrill manifold these considerations apply and the fundamental period is generating function for walks in the $A_4$~lattice. The Hulek-Verrill manifold fits into a closely related sequence of manifolds that correspond to taking $n{\=3,4,5,\ldots}$ in \eqref{eq:Verrilln}. Verrill~\cite{verrill1996} examined this sequence and noted, for the case of the K3 manifold, corresponding to $n{\=}4$, that the fundamental period is the generating function for lattice walks in the $A_3$ lattice.

The study of lattice walks and of Feynman diagrams such as the banana graph leads naturally to integrals of products of Bessel functions, so the Hulek-Verrill manifold has appeared also in this context, see for example~\cite{Bailey:2008ib}.

\subsection{The attractor mechanism}\label{sec:AttractorMechanism}
\vskip-10pt
One may construct four dimensional $\mathcal{N}=2$ black holes by compactifying IIB supergravity on a Calabi-Yau threefold $X$ with complex structure parameter $\varphi$. The charges of the black hole are determined by a 3-cycle $\g {\,\in\,} H_3(X,\IZ)$, which is viewed as being wrapped by D3-branes. 

Infinitely far from the horizon of the black hole, space-time is flat and the value of $\varphi$ is unconstrained. However, as one moves towards the horizon of the black hole, $\varphi$ must evolve in a manner dictated by the {\em attractor mechanism}. Moreover, the value of $\varphi$ at the horizon of the black hole is an {\em attractor point} that (for small enough perturbations) is independent of the value of $\varphi$ at infinity and is only determined by a choice of $\g {\,\in\,} H_3(X,\IZ)$.

The four dimensional black hole is assumed to be spherically symmetric with a metric
of the form
\beq
\dd s^2 \;= -\ee^{2U(r)}\dd t^2 + \ee^{-2U(r)}\dd\vec{x}^{\,2},
\notag\eeq
where $r$ is a radial coordinate that is taken to vanish at the horizon.
In the supergravity approximation, the preservation of supersymmetry requires 
that the complex structure of $X$ varies with the radius in a manner governed by  differential equations, which are written most simply in terms of a new variable $\r{\=}\frac{1}{r}$,
\beq\begin{split}
\frac{\dd U(\r)}{\dd\r}&\;= -\phantom{2}\ee^{U(\r)}|Z_\g(\vph)|,\\[8pt]
\frac{\dd\vph(\r)}{\dd\r}&\;= -2e^{U(\r)}g^{\vph\bar{\vph}}\,\partial_{\bar{\vph}}|Z_\g(\vph)|~.
\label{eq:attractorequations}
\end{split}\eeq

We use the initial condition $U{\;=\;}0$ when $\r{\;=\;}0$, appropriate to an asymptotically flat space-time. In the above formula, the quantity
\[ 
Z_\g(\vph)\= \ee^{K/2}\int_{\g}\Omega 
\]
denotes the {\em central charge} and $K$ denotes the {\em K\"ahler potential\/} of the special geometry metric on moduli space. By a change of variables, these equations can be recast as a {\em gradient flow} of the function $|Z_\g(\vph)|$ with respect to this metric.

If we pick a symplectic basis $\{A^a, B_b\}$ of $H_3(X,\IZ)$, we can write the cycle $\g$ as
\beq
\g\= q_a A^a - p^a B_a \;\in\; H_3(X,\IZ)
\notag\eeq
and the black hole will have electric and magnetic charges given by the charge vector
\beq
Q\= \begin{pmatrix}q_a \\[3pt] p^b  \end{pmatrix}\,.
\notag\eeq

For the basis $\{\a_a,\b^b\}$ of $H^3(X,\IZ)$, dual to the symplectic basis $\{A^a, B_b\}$, we have
\beq
\int_{A^b}\a_a \; = -\int_{B_a}\b^b \= \int_{X_\vph}\a_a\wedge\b^b \= \d_a{}^b,
\notag\eeq 
so that the dual in cohomology of the cycle $\g$  is given by
\beq
\G\= p^a \a_a - q_a\b^a~ 
\notag\eeq
and the central charge can be written as
\beq
Z_\g(\vph)\= \ee^{K/2}\int_X \G\wedge\O \= \frac{Q^T\S\P}{(-\ii \P^{\dagger}\S\P)^\frac12},
\notag\eeq
where $\P$ is the vector of periods in an integral symplectic basis and $\S$ the matrix of the
symplectic form on $H^3(X,\IZ)$. For a concise review of special geometry and our conventions see Appendix~\ref{section: special geometry}. In \sref{sec:periods} we give precise details on these matters for the family we consider~here. 

\begin{figure}[!b]
\begin{center}
\includegraphics[width=0.6\textwidth]{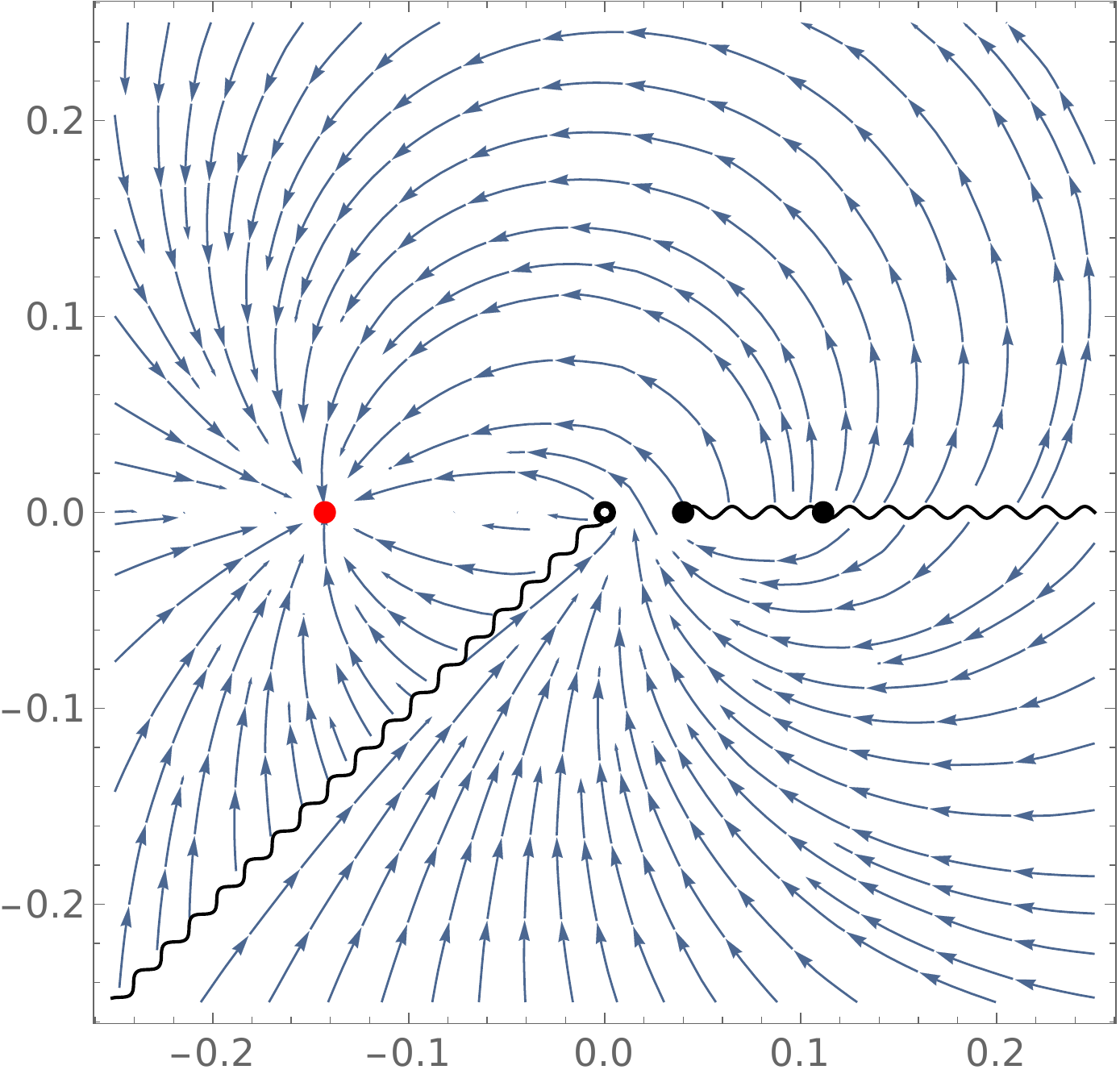}
\vskip 5pt
\capt{6.1in}{fig:full4m15m50}
{Attractor flow associated to the charge vector $Q=(4,-15,-5,0)$ in the $\vph$-plane. The red dot represents the attractor point $\vph=-1/7$, the hollow black dot is the large complex structure point $\varphi=0$ and the solid black dots represent the two nearest conifolds at $\varphi=1/25$ and $\varphi=1/9$. The flow lines are discontinuous across branch cuts which illustrates the fact that the flow takes place on a Riemann surface that is a multi-sheeted cover of the $\vph$-plane.}
\end{center}
\end{figure}

It follows from the gradient nature of the flow that, for a given $\g \in H_3(X,\IZ)$, the `end point' $\vph_{*}=\vph_{*}(\g)$ of the flow is a minimum of $|Z_\g|$ and is independent of the starting point $\vph_{\infty}$, at least under small variations of $\vph_{\infty}$, and thus will only depend on the charges~$Q$. This is the origin of the name attractor point. Note however, that due to the multi-valuedness caused by the monodromy around the singular points, the flow really takes place on a Riemann surface covering the $\vph$-plane. We give an example of the attractor flow for a specific charge vector leading to the attractor point $\vph{\;=}-1/7$ in Figure~\ref{fig:full4m15m50}.

It follows from \eqref{eq:attractorequations} that the black hole metric near the horizon is asymptotic to that corresponding to $AdS_2\times S^2$ and the area of the horizon is given by
\begin{equation}
A \= 4\pi |Z_{\g}(\vph_{*})|^2
\label{eq:areaformula}
\end{equation}
and this determines the entropy of the black hole in the limit of large charges.

The attractor points have a number of special properties. Firstly, as already mentioned, attractor points are critical points of the absolute value of the central charge function
$|Z_\g(\vph)|$, as can be seen from \eqref{eq:attractorequations}. 
Secondly, with a bit more work, it can be shown that the complex structure at an attractor point $\vph{\=}\vph_{*}$ is such that the dual of the charge vector satisfies the relation
\beq
\G\;\in\; H^{3,0}\oplus H^{0,3}\;\; \textup{or equivalently}\;\;\; \G^{2,1} \= \G^{1,2} \=0~.
\label{eq:LambdaPlane}\eeq

The condition that \eqref{eq:LambdaPlane} imposes on $\vph$ can be expressed more geometrically in the following way. The space $V(\vph){\=}H^{3,0}\oplus H^{0,3}$ is a plane, generated by $\O$ and $\overline\O$, in the space $H^3(X,\IZ){\,\otimes\,}\IC{\=}H^3(X,\IC)$. The intersection with the real four dimensional space $H^3(X,\IZ){\,\otimes\,}\IR{\=}H^3(X,\IR)$ is the 2-plane $V_ {\IR}(\vph)$ spanned, over $\IR$, by $\text{Re}\, \Omega$ and $\text{Im}\, \Omega$. Inside the vector space $H^3(X,\IR)$ we have the lattice of dual charge vectors $H^3(X,\IZ)$. This lattice is fixed, but the plane $V_{\IR}(\vph)$ moves with respect to this lattice as $\vph$ varies. There are three possibilities:

\begin{enumerate}
\setcounter{enumi}{-1}
\item The plane $V_{\IR}({\vph})$ intersects $H^3(X,\IZ)$ only in $0$. This is the generic case and $\vph$ is
not an attractor point.\smallskip
\item The intersection $V_\IR({\vph}) \cap H^3(X,\IZ)$ is a {\em lattice line}, i.e.\ a copy of $\IZ$. 
The point $\vph$ is attractor point for any non-zero $\Gamma \in V_\IR({\vph}) \cap H^3(X,\IZ)$. In this
case $\vph$ is an {\em attractor point of rank one}.\smallskip
\item The intersection $\L:=V_\IR({\vph}) \cap H^3(X,\IZ)$ is a {\em lattice plane}, i.e.\ a copy of $\IZ^2$. 
In this case one can find two independent charges $\G_1$ and $\G_2$ in $\L$, which have symplectic 
product $\langle \G_1,\G_2 \rangle \neq 0$. In this case $\vph$ is an {\em attractor point of rank two}.
\end{enumerate}
\begin{figure}[!ht]
\begin{center}
\framebox[\textwidth]{\begin{minipage}[c]{0.98\textwidth}
\centering
\vspace{5pt}
\includegraphics[width=0.4\textwidth]{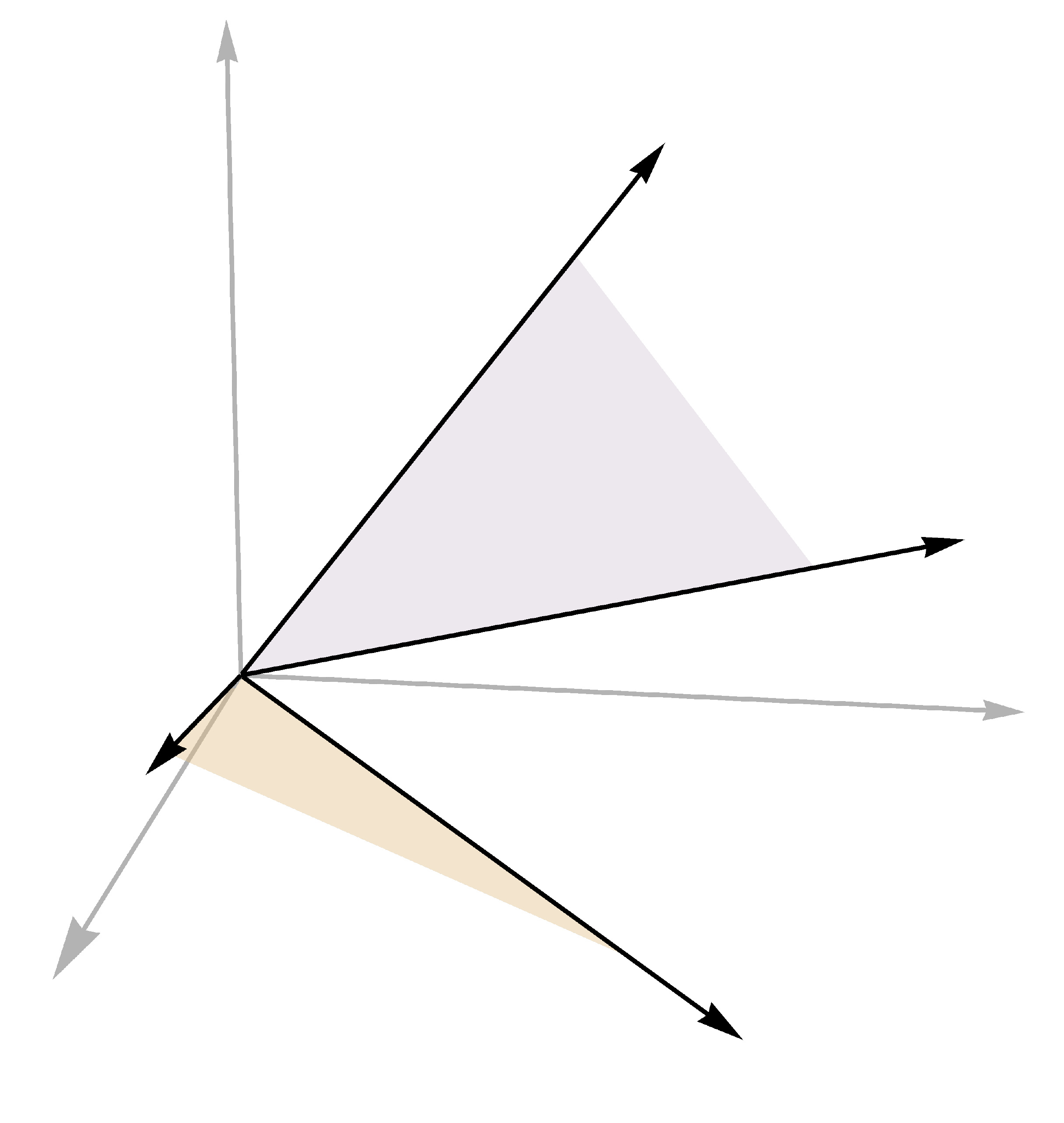}
\end{minipage}}
\vskip0pt 
\place{4.35}{1.45}{$\text{Im}\,\O$}
\place{3.6}{2.5}{$\text{Re}\,\O$}
\place{2.15}{0.82}{$\G_1$}
\place{3.85}{0.17}{$\G_2$}
\vskip10pt
\capt{5.8in}{fig:Hthree}{A sketch of the (four dimensional) space $H^3(X,\IR)$ for generic $\vph$, showing the two planes generated by $\text{Re}\,\Omega$ and $\text{Im}\,\Omega$ and by charge vectors $\G_1$ and $\G_2$. As $\vph$ varies, the plane generated by $\text{Re}\,\Omega$ and $\text{Im}\,\Omega$ moves and, when $\vph{\;=\;}\vph_*$ is an attractor point of rank two, the two planes coincide.}
\end{center}
\end{figure}

As we are dealing with the geometry of 2-planes in a four dimensional vector space, it is natural to formulate equation \eqref{eq:LambdaPlane} in terms of the
Grassmanian $\text{Gr}(2,\IC^4)$, which by the Pl\"ucker embedding 
\[\text{Gr}(2,\IC^4) \hookrightarrow \IP^5(\IC)\]
can be identified with the Pl\"ucker quadric.
The natural map 
\[ \vph \mapsto V(\vph)=H^{3,0}\oplus H^{0,3} \subset H^3(X,\IC)\]
from the complex structure moduli space to the Grassmanian can be composed with
the Pl\"ucker embedding. 
Since $H^{3,0}\oplus H^{0,3}$ is spanned by the cohomology classes of $\text{Re}\,\O$ and $\text{Im}\,\O$ the resulting map can be identified with the map
\beq
\vph \mapsto \ccP\= \big(\text{Re}\,\Pi,\text{Im}\,\Pi\big)
\mapsto [\p_{12},\,\p_{13},\, \p_{14},\, \p_{23},\, \p_{24},\, \p_{34}]\in \IP^5(\IR) \subset \IP^5(\IC)
\notag\eeq
where $\p_{ij}$ is the minor formed by the $i^{\text{th}}$ and $j^{\text{th}}$ rows of $\ccP$. 
The rows of $\ccP$ form a basis of $H^{3,0}{\,\oplus\,} H^{0,3}$ and any other basis is related to this one by $\ccP\mapsto  \ccP g$ for some $g\in \text{GL}(2,\IC^4)$ which simply multiplies each $\p_{ij}$ by $\det(g)$, so the image in $\IP^5(\IC)$ is left unchanged. One also sees that the map does not depend on the normalization of $\Omega$ and that the Grassmannian is given by the Pl\"ucker quadric

\beq
 \p_{12}\, \p_{34} - \p_{13}\, \p_{24} + \p_{14}\, \p_{23}\= 0~
\notag\eeq

which the moduli space maps into.

The equation \eqref{eq:LambdaPlane} characterising attractor points is more commonly written as
\beq
Q\;= -2\,\text{Im}\Big(\,\overline{Z}_\g(\vph_{*})\;\Pi(\vph_{*})\Big).
\label{eq:attractorequationsforcharge}\eeq
Given $\g \in H_3(X,\IZ)$, one can solve the Picard-Fuchs equation and the attractor equations numerically and find the attractor point $\vph_{*}(\g)$ that makes $\g$ the $(2,1)$ part and $(1,2)$ part of~$\G$ vanish to high
precision.  Conversely, at an arbitrary point $\vph_{*}$, we can solve Eqs.~\eqref{eq:attractorequationsforcharge} for the charges $Q$ for which $\vph_{*}$  
would be an attractor point. By a simple computation we find that the charges 
are given by
 \beq
 Q \=\Big(\frac{\p_{14}}{\p_{34}}\, p^0+\frac{\p_{31}}{\p_{34}}\, p^1,\;\frac{\p_{24}}{\p_{34}}\, p^0 +\frac{\p_{32}}{\p_{34}}\,p^1,\; p^0,\; p^1\Big)^T~.
\label{eq:Qatarbitraryphi}\eeq
However, this charge vector $Q$ will, generically, not be integral. 

At a rank one attractor, the first two components of $Q$ are integral for some choice of $p_0$ and $p_1$ unique up to an overall scale. However, at a rank two attractor, we require that each of the four ratios in Equation~\ref{eq:Qatarbitraryphi} are rational. This is much more constraining and explains the scarcity of rank two attactors.

In other words, the rank two attractors are precisely the $\IQ$-rational points on the moduli space in $\text{Gr}(2,\IR^4)$.

We now concentrate on the case of a rank two attractor point. It follows 
from the above discussion that
\beq
\L \otimes\IC \= H^{3,0}\oplus H^{0,3}~.
\notag\eeq
\vskip10pt
The lattice $\L^\perp\subset H^3(X,\IZ)$ that is orthogonal to $\L$ under the symplectic product $\S$ has the property that
\beq
\L^\perp\otimes\IC \= H^{2,1}\oplus H^{1,2}~.
\notag\eeq

The fact that the spaces $H^{3,0}{\oplus} H^{0,3}$ and $H^{2,1}{\oplus} H^{1,2}$ are spanned by lattice planes in this way is very remarkable.\footnote{ It should be noted that the elements of $\Lambda^{\perp}$ lead the same attractor point as those in $\Lambda$. However, the central charge at the attractor point vanishes for any charge in $\Lambda^{\perp}$ because one ends up integrating a $(2,1)$+$(1,2)$ form againts a $(3,0)$ form. As a result, the ``black hole" will have zero mass. We will have more to say about this in the conclusion.} We note that the the sum of these two lattices
\[ \L \oplus \L^\perp \subset H^3(X,\IZ)\]
has finite index, so after extension of coefficients to $\IQ$ we obtain an isomorphism
\[ H^3(X,\IQ)=\L_{\IQ} \oplus \L_{\IQ}^\perp\, ,\]
which then can interpreted as saying that at a rank two attractor point we have a {\em splitting} 
of the of the Hodge structure $H^3(X,\IQ)$ into two sub-Hodge structures, where $\L_{\IQ}$ has 
Hodge numbers  $(3,0), (0,3)$ and $\L^\perp_{\IQ}$ with Hodge numbers $(2,1), (1,2)$, so that on the level of Hodge vectors we have:
\[ (1,1,1,1)=(1,0,0,1)+(0,1,1,0) .\]

We note that the real plane $\L \otimes \IR$ spanned by $\text{Re}\,\O$ and
$\text{Im}\,\O$ can be identified with the one dimensional complex space 
$H^{3,0}$ and similarly $\L^\perp \otimes \IR$ can be identified with the
one-dimensional complex space $H^{2,1}$. So associated with the splitting there
are also two one-dimensional complex tori
\[T_\L=H^{3.0}/\L,\;\;\;T_{\L^\perp}=H^{2,1}/\L^\perp .\] 
In fact, the Hodge structure $\L^{\perp}$ is the {\em Tate-twist} of a the 
Hodge structure of weight one of the elliptic curve $E:=T_{\L^\perp}$:
\[ H^1(E,\IQ)(-1) \cong \L_{\IQ}^\perp \subset H^3(X,\IQ) .\]
 
In the seminal paper~\cite{Moore:1998pn}, G. Moore speculated on the arithmetical nature of the parameter values $\varphi_*$ of attractor points and the associated varieties $X_{\varphi_{*}}$. He analysed these in detail for families related to $K3$-surfaces. Furthermore, he identified three examples of attractor points in one-parameter models. The varieties in question are Fermat points and lead to (apparent) singularities of the associated Picard-Fuchs equation. Attractor points and associated lines of marginal stability on the mirror quintic famliy have been investigated in papers by Denef et al.~in~\cite{Denef_2000,Denef_2001}.

Below we describe how attractor points of rank two can be found by an arithmetic method. 

\subsection{The arithmetic of $X_{\varphi_{*}}$}\label{sec:Arithmetic}
\vskip-10pt     
Any projective variety $X$ defined over $\IQ$ can be defined by polynomial
equations with integral coefficients. For any prime $p$ we may then
ask how many solutions these equations have over $\IF_{\!p^r}$, the field with
$p^r$ elements. Let $N_r$ be this number. These numbers are collected into 
the generating function
\beq
\z(T)\=\exp\left(\sum_{r=1}^{\infty}N_r \frac{T^r}{r}\right)~,
\notag\eeq
known as the {\em Artin-Weil Zeta Function}. Of course, it also depends on $p$,
but we suppress this dependence from the notation.
The form of $\z(T)$ is governed by the (now proved) Weil Conjectures. 
We will not state these in full, but simply note that the first of these asserts that $\z(T)$ is a rational function of $T$. If the reduction modulo $p$ of $X$ is
smooth of dimension $n$, $\z(T)$ has a factorisation of the form
\beq
\z(T) \=\frac{R_1 R_3 \ldots R_{2n-1}}{R_0 R_2 \ldots R_{2n}}~,
\label{eq:superdeterminant}\eeq
where the polynomials $R_k$, $k{\=}0,1,\ldots,2n$ have a cohomological origin. We pause to explain this in rather greater detail and to recall the basic facts pertaining to the Frobenius map. 

For $c$ an integer, recall Fermat's Little Theorem that 
\beq
c^p\= c \mod p~.
\notag\eeq
So if we think of $c$ as a number in $\IF_{\! p}$ we have $c^p{\=}c$. If however $c$ is in a higher field $\IF_{\!p^r}$ then $c^p{\,\neq\,}c$, in general, since the analogous identity is $c^{p^r}{\=}c$. Now take $c_1$ and $c_2$ to be numbers in $\IF_{\! p^r}$, for some $r$, and note the identity
\beq
(c_1 + c_2)^p \= c_1^p + c_2^p~,
\notag\eeq
since all the intermediate terms in the binomial expansion are divisible by $p$. 

Suppose now that a manifold is defined by a polynomial
\beq
F(x) \= \sum_m c_m x^m
\label{eq:PolyF}\eeq
where we use a multi-index notation and $x^m{\=}x_1^{m^1}\ldots x_n^{m^n}$. Let us further suppose that the coefficients $c_m$ are in $\IF_{\! p}$, while the coordinates $x$ are in some higher field $\IF_{\! p^r}$. Then we have
\beq\begin{split}
F(x)&\= 0\\[3pt]
\Rightarrow\quad F(x)^p&\=0\\[3pt]
\Rightarrow\quad F(x^p)&\= 0~.
\end{split}\notag\eeq
The map $x\to x^p$ is the \emph{Frobenius map}, which we shall denote by $\Frob$. It would be more correct to denote the map by $\Frob_p$, but we shall drop the suffix $p$ in the following. What we have seen is that $\Frob$ is an automorphism that every manifold defined over $\IQ$ has. The fixed points of the map are of interest. These correspond to the points for which
\beq
x^p\= x
\notag\eeq
and this relation picks out the the points that are defined in $\IF_{\! p}\subset\IF_{\! p^r}$. So another way to look at $N_1$ is as the number of fixed points of the Frobenius map; more generally $N_k$ counts the number of fixed points of $\Frob^k$. It can also be shown that the Frobenius map generates the Galois group of the polynomial \eqref{eq:PolyF}. 
If suitable cohomology groups are defined, then the action of $\Frob$ extends to cohomology. It was Dwork~\cite{Dwork1960Rationality} who showed that the $\z$-function is a rational function which decomposes as in \eqref{eq:superdeterminant} by showing that the $\z$-function is a superdeterminant, though Dwork did not use this term, which decomposes into factors corresponding to the different cohomology groups with
\beq 
R_k(T)\= \det(1-T\,\Frob_k^{-1}) \in \IZ[T]~,\qquad\Frob_k:~H^k(X) \longrightarrow H^k(X)~,
\notag\eeq
where $H^k$ can be any Weil-cohomology, for example $\ell$-adic cohomology, ($\ell \neq p$).  In particular, the degree of $R_k$ is equal to the $k$-th Betti-number $b^k$ of the complex variety defined by $X$. A textbook account is given in~\cite{KoblitzPadicNumbers} and one in the style of the present work is given in~\cite{Candelas:2007mb}, which also gives more detailed references to the original literature.

For the situation of Calabi-Yau threefolds with $h^{21}=1$ considered here,
$\z(T)$ is further constrained and assumes the~form
\beq
\z(T)\= \frac{R(T)}{(1-T) (1-p T)^{h^{11}} (1-p^2 T)^{h^{11}} (1-p^3 T)}
\notag\eeq
The denominator in this expression gives the form of the product $R_0R_2R_4R_6$, while, in the numerator,
      the factors $R_1$ and $R_5$ are trivial, corresponding to the fact that $b^1{\=}b^5{\=}0$, so we are left with $R_3$ and we henceforth dispense with the suffix. The polynomial $R(T)$ has integer coefficients and is of degree four if the reduction mod $p$ of $X$ is smooth, and we will refer to it as the {\em Frobenius polynomial}. It is of the form
\beq
R(T)\=1 + a T + b p T^2 + a p^3 T^2 + p^6 T^4~,
\notag\eeq
and so is determined by two integers $a$ and $b$, that depend on $p$ and, of course, the manifold~$X$. When
the manifold $X_\vph$ lies in a family they depend on the parameter $\vph$.
 
Now, if $X=X_{\vph_*}$ is a rank two attractor variety, the third cohomology group {\em splits as a Hodge structure}:
\[H^3(X,\IQ) =\L_{\IQ} \oplus \L^\perp_{\IQ}.\] 
By the Hodge Conjecture, such a splitting is supposed to have a geometrical 
origin. To be more precise, let 
\[\sigma: H^3(X,\IQ) \to H^3(X,\IQ)\]
be the projection $(\sigma \circ \sigma=\sigma)$ with image $\L_{\IQ}$ and kernel $\L^\perp_{\IQ}$. Writing $H^3:=H^3(X,\IQ)$, the
element $\sigma$ can be considered as an element of the space
\[
\text{Hom}(H^3,H^3)\= H^{3*}\otimes H^3 \= H^3 \otimes H^3\; \subset\; H^6(X{\times}X,\IQ)~,
\]
where we used Poincar\'e duality and the K\"unneth-formula. In fact, as $\sigma$
is a morphism of Hodge structures, it can be checked that 
\[
 \sigma \in H^{3,3}(X{\times} X,\IQ)~, 
\]
which, according to the Hodge Conjecture, can be represented by a $3$-cycle $S$
on the product space $X {\,\times\,} X$. 

If the cycle $S$ is defined over $\IQ$, this gives a splitting of the {\em $\IQ$-motive} $H^3(X)$ into two rank two $\IQ$-submotives (we will not give a formal definition of a motive, one can think of this, informally, as an algebraically defined part of the cohomology). As a consequence, the cycle $S$ induces a similar decomposition on any Weil-cohomology. In particular, the matrix of $\Frob$, expressed in a suitable basis, will appear in block-diagonal form and consequently its characteristic polynomial $R(T)$ {\em factors over $\IZ$} into two quadratic factors as 
\beq
R(T)\= (1 - \a pT + p^3T^2) (1 - \b T + p^3T^2)~.
\label{eq:factorisationR}\eeq
The first factor comes from $H^{2,1}\oplus H^{1,2}$ and there is an `extra' factor of $p$ that accompanies the coefficient $\a$. This corresponds to the
Tate-twist refered to above and has the effect that the first factor can be rewritten as
\beq
1 - \a (pT) + p (pT)^2~,
\notag\eeq
which has the form of the numerator of the $\z$-function for an elliptic curve.
In fact, this elliptic curve is just $\cE^\perp{\=}T_{\L^\perp}$, which is defined over 
$\IQ$ if $S$ is, and the polynomial
\beq
1 - \a T + p T^2~,
\notag\eeq
is identified with the factor corresponding to $H^1$ of this elliptic curve. 

The second factor has the form of the numerator of the $\z$-function of a rigid Calabi-Yau manifold; the torus $T_{\L}$ cannot be expected to be defined over $\IQ$ or even over a number field.

\begin{table}[!t]
\begin{center}
\renewcommand{\arraystretch}{1.3}
\begin{tabular}{| >{\footnotesize$~} c <{~$} | >{~\footnotesize} l <{~} 
| >{\centering\footnotesize$}p{1in}<{$} |>{\centering\footnotesize $}p{3in}<{$} |}\hline
\multicolumn{4}{|c|}{\vrule height 13pt depth8pt width 0pt \small $p=19$}\tabularnewline[0.5pt] \hline
\vph & smooth/sing. & $singularity$ & R(T)\tabularnewline[0.5pt] \hline\hline
1	 &	singular  &	1&	(1-p T) (1-20 T+p^3 T^2)
\tabularnewline[0.5pt]\hline				
2	 &	smooth    &	          &	1+4 p T+2 p T^2+4 p^4 T^3+p^6 T^4
\tabularnewline[0.5pt]\hline				
3	 &	smooth    &	          &	1-8 T+242 p T^2-8 p^3 T^3+p^6 T^4
\tabularnewline[0.5pt]\hline				
4	 &	smooth    &	          &	(1+4 p T+p^3 T^2)(1-60 T+p^3 T^2)
\tabularnewline[0.5pt]\hline				
5	 &	smooth    &	          &	(1+4 p T+p^3 T^2)(1-60 T+p^3 T^2)
\tabularnewline[0.5pt]\hline				
6	 &	smooth    &	          &	1+8 T-318 p T^2+8 p^3 T^3+p^6 T^4
\tabularnewline[0.5pt]\hline				
7	 &	smooth    &	          &	1-44 T-238 p T^2-44 p^3 T^3+p^6 T^4
\tabularnewline[0.5pt]\hline				
8	 &	smooth    &	          &	(1-2 p T+p^3 T^2)(1-80 T+p^3 T^2)
\tabularnewline[0.5pt]\hline				
9	 &	smooth    &	          &	(1+4 p T+p^3 T^2)(1-160 T+p^3 T^2)
\tabularnewline[0.5pt]\hline				
10	 &	smooth    &	          &	1+12 T+562 p T^2+12 p^3 T^3+p^6 T^4
\tabularnewline[0.5pt]\hline				
11	 &	smooth    &	          &	(1+4 p T+p^3 T^2)(1-140 T+p^3 T^2)
\tabularnewline[0.5pt]\hline				
12	 &	smooth    &	          &	1+12 T+82 p T^2+12 p^3 T^3+p^6 T^4
\tabularnewline[0.5pt]\hline				
13	 &	smooth    &	          &	1+178 T+1082 p T^2+178 p^3 T^3+p^6 T^4
\tabularnewline[0.5pt]\hline				
14	 &	smooth    &	          &	1+12 T-158 p T^2+12 p^3 T^3+p^6 T^4
\tabularnewline[0.5pt]\hline				
15	 &	smooth    &	          &	1+42 T-2 p^2 T^2+42 p^3 T^3+p^6 T^4
\tabularnewline[0.5pt]\hline				
16	 &	singular  &	\frac{1}{25}&	(1-p T) (1+76 T+p^3 T^2)
\tabularnewline[0.5pt]\hline				
17	 &	singular  &	\frac{1}{9}&	(1-p T) (1-20 T+p^3 T^2)
\tabularnewline[0.5pt]\hline				
18	 &	smooth    &	          &	1-54 T+322 p T^2-54 p^3 T^3+p^6 T^4
\tabularnewline[0.5pt]\hline				
\end{tabular}
\capt{5.0in}{tab:p19}{The $R$-factors for $\vph{\;\in\;}\IF_{19}$. Note the factorisations into two quadrics for the five values $\vph{\;=\;}4,5,8,9,11$.}
\end{center}
\end{table}

The arithmetic information of the Frobenius transformations for
various $p$  can conveniently be packed into what is called a
 {\em Galois representation}
\[ 
\rho: \text{Gal}(\overline{\IQ}/\IQ) \to \text{GL}_4(\IQ_{\ell})
\]
that maps a Frobenius element at $p$ to the matrix $\Frob$.
In case of a splitting, we end up with two 2-dimensional representations
\[ 
\rho: \text{Gal}(\overline{\IQ}/\IQ) \to \text{GL}_2(\IQ_{\ell})~,
\]
which is the subject of {\em Serre's conjecture\/}\cite{AST_1975__24-25__109_0,serre1987}. This asserts that such representations are attached to modular forms of specific weight and conductor and can as such be seen as a generalisation of the Taniyama-Weil conjecture, which, following on from the work of Wiles~\cite{10.2307/2118559} and Wiles and Taylor~\cite{10.2307/2118560},  was proved by Breuil, Conrad, Diamond and Taylor~\cite{10.2307/827119}. Further work by Taylor, and many others, led to a complete proof of the Serre conjecture by Dieulefait~\cite{dieulefait2009modularity}, Khare and Wintenberger~\cite{Khare2009article1,Khare2009article2} and Kisin~\cite{kisin2007}. As a result of this important development in number theory, there is now very good arithmetic control over 2-dimensional Galois representations coming from geometry. Gouv\^{e}a and Yui~\cite{GuveaYui} have shown the modularity of rigid Calabi-Yau threefolds defined over $\IQ$ can be derived from it. But also for non-rigid varieties defined over $\IQ$, that split in the above way the modularity has been {\em proved}, which means that the coefficients $a_p$ and $b_p$ are Fourier coefficients of cusp forms of weight $2$ and $4$ for some congruence group $\G_0(N)$ of the modular group. So, for an attractor point of rank two, we expect a factorisation into two quadratic factors, giving rise to modular forms of weights $2$ and $4$. 

If the variety $X$ (or the cycle $S$ producing the splitting) is {\em not\/} defined over $\IQ$ but over some number field $\IK$, the situation is more complicated, as we are then dealing with representations of $\text{Gal}(\overline{\IQ}/\IK)$. But the Chebotar{\"e}v density theorem~\cite{Stevenhagen1996} implies that in such cases one still has such a splitting of $R(T)$ for infinitely many and in fact a positive fraction of primes $p$. In the case of totally real fields one in general expects  Hilbert modular forms. However, in the cases we encounter
      here, we find classical modular forms for $\G_1(N)$.

\subsection{The strategy}\label{sec: Strategy}
\vskip-10pt
We consider a $1$-parameter family $X_\vph$ of Calabi-Yau threefolds with $h^{2,1}=1$, defined by a polynomial 
       equation
\[ 
P(x,\vph)\= 0
\]
with integral coefficients. In the light of the previous discussion, the strategy to find rank two attractor points $\vph_*$ is now quite clear: we compute the polynomial 
\[
R(T)\= 1 + a T + b p T^2 + a p^3 T^2 + p^6 T^4 
\] 
for many $p$ and $\varphi$ and look for persistent factorisations into a product of two quadratic factors. By this we mean that the factorisations occur whenever $\vph$ is the root of some algebraic equation $G(\vph)$ defined over $\IQ$, without any reference to a particular prime.
 
For this to be feasible, we need an efficient way to compute $R(T)$. The coefficients $a$ and $b$ can, in principle, be determined by directly counting the number of points of $X$ over~$\IF_{\! p^r}$, in fact it is sufficient to count points over $\IF_{\! p}$ and $\IF_{\! p^2}$. This however quickly becomes impractical as $p$ is increased. Sometimes even for small $p$, it is onerous to count the $\IF_{\! p^r}$-points of a manifold, for example if $X$ is defined as a quotient by a group, since these `points' are then group-orbits that are defined over $\IF_{\! p^r}$, not the orbits of group-invariant points, and there are frequently orbits without any points, for example. 

Fortunately, there are much better ways to compute $R(T)$. It was discovered by Dwork and developed further by Lauder~\cite{lauder2004} that the $\z$-function can be calculated from a $p$-adic computation of the periods, using the Picard-Fuchs equations. This goes under the name {\em deformation method}. A more detailed discussion of this fascinating process,  pertaining to the $\z$-function of one-parameter families of Calabi-Yau manifolds with a point of maximal unipotent monodromy which is taken as expansion point, may be found in~\cite{LocalZetaFunctionsI}.

Using these methods, the quantities $R(T)$ were calculated, in~\cite{LocalZetaFunctionsI}, for $\vph{\=}1,...,p-1$ for the 500 values $p{\=}5,...,3467$, for a family first described by H.~Verrill~\cite{verrill1996, verrill2004sums} and that is number 34 in the AESZ list~\cite{almkvist2005tables}. In the following, we will often refer to this manifold as AESZ34.

For example for $p{\=}19$ we have \tref{tab:p19} and we see that $R(T)$ factors in the form indicated for the five values $\vph{\=}4,5,8,9,11$. At the conifold points $R(T)$ degenerates to a cubic, and factorises into a linear factor and a quadric. These cases are also very interesting, not least because they also exhibit modular behaviour  and can be thought of as corresponding to massless black holes. We will not however pursue the factorisations due to the conifolds~here.

We do not want to assert that every factorisation of the form \eqref{eq:factorisationR} corresponds to a rank two attractor point. However, there is a form of converse statement that we do expect. Let us suppose that, as conjectured by Moore~\cite{Moore:1998pn}, the rank two attractor points are algebraic, in the sense that there is a polynomial $G(\vph)$ with rational (so integer) coefficients, whose roots are the rank two attractor points. If this is so, then it makes sense to reduce $G(\vph)$ mod $p$ and the roots will exist in $\IF_{\! p}$ for some, and in fact for infinitely many, $p$. For these $p$ we expect $R(T)$ to factorise. By assuming that there is a single polynomial $G(\vph)$, whose roots are the rank two attractor points, we are assuming, not only that the rank two attractor points are algebraic, but also that there are finitely many such points. These comments are made for the case that there is one parameter. If there are more parameters, we would expect the rank two attractor points to lie on algebraic submanifolds of the parameter space.

\begin{figure}[!t]
\begin{center}
\includegraphics[keepaspectratio,height=9.5cm]{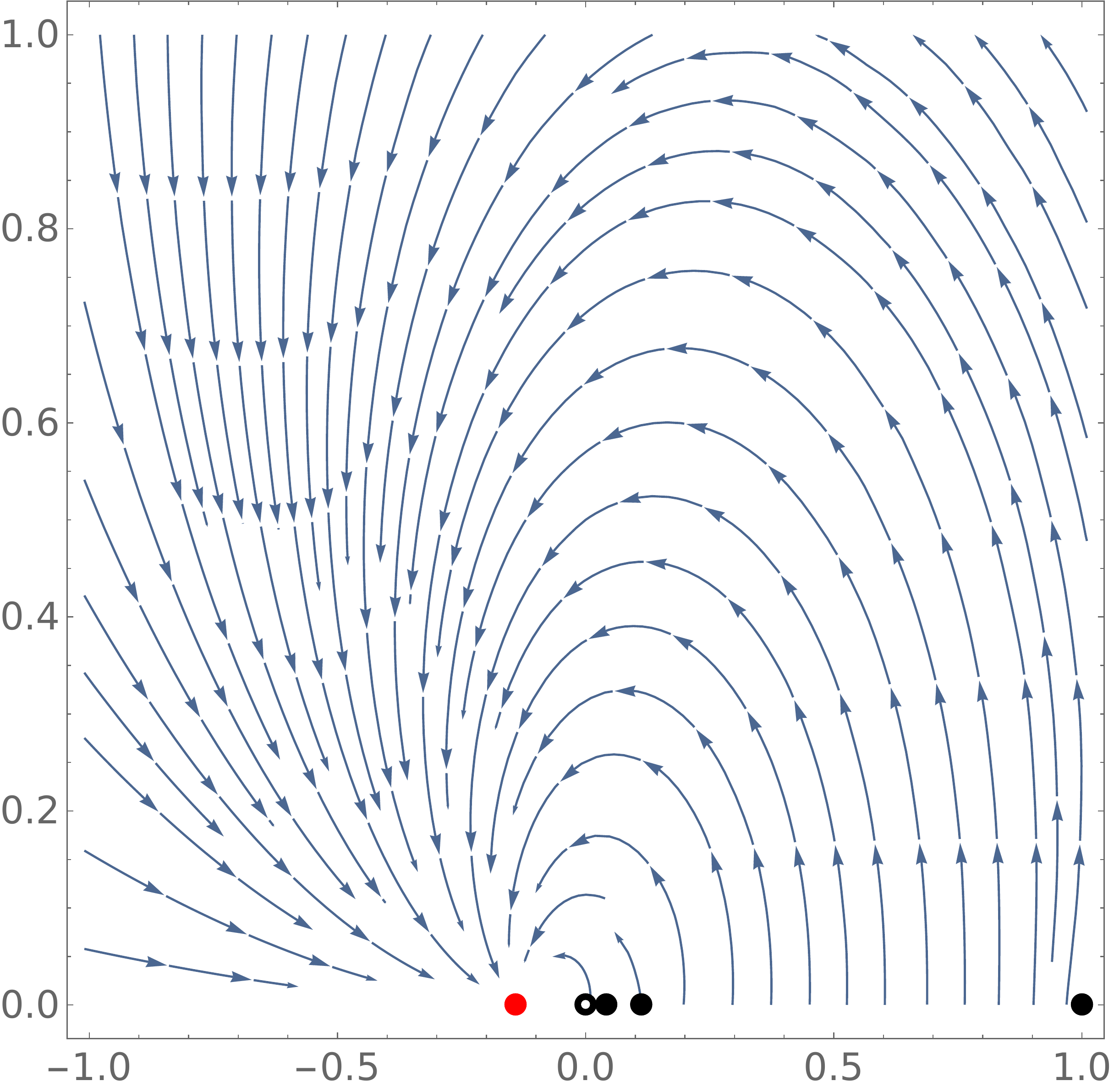}
\vskip12pt
\includegraphics[keepaspectratio,height=9.5cm]{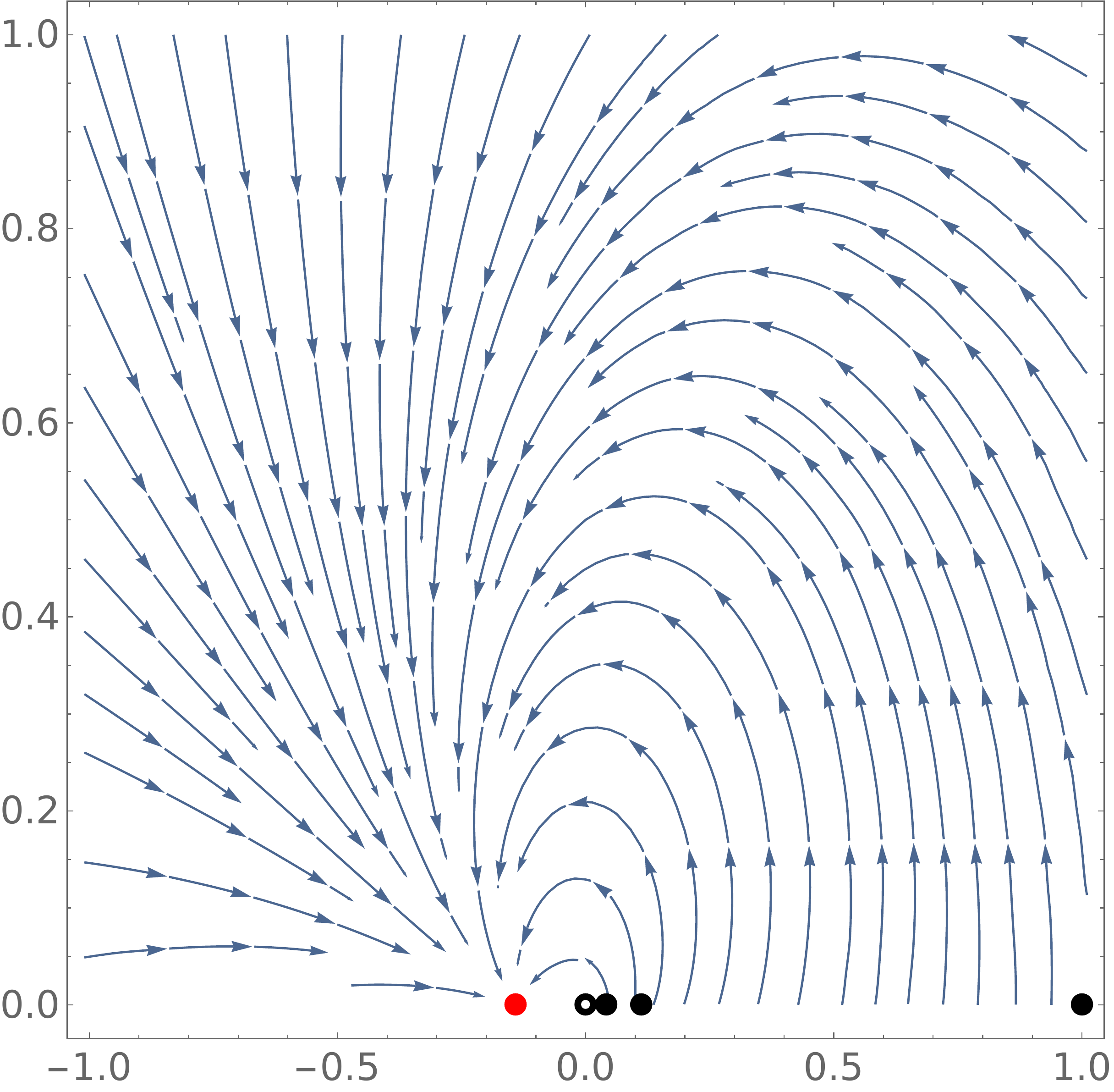}
\vskip7pt
\capt{6.3in}{fig:flowplotsfor-1/7}{The flows for $\varphi=\varphi(\rho)$ for the charges $Q=(0,0,2,1)$ (above) and $Q=(-4,15,5,0)$ (below) leading to attractor point at $\varphi{\;=\;}-1/7$. The point of maximal unipotent monodromy at $\varphi=0$ is indicated by a hollow black dot while the solid black dots represent conifold singularities.}
\end{center}
\end{figure}

In \SS8 of~\cite{Moore:1998pn} Moore has made several conjectures as to the arithmetical nature of  the attractor points, making a distinction between strong and weak versions of these conjectures, according to whether they apply to all attractor points, or only to the rank two attractors. In particular, the Attractor Conjecture \SS8.2.2 of~\cite{Moore:1998pn} asks if rank two attractor points $\vph_*$ are algebraic, and hence whether the corresponding varieties $X_{\vph_*}$ are defined over a number field. In \SS3.6.2 of~\cite{Moore:2004fg} it is stated that (according to Nori) this actually follows from the Hodge conjecture, but no details are given. Indeed, for any projective family of varieties over a (possibly higher dimensional) base $S$ defined over a number field, the locus of points $s \in S$, where $X_s$ carries an algebraic cycle in a specific homology class, is an algebraic subvariety of $S$ that is defined over a number field. Taking the union over all possible homology classes gives a countable union of such sub-varieties\footnote{Independent of the Hodge conjecture, it follows from general results of Cattani, Deligne and Kaplan~\cite{cattanidelignekaplan} that the locus of Hodge cycles is a countable union of algebraic varieties in $S$, and if we knew that ``Hodge cycles are absolute Hodge cycles'', then these varieties would be algebraic varieties defined over a number~field.}.

As a rank two attractor point can be seen as a special kind of Hodge class in $X_{\vph_*} {\times\,} X_{\vph_*}$, the Hodge conjecture implies that the rank two attractor points belong to a countable union of algebraic sub-varieties in $S$, that are defined over a number field. So, in particular, for a one-parameter family (always assuming that {\em not all} points are rank two attractor points), the Hodge conjecture implies that the parameter-values $\vph_*$ for rank two attractor points are {\em algebraic}. Based on our searches, we are tempted to strengthen  Moore's Attractor Conjecture 8.2.2 and conjecture that the rank two attractor points are contained in an algebraic sub-variety defined over a number field, rather than a countable union of them. For a one-parameter family this would mean that the set of rank two attractor points is {\em finite} and hence to be found among the solutions to a single polynomial equation
\[ 
G(\vph)\= c_n \vph^n+c_{n-1}\vph^{n-1}+\ldots+c_1 \vph+c_0
\] 
where the coefficients $c_k$ are integers.

The crudest summary of the tables produced in~\cite{LocalZetaFunctionsI} is to count how many times $R(T)$ factorises in the indicated way for each prime $p$. We have just seen that for $p{\=}19$ it factorises 5 times. This leads to the two plots in \fref{fig:AESZ34andQuintic}. The first gives the data for the manifold AESZ34, while the second gives the analogous data for the mirror of the quintic threefold and is presented for comparison. Clearly $R(T)$ for AESZ34 factorises much more often than for the mirror quintic. Notice also that while for the mirror quintic there are many primes for which $R(T)$ does not factorise, for AESZ34 the polynomial $R(T)$ factorises at least once for each $p$. This suggests that, for AESZ34, the polynomial $G(\vph)$ has a linear factor\footnote{We are grateful to Noam Elkies for this elementary but important observation.}, since a linear equation
\beq
c_1 \vph + c_0 \= 0
\notag\eeq
has a solution mod $p$ for all $p$, apart from primes that divide $c_1$. 

By looking first at the primes for which $R(T)$ factorises precisely once, and using a variant of the Chinese Remainder Theorem, or by simply performing a computer search over integers $c_0$ and $c_1$, we find that (apart from the case $p{\=}7$) the polynomial $R(T)$ always factorises~when
\beq
\vph\;=-1/7~.
\notag\eeq
\emph{[In\/ $\IF_{\!p}$, $\vph{\;=}-1/7$ is the integer that satisfies the relation $7\vph{\;+\;}1{\;=\;}0$. For $p{\;=\;}19$, for example, we have $7{\;\times\;}8{\;=}-1$, so $-1/7{\;=\;}8$ in $\IF_{\!19}$ and this indeed is one of the values for which factorisation of the desired form occurs in \tref{tab:p19}.]}

It is easy to check that, considered as a point of $\IC$, $\vph{\;=}-\frac17$ is indeed a rank two attractor point. By this, we mean that we solve the Picard-Fuchs equation around $\varphi=0$ and, by numerical integration, evaluate it at $\vph=-\frac17$ to 1000 decimal places. We then check that, the ratios in Equation~\ref{eq:Qatarbitraryphi} are rational to this precision. We will later, in \sref{sec:periodsandderivatives}, propose identities between the periods at $\vph=-\frac17$ and critical $L$-values that will also be verified to at least 1000 decimal places. Although not a proof, these observations leave little doubt that $\vph=-\frac17$ is indeed a rank two attractor.

\begin{figure}[!t]
\begin{center}
\includegraphics[width=\linewidth]{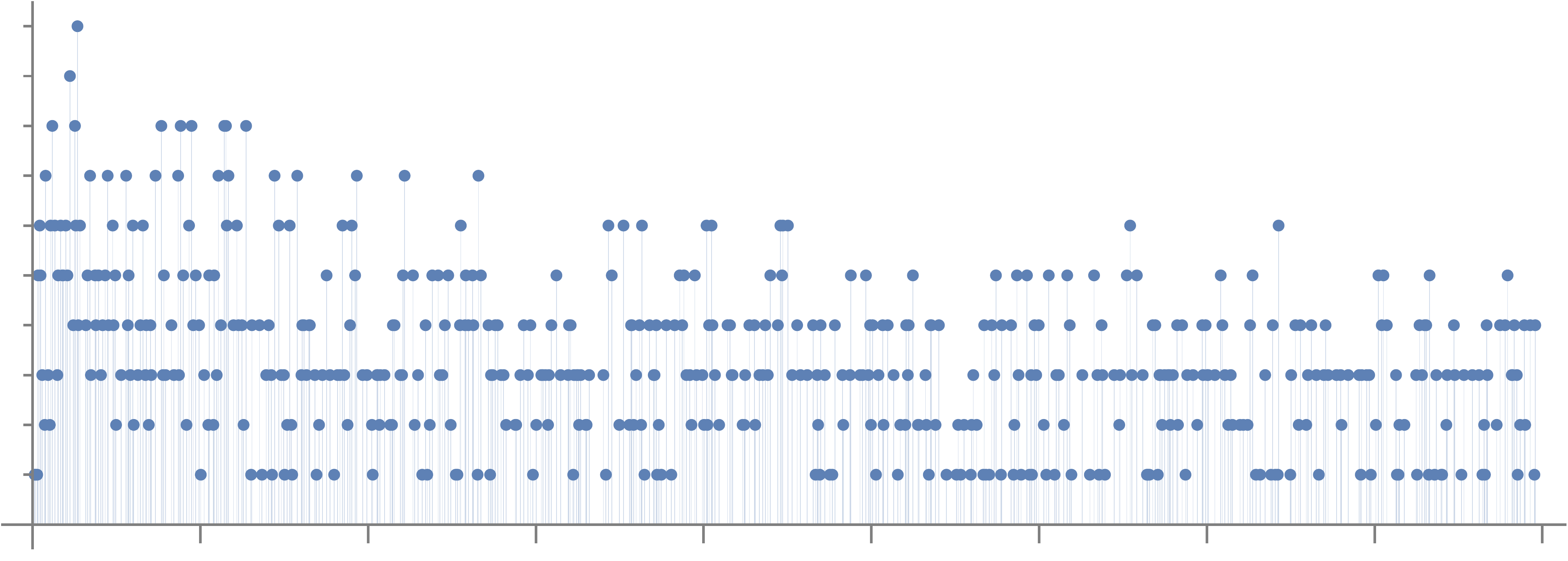}
\vskip0pt 
\place{-0.05}{0.54}{\scriptsize 1}
\place{-0.05}{0.75}{\scriptsize 2}
\place{-0.05}{0.96}{\scriptsize 3}
\place{-0.05}{1.16}{\scriptsize 4}
\place{-0.05}{1.37}{\scriptsize 5}
\place{-0.05}{1.57}{\scriptsize 6}
\place{-0.05}{1.78}{\scriptsize 7}
\place{-0.05}{1.99}{\scriptsize 8}
\place{-0.05}{2.20}{\scriptsize 9}
\place{-0.07}{2.41}{\scriptsize 10}
\place{0.73}{0.15}{\scriptsize 400}
\place{1.43}{0.15}{\scriptsize 800}
\place{2.10}{0.15}{\scriptsize 1200}
\place{2.80}{0.15}{\scriptsize 1600}
\place{3.49}{0.15}{\scriptsize 2000}
\place{4.19}{0.15}{\scriptsize 2400}
\place{4.89}{0.15}{\scriptsize 2800}
\place{5.59}{0.15}{\scriptsize 3200}
\place{6.29}{0.15}{\scriptsize 3600}
\place{6.6}{0.3}{$p$}
\vskip0pt
\includegraphics[width=\linewidth]{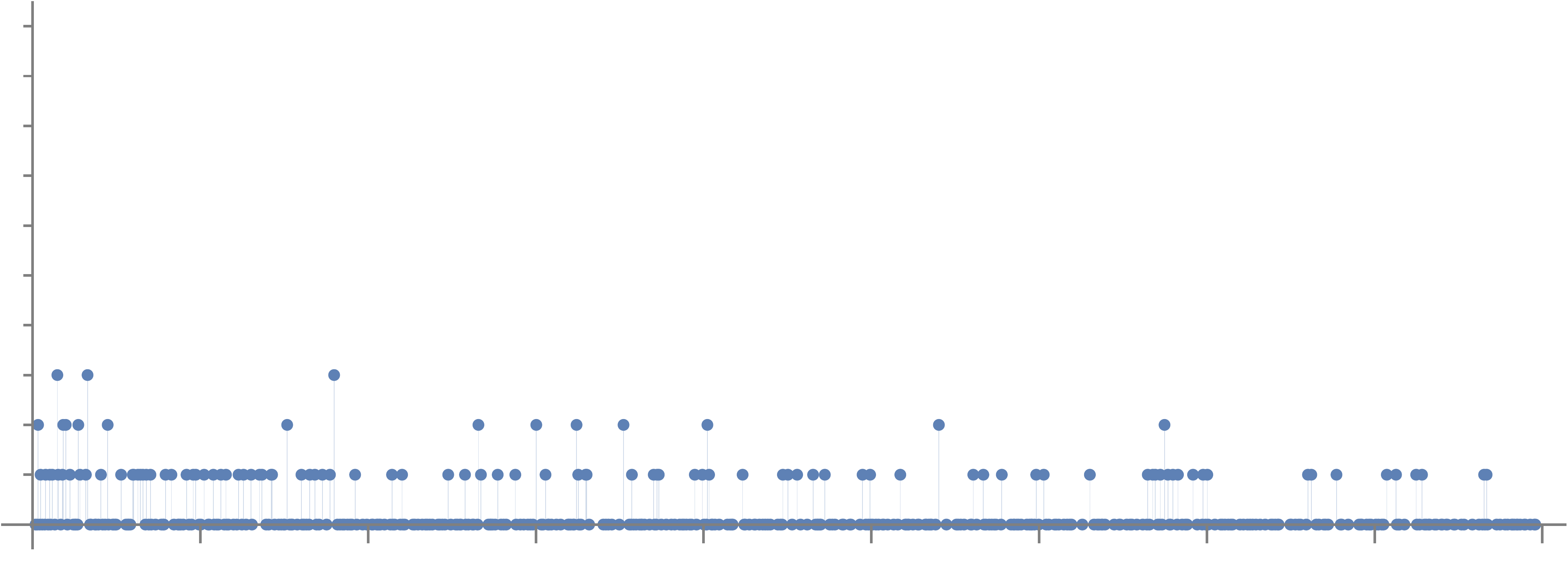}
\vskip0pt 
\place{-0.05}{0.54}{\scriptsize 1}
\place{-0.05}{0.75}{\scriptsize 2}
\place{-0.05}{0.96}{\scriptsize 3}
\place{-0.05}{1.16}{\scriptsize 4}
\place{-0.05}{1.37}{\scriptsize 5}
\place{-0.05}{1.57}{\scriptsize 6}
\place{-0.05}{1.78}{\scriptsize 7}
\place{-0.05}{1.99}{\scriptsize 8}
\place{-0.05}{2.20}{\scriptsize 9}
\place{-0.07}{2.41}{\scriptsize 10}
\place{0.73}{0.15}{\scriptsize 400}
\place{1.43}{0.15}{\scriptsize 800}
\place{2.10}{0.15}{\scriptsize 1200}
\place{2.80}{0.15}{\scriptsize 1600}
\place{3.49}{0.15}{\scriptsize 2000}
\place{4.19}{0.15}{\scriptsize 2400}
\place{4.89}{0.15}{\scriptsize 2800}
\place{5.59}{0.15}{\scriptsize 3200}
\place{6.29}{0.15}{\scriptsize 3600}
\place{6.6}{0.3}{$p$}
\capt{6in}{fig:AESZ34andQuintic}{The upper plot shows the number of factorisations into two quadrics as $\vph$ varies over each $\IF_p$, $7\leq p\leq3583$, for the manifold AESZ34. For comparison, the lower plot provides the same information for the mirror of the quintic which explains why it is difficult to find rank two attractor points on this family.}
\end{center}
\end{figure}

Encouraged by finding a linear factor of $G(\vph)$, we search for a quadratic factor
\beq
c_2\vph^2 + c_1\vph + c_0 \=0
\notag\eeq
and find that $R(T)$ always factorises when $\vph^2 - 66\,\vph + 1 {\=}0$ and so when
\beq
\vph \= \vph_{\pm} \= 33 \pm 8\sqrt{17}
\notag\eeq
exists in $\IF_{\!p}$. This occurs when 17 is a square mod $p$, and so, by quadratic reciprocity, when p is a square mod 17. 

\emph{ [Pursuing our example for $p{\=}19$, note that $17{\=}6^2$ in $\IF_{\!19}$ so $\vph_{\pm}{\=}4,\,5$ and the desired factorisations also occur for these values of $\vph$ in \tref{tab:p19}.]}

Again, if we take $\vph_{\pm}$ to be points in $\IC$, then it is straightforward to check numerically that these values correspond to rank two attractor points. These flow plots are presented in \fref{fig:flowplotsforphiplus1} and
\fref{fig:flowplotsforphiplus2}.

\begin{figure}[!p]
\begin{center}
\includegraphics[keepaspectratio,height=9.5cm]{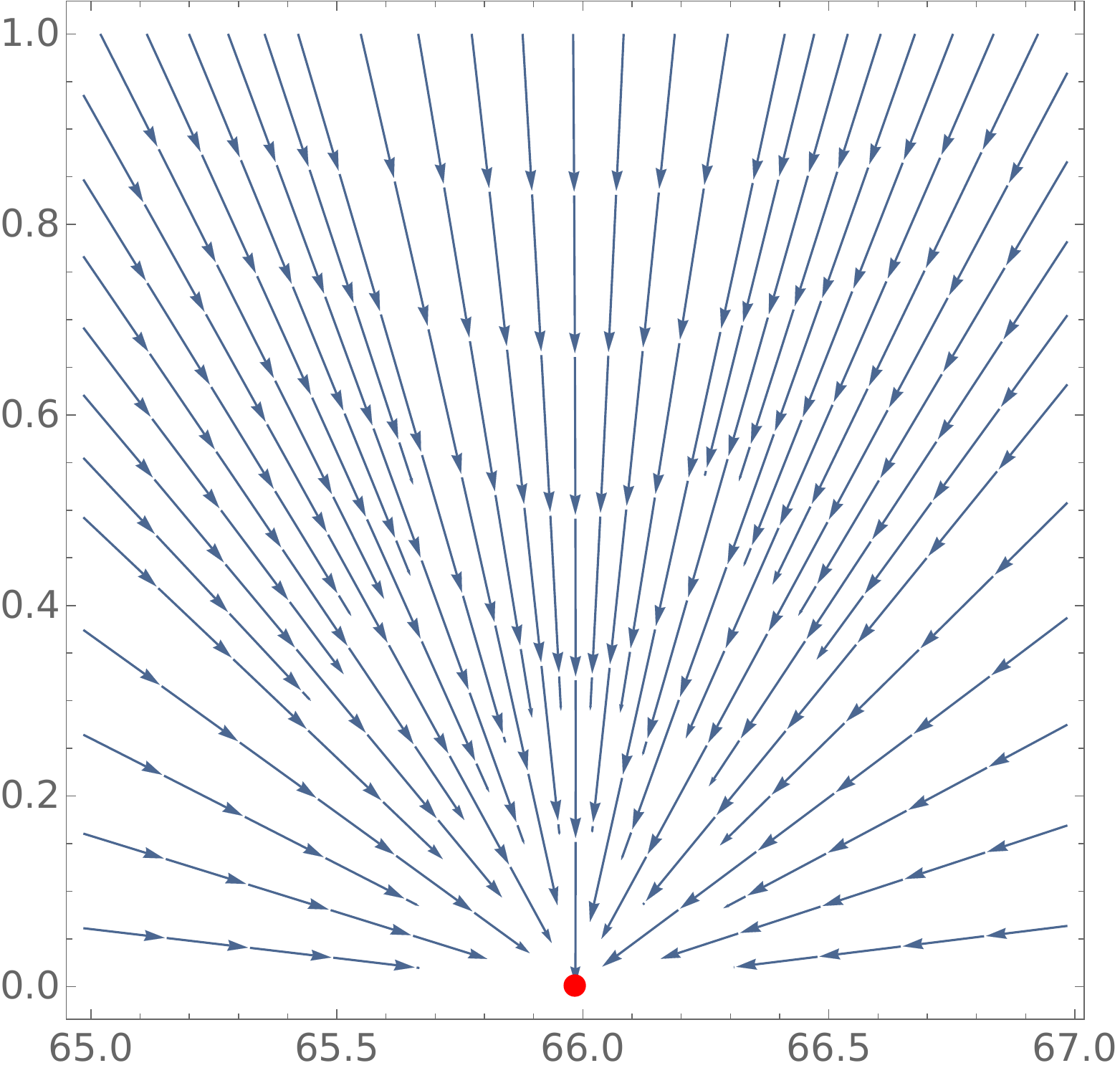}
\vskip12pt
\includegraphics[keepaspectratio,height=9.5cm]{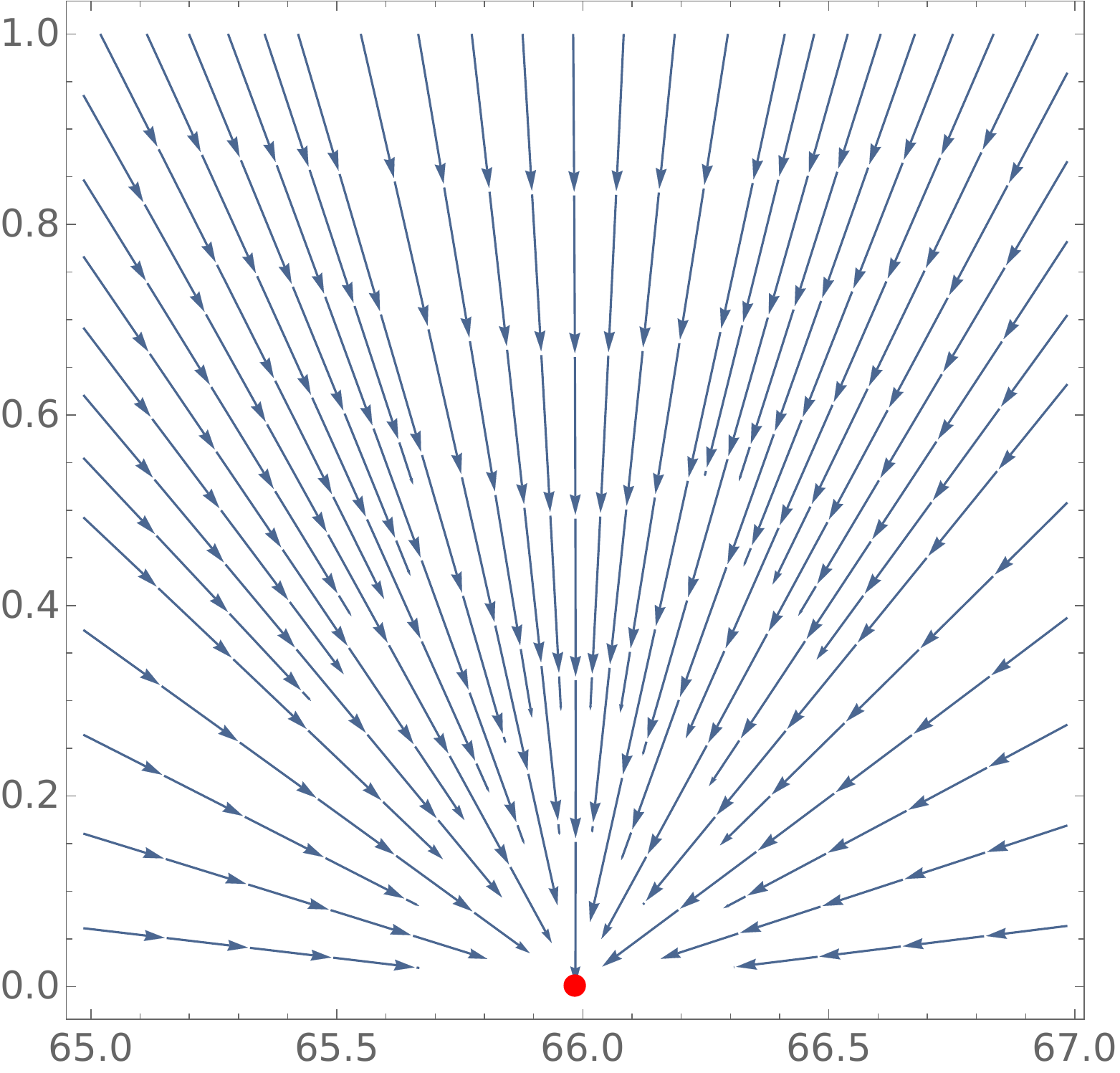}
\vskip7pt
\capt{5.5in}{fig:flowplotsforphiplus1}{The flows for $\varphi=\varphi(\rho)$ for the charges $Q=(4,-9,7,4)$ (above) and $Q=(4,-30,-30,-5)$ (below) leading to attractor point at $\varphi{\;=\;}33+8\sqrt{17}$}
\end{center}
\end{figure}
\begin{figure}[!p]
\begin{center}
\includegraphics[keepaspectratio,height=9.5cm]{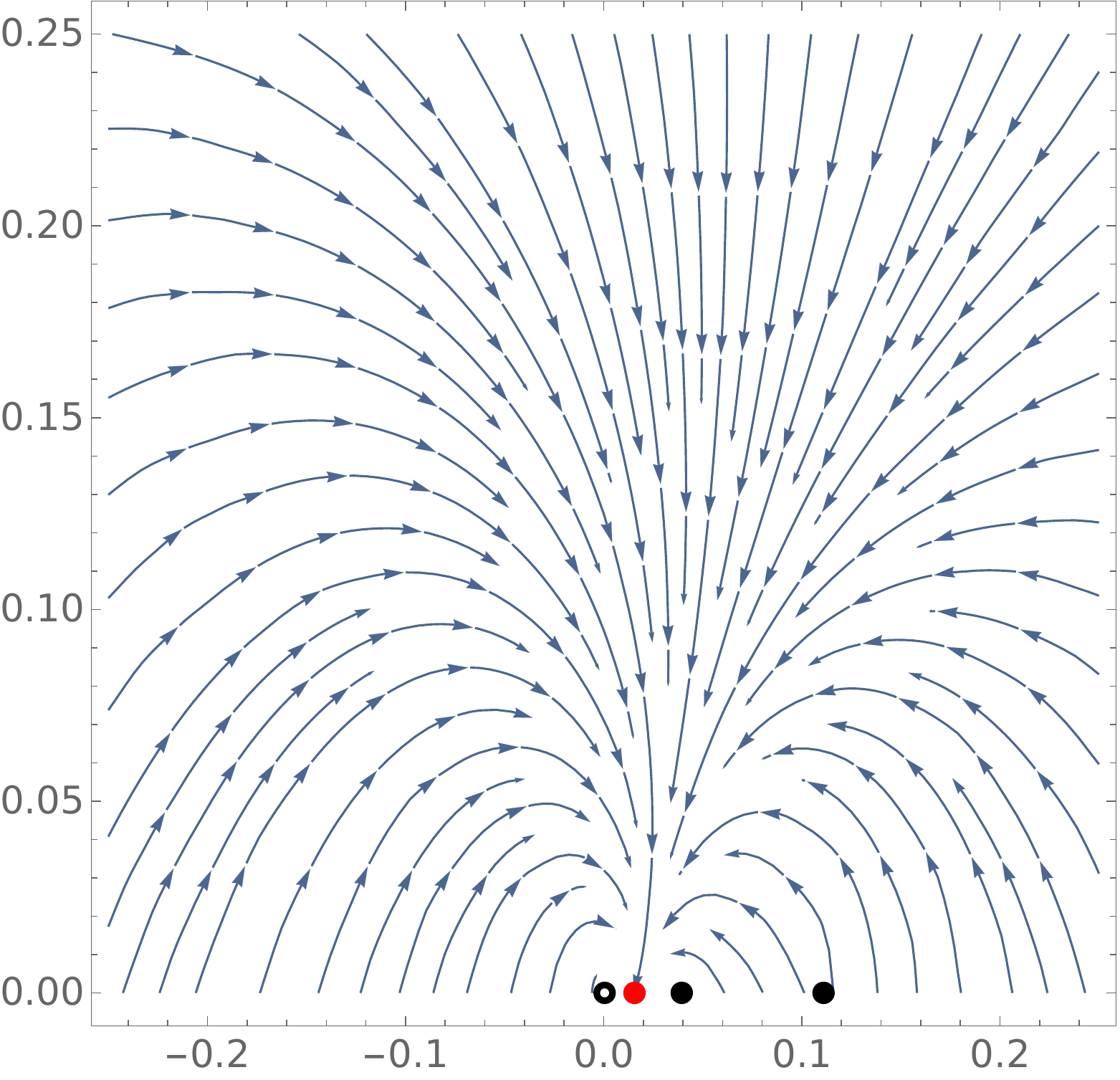}
\vskip12pt
\includegraphics[keepaspectratio,height=9.5cm]{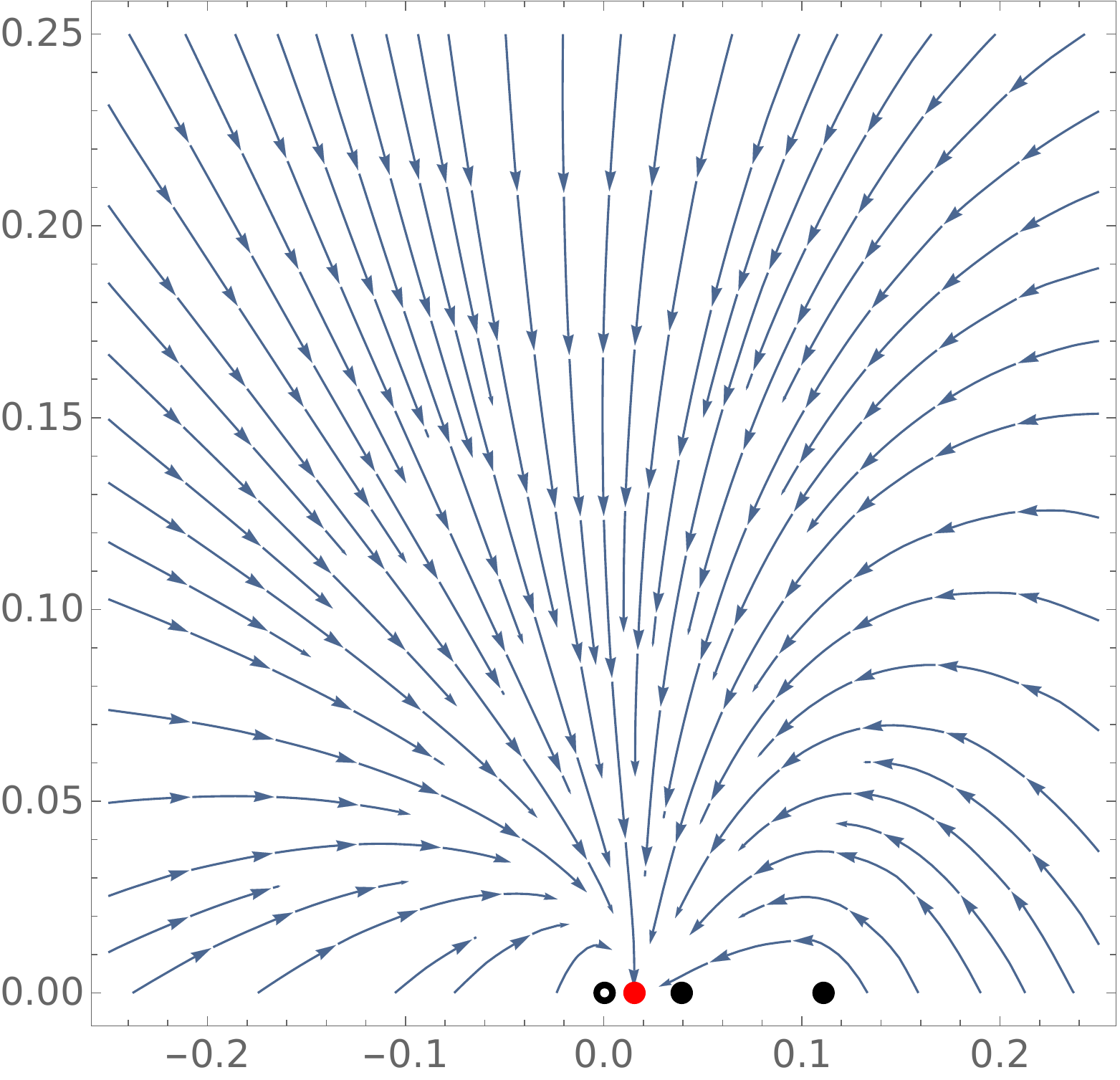}
\vskip7pt
\capt{5.5in}{fig:flowplotsforphiplus2}{The flows for $\varphi=\varphi(\rho)$ for the charges $Q=(-2,0,0,5)$ (above) and $Q=(0,3,1,0)$ (below) leading to attractor point at $\varphi{\;=\;}33-8\sqrt{17}$. The point of maximal unipotent monodromy at $\varphi=0$ is indicated by a hollow black dot while the solid black dots represent conifold singularities.}
\end{center}
\end{figure}

The tables of the Frobenius polynomials $R(T)$ contain much more information than that 
shown in \fref{fig:AESZ34andQuintic}. For example, let us consider the coefficients
$\a$ and $\b$ for the attractor points, as $p$ varies. For $\vph{\;=}-1/7$ we list primes 
$5\leq p\leq 137$. While for $\vph{\;=\;}33\pm 8\sqrt{17}$ we list primes $5\leq p\leq349$ 
such that 17 is a square mod $p$. A first remark is that $R(T)$ is the same for $\vph{\=}\vph_{\pm}$ so we need only present a single table for these parameter values.

For $\vph{\;=}-1/7$ we observe that the $\a$'s are the p$^\text{th}$ coefficients of a weight 2 modular form, with LMFDB designation {\bf 14.2.a.a}  for the group $\G_0(14)$. The coefficients $\b$ are similarly the p$^\text{th}$ coefficients of a weight four modular form, with designation {\bf 14.4.a.a}, also for $\G_0(14)$. 

The appearance of modular forms was anticipated by mathematicians, but for a physicist these have appeared, seemingly out of nowhere.

Conventions differ between references, so we pause to state these. We understand a modular form of weight $k$ to satisfy the relation
\beq
f(\g\t) ~=~ (c\t+d)^k f(\t) ~~~\text{for all}~~~
\g=
\left(\begin{matrix} a&b\\ c&d\\ \end{matrix}\right) \in \G_0(N) ~.
\notag\eeq
There is also a further relation, that is not a special case of the above, if $N\neq1$:
\beq
f\left(-\frac{1}{N\t}\right) \= \e N^{k/2} \t^k f(\t)~,
\label{inversion}\eeq
where $\e=\pm1$ is a sign that depends on the particular modular form $f$. The group $\G_0(N)$ is the subgroup of matrices
\beq
\left(\begin{matrix} a&b\\ c&d\\ \end{matrix}\right) \subset \text{SL}(2,\IZ)~~~\text{with}~~~
c \equiv 0 \mod N~.
\notag\eeq

For the modular forms for $\G_0(14)$, that we need, the weight 2 form admits a representation in terms of the Dedekind $\eta$-function 
\beq\begin{split}
f_{\bf 14.2.a.a}(\t) &\= \eta(\t) \eta(2\t) \eta(7\t) \eta(14\t) \\[5pt]
&\= q-q^2-2 q^3+q^4+2 q^6+q^7-q^8+q^9-2 q^{12}-4 q^{13}-q^{14}+q^{16}+6 q^{17}-\\
&\hskip30pt q^{18}+2q^{19}+\ldots~.\\[-20pt]
\end{split}\notag\eeq

For the weight 4 form we do not know of an analogous expression, however the LMFDB provides the expansion
\beq\begin{split}
f_{\bf 14.4.a.a}(\t) \;
&=\; q-2 q^2+8 q^3+4 q^4-14 q^5-16 q^6-7 q^7-8 q^8+37 q^9+28 q^{10}-28 q^{11}+\\
&\hskip30pt  32 q^{12}+18 q^{13}+14 q^{14}-112 q^{15}+16 q^{16}+74
   q^{17}-74 q^{18}+80 q^{19}+\ldots~.\\[-20pt]
\end{split}\notag\eeq

For $\vph{\=}33\pm 8\sqrt{17}$, with the exception of $p{\=}17$, the correspondence is for primes such that  17 is a square mod $p$. For these primes, the $\a$'s are the p$^{\text{th}}$ coefficients of the weight two modular form, with designation {\bf 34.2.b.a} and the $\b$'s are the p$^{\text{th}}$ coefficients the weight 4 modular form {\bf 34.4.b.a}, both for the congruence subgroup~$\G_1(34)$. 

The group $\G_1(N)$ is the subgroup of $\text{SL}(2,\IZ)$
\beq
\left(\begin{matrix} a&b\\ c&d\\ \end{matrix}\right) \subset \G_0(N)~~~\text{with}~~~
\left(\begin{matrix} a&b\\ c&d\\ \end{matrix}\right) \equiv  
\left(\begin{matrix} 1&0\\ 0&1\\ \end{matrix}\right)\mod N~.
\notag\eeq
In this case, a modular form of weight $k$ satifies
\[
f(\g\t) ~=~ \chi(d)\, (c\t+d)^k f(\t) ~~~\text{for all}~~~
\g=
\left(\begin{matrix} a&b\\ c&d\\ \end{matrix}\right) \in \G_0(N) ~,
\]
where $\chi$ is a Dirichlet character of modulus $N$.  

For $\vph{\=}33\pm 8\sqrt{17}$, the $\a$'s appear as the coefficients of $q^p$ in the $q$-expansion of the weight 2 modular form for $\G_1(34)$ with LMFDB designation {\bf 34.2.b.a} and Fourier expansion
\beq\begin{split}
f_{\bf 34.2.b.a} &= q - q^2 + 2\ii\sqrt{2} q^3 + q^4 - 2\ii\sqrt{2} q^5 - 2\ii\sqrt{2} q^6 - q^8 - 5 q^9 + 2\ii\sqrt{2} q^{10} - 2\ii\sqrt{2} q^{11} +\hskip30pt{} \\[3pt]
&\hskip30pt  2\ii\sqrt{2} q^{12} + 2 q^{13} + 8 q^{15} + q^{16} - \big(3-2\ii\sqrt{2}\big) q^{17} + 5q^{18} - 4 q^{19} + \ldots~.
\end{split}\label{eq: Fourier expansion of 34.2.b.a}\raisetag{18pt}\eeq
The $\b$'s appear as the the coefficients of $q^p$ of the weight 4 modular form for $\G_1(34)$ with LMFDB designation {\bf 34.4.b.a} and Fourier expansion
\beq\begin{split}
f_{\bf 34.4.b.a} &= q - 2q^2 + 2\ii q^3 + 4q^4 + 8\ii q^5 - 4\ii q^6 + 34\ii q^7 - 8 q^8 + 23 q^9 -
    16\ii q^{10} - 30\ii q^{11} + 8\ii q^{12} - \\[3pt]
&\hskip30pt 42 q^{13}-68 \ii q^{14}-16 q^{15}+
    16 q^{16}+(17-68\ii) q^{17}-46 q^{18}+60 q^{19}+\ldots ~.
\end{split}\label{eq: Fourier expansion of 34.4.b.a}\raisetag{18pt}\eeq
At first sight, these last two $q$-series are surprising since the coefficients are not all integers. However, the coefficients we need to compare with the $\a$'s and $\b$'s are those of terms $q^p$ for primes such that 17 is a square mod $p$, and for these the coefficients \emph{are} integers. The coefficients in these expansions that are not integral are complex so there is a choice that has been made in defining the forms
      $f_{\bf 34.2.b.a}$ and $f_{\bf 34.4.b.a}$ above, since the complex conjugates of these forms are also modular forms of the same weight for $\G_1(34)$. 
      
\emph{[Returning, once again, to the case $p{\=}19$, notice that the coefficients of $q^{19}$ in the modular forms above are -4 and 60 and that these are the $\a$ and $\b$
      coefficients that appear for $\vph{\=}4,\,5$ in \tref{tab:p19}.]}
      
\subsection{Outline of the paper}
\vskip-10pt
In outline, the rest of this paper is as follows. We recall the essential features of the Hulek-Verrill manifold in \SS2. This manifold admits a freely acting symmetry group that is abstractly $\IZ/10\IZ$ and taking the quotient by this group, or by the $\IZ/5\IZ$ subgroup, yields manifolds with one complex structure parameter. These one parameter families of manifolds are the subject of our investigation. In \SS3 we set out the Picard-Fuchs equation, define bases of periods that satisfy this equation and explain the relations between these. Having set out our conventions, we proceed in \SS4 to calculate the periods, and their first three covariant derivatives, at the rank two attractor points. Since the attractor points are of rank two, we expect, and duly find, two $\IQ$-linear relations between the periods, at each attractor point. We are also able to evaluate the periods and their covariant derivatives, at the attractor points, in terms of critical $L$-values, for the modular groups together with the $\t$-parameter of the $H^{2,1}{\oplus}H^{1,2}$ lattice. The principal results, in this direction, are recorded in \tref{tab:xivals} and \tref{tab:xivalsplusminus}. Give the periods, we are able to evaluate also the central  charge and so the area of the horizon of the black hole in terms of ratios of $L$-values.  

It was pointed out, already by Moore, that if an attractor point occurs for a parameter value that is within the region of convergence of the instanton sum for the Yukawa coupling, then there will be interesting identities that involve the instanton numbers. The attractor point at $\vph{\=}33-8\sqrt{17}$ is just such a point and we write out the simplest of these identities in \SS5. These identities are morally like the identities \eqref{eq:FundamentalPeriodIdentity1} and \eqref{eq:FundamentalPeriodIdentity2}, except that they involve the special geometry coordinate $t$ and the prepotential $\cF$, and so the instanton numbers. We note also that as with the relation \eqref{eq:FundamentalPeriodIdentity2} interesting identities exist even outside the region where the instanton sums converge.

We have come to the attractor points and the consequent splitting of the Hodge structure by an indirect means. The Hodge Conjecture requires, as we have noted, that there should be a geometrical reason for this splitting. We have not observed this directly in the geometry of the manifold, but speculate in \SS5 how this may come about. We speculate also with regard to the physics interpretation of our results. Prominent among these are how to interpret the infinite number of cycles with vanishing central charge, corresponding to points of the $H^{2,1}{\oplus}H^{1,2}$ lattice, that become massless at the attractor point. We discuss this in \SS6.

Three appendices deal with ancillary matters. In Appendix A we discuss the toric polyhedron associated to AESZ34 and its dual. In Appendix B we discuss the likelihood that there are further rank two attractor points in the moduli space of AESZ34. As part of this discussion we ask how many factorisations of the Frobenius polynomial can be expected to occur `at random'. This number turns out to be much smaller than the number of factorisations that do occur. From the statistics of the distribution of the coefficients of the Frobenius polynomial, we are also led to conjecture that these are distributed according to the statistics of random $\text{USp}(4)$ matrices. Appendix C is a telegraphic review of special geometry, included largely to set our conventions.
\begin{table}
\begin{center}
\begin{tabular}[!t]{lr}
\begin{minipage}[t]{2.2in}
\begin{center}
\renewcommand{\arraystretch}{1.0}
\begin{tabular}[H]{|c|c|c|}
\hline
\multicolumn{3}{|c|}{\vrule height16pt depth10pt width0pt$\vph\;= -\frac17$}\\
\hline
\vrule height13pt depth8pt width0pt$~~p~~$ & $~~\a~~$ & $\b$ \\\hline\hline
 5 & 0 & -14\+ \\\hline 
 7 &  &   \\\hline 
 11 & 0 & -28\+ \\\hline  
 13 & -4\+ & 18 \\\hline  
 17 & 6 & 74 \\\hline  
 19 & 2 & 80 \\\hline  
 23 & 0 & -112\+ \\\hline  
 29 & -6\+ & 190 \\\hline  
 31 & -4\+ & 72 \\\hline  
 37 & 2 & -346\+ \\\hline  
 41 & 6 & 162 \\\hline  
 43 & 8 & -412\+ \\\hline  
 47 & -12\+ & 24 \\\hline  
 53 & 6 & 318 \\\hline  
 59 & -6\+ & -200\+ \\\hline  
 61 & 8 & -198\+ \\\hline  
 67 & -4\+ & -716\+ \\\hline  
 71 & 0 & 392 \\\hline  
 73 & 2 & 538 \\\hline  
 79 & 8 & 240 \\\hline  
 83 & -6\+ & -1072\+ \\\hline  
 89 & -6\+ & 810 \\\hline  
 97 & -10\+ & 1354 \\\hline  
 101 & 0 & -1358\+ \\\hline  
 103 & -4\+ & -832\+ \\\hline  
 107 & 12 & 444 \\\hline  
 109 & 2 & 1870 \\\hline  
 113 & 6 & 1378 \\\hline  
 127 & -16\+ & 1944 \\\hline  
 131 & 18 & -848\+ \\\hline  
 137 & 18 & -2966\+ \\\hline 
 \end{tabular}
\end{center}
\end{minipage}
\hskip0.75in
&
\hskip0.75in
\begin{minipage}{2.2in}
\begin{center}
\renewcommand{\arraystretch}{1.0}
\begin{tabular}[H]{|c|c|c|}
\hline
\multicolumn{3}{|c|}{\vrule height16pt depth10pt width0pt$\vph\;=\; 33\pm 8\sqrt{17}$}\\
\hline
\vrule height13pt depth8pt width0pt$~~p~~$ & $~~\a~~$ & $\b$ \\\hline\hline
 13 & 2 & -42\+ \\\hline
 17 & -6\+ & 34 \\\hline  
 19 & -4\+ & 60 \\\hline  
 43 & -4\+ & 508 \\\hline  
 47 & 0 & -136\+ \\\hline  
 53 & 6 & 318 \\\hline  
 59 & 12 & 300 \\\hline  
 67 & -4\+ & -676\+ \\\hline  
 83 & -12\+ & -1132\+ \\\hline  
 89 & 6 & -350\+ \\\hline  
 101 & -6\+ & -1218\+ \\\hline  
 103 & 8 & 8 \\\hline  
 127 & -16\+ & -1216\+ \\\hline  
 137 & -18\+ & 1954 \\\hline  
 149 & 6 & -1010\+ \\\hline  
 151 & 8 & -968\+ \\\hline  
 157 & 14 & 1654 \\\hline  
 179 & 12 & -980\+ \\\hline  
 191 & 0 & 952 \\\hline  
 223 & -16\+ & -712\+ \\\hline  
 229 & -22\+ & 5230 \\\hline  
 239 & 0 & 2040 \\\hline  
 251 & -12\+ & -5868\+ \\\hline  
 257 & 6 & -4646\+ \\\hline  
 263 & 24 & -6472\+ \\\hline  
 271 & -16\+ & 8312 \\\hline  
 281 & 18 & -518\+ \\\hline  
 293 & 6 & -6402\+ \\\hline  
 307 & 20 & -3516\+ \\\hline  
 331 & -4\+ & 2892 \\\hline  
 349 & -34\+ & 5270 \\\hline 
\end{tabular}
\end{center}
\end{minipage}
\end{tabular}
\vskip1cm
\capt{6.2in}{tab:AttractorCoeffs}{The $(\a,\,\b)$-coefficients for the attractor points $\vph{\;=}-\frac17$ and $\vph{\;=\;}33{\;\pm\;}8\sqrt{17}$.}
\end{center}
\end{table}
\newpage
\section{AESZ34: A Quotient of a Hulek--Verrill Manifold}
\label{section: A quotient of a Hulek--Verrill manifold}
\vskip-10pt
Hulek and Verrill in~\cite{hulek_verrill_2005} consider a family of \cys that are birational to a  variety defined on $\IT\= \IP^4\setminus\{X_1X_2X_3X_4X_5\=0\}$ by the equation
\beq
(X_1+X_2+X_3+X_4+X_5)
\left(\frac{\m_1}{X_1}+\frac{\m_2}{X_2}+\frac{\m_3}{X_3}+\frac{\m_4}{X_4}+\frac{\m_5}{X_5}\right)
\= \m_6~.
\label{eq:SingVar}\eeq
For generic parameters $\m_1,...,\m_6$, the variety $X^\sharp$ that is defined by this equation is smooth on $\IT$, however there are 30 nodes where a subset of the coordinates $X_j$ vanish. Three nodes lie on each of ten surfaces. The singularities can be simultaneously resolved by blowing up each of these these ten surfaces yielding a smooth \cy manifold $\widehat{X}$.

A multiplication of the coefficients $\m_j,\,j{\=}1,...,6$ in \eqref{eq:SingVar} by a common scale has no effect, so superficially this equation defines a five parameter family of manifolds. The equation defines a reflexive polyhedron, in the sense of Batyrev. Analysis of the polyhedron and the resolution just described reveals that the superficial count of complex structure parameters is in fact correct and that the Hodge numbers for a generic member of the family are given~by
\beq
h^{pq}\big(\widehat{X}\big)\=\begin{array}{ccccccc}
& & &1& & &\\
& &0&&0& & \\
&0&&45&&0& \\ 
1&&5&&5&&\hskip1.5pt 1~. \\
&0&&45&&0& \\ 
& &0&&0& & \\
& & &1& & &\\
\end{array}
\notag\eeq
Thus $\chi\big(\widehat{X}\big){\=}2(h^{11}-h^{21}){\=}80$. 

We consider now a 1-parameter subfamily where $\m_j{\=}1,\;j{\=}1,\ldots,5$ and $\m_6{\=}1/\vph$ then the manifold admits a symmetry isomorphic to $\IZ/10\IZ$ with generator
\beq
X_i \;\to\; \frac{1}{X_{i+1}}~,
\notag\eeq
with the indices understood mod 5. This symmetry is fixed point free if 
$\vph{\;\not\in\;}\{1,\,\frac19,\,\frac{1}{25},\infty\}$.  This is easy to see for points of the singular variety $X^\sharp$. For the resolution $\widehat{X}$, we note that, since it is a resolution, there is a projection $\widehat{X}\to X^\sharp$ and so a fixed point of $\widehat{X}$ would project to a fixed point of $X^\sharp$, and these do not exist unless $\vph$ takes one of the values $\{1,\,\frac19,\,\frac{1}{25},\,\infty\}$.
      Taking the quotient by either the $\IZ/10\IZ$, or the $\IZ/5\IZ$ subgroup with generator $X_i\to X_{i+2}$, yields a family of smooth manifolds, that we shall denote by $X$, with one complex structure parameter and the following Hodge numbers:

\beq
h^{pq}(X)\=\begin{array}{ccccccc}
& & &1& & &\\
& &0&&0& & \\
&0&&\hskip-8pt 4\k{+}1\hskip-8pt{}&&0& \\ 
1&&1&&1&&1~,\\
&0&&\hskip-8pt 4\k{+}1\hskip-8pt{}&&0& \\ 
& &0&&0& & \\
& & &1& & &\\
\end{array}
\notag\eeq
where $\k{\=}1,\,2$ according as the quotient is taken by a group of order 10 or 5.

We wish to describe the singular members of the family $X_\vph$ and how the symmetries act on these in somewhat greater detail. We will restrict attention to points $X_j$ in $\IT$ since the discussion of the points not in $\IT$ is part of the story of how the non-compact manifold described by \eqref{eq:SingVar} is compactified so as to yield a \cym.

A first remark is that the manifold defined by \eqref{eq:SingVar} can be regarded as arising from two linear equations in six variables
\beq
\sum_{i=1}^6 X_i\= 0~~~\text{and}~~~\sum_{i=1}^6\frac{\m_i}{X_i} \= 0~,
\label{eq:TwoEqs}\eeq
since eliminating $X_6$ between these two equations returns us to \eqref{eq:SingVar}.

Let $P(X)$ denote the defining equation
\beq
P(X)\= \left(\sum_{i=1}^5 X_i\right)\left(\sum_{i=1}^5\frac{1}{X_i}\right) - \frac{1}{\vph}~.
\notag\eeq
The partial derivatives of $P$ vanish at a singularity, yielding the conditions
\beq
\left(\sum_{i=1}^5\frac{1}{X_i}\right) - \frac{1}{X_j^2} \left(\sum_{i=1}^5 X_i\right) \= 0~,
\label{eq:SingConds}\eeq
for $j{\=}1,...,5$. It follows, since we are assuming that $X_j$ does not vanish, that if either $\sum_{i=1}^5 X_i$ or $\sum_{i=1}^5\frac{1}{X_i}$ vanish, then both sums vanish. This can only happen when $\vph{\=}\infty$, but in this case \eqref{eq:SingConds} provides no further constraints so the singular set is a two-dimensional surface described by the equations
\beq
\sum_{i=1}^5 X_i\= 0~~~\text{and}~~~\sum_{i=1}^5\frac{1}{X_i} \= 0~,
\notag\eeq
analogous to \eqref{eq:TwoEqs} but with five variables, instead of six. This being so, we expect the singular set to be a K3 surface.

If now neither sum in \eqref{eq:SingConds} vanishes, then the $X_j^2$ are all equal and by choice of scale can all be set to unity. Let us suppose that $r$ of the $X_j$ take the value $-1$ and the remaining $5{\;-\;}r$ take the value $1$. So, up to permutation of the coordinates, the singular points are given by
\beq
X_j \= (\,\underbrace{1,\ldots,1}_{5-r},\underbrace{-1,\ldots,-1}_{r}\,)~,
\label{eq:Z2FixedPoint}\eeq
and we may assume that $r{\=}0,1$ or $2$. Such a point lies on the manifold with $\vph{\=}(5-2r)^{-2}$.

\underline{$r{\=}0$}\\
In this case $\vph{\=}1/25$ and there is one singular point $X_j{\=}(1,1,1,1,1)$. This point is fixed by both the $\IZ/5\IZ$ and $\IZ/2\IZ$ symmetry generators and so gives rise to a single point that is fixed by either $\IZ/10\IZ$ or $\IZ/5\IZ$ on the respective quotient manifolds.
     
\underline{$r{\=}1$}\\
This case corresponds to $\vph{\=}1/9$ and $X_j{\=}(1,1,1,1,-1)$, up to cyclic permutation. These five points are fixed by the $\IZ/2\IZ$ generator and give rise to a single point on the $\IZ/5\IZ$ quotient and a single point that is fixed by $\IZ/2\IZ$ on the $\IZ/10\IZ$ quotient.

\underline{$r{\=}2$}\\
The last case corresponds to $\vph{\=}1.$ Now there are ten points, which are the cyclic permutations of $(1,1,1,-1,-1)$ and $(1,1,-1,1,-1)$. These points give rise to two points in the $\IZ/5\IZ$ quotient and two points fixed by a $\IZ/2\IZ$ action, in the $\IZ/10\IZ$ quotient.

It is easy to see that a point that is fixed by an element of $\IZ/10\IZ$ must be fixed by either, or both of, the $\IZ/2\IZ$ or the $\IZ/5\IZ$ generators. A point fixed by the $\IZ/2\IZ$ generator is, up to permutation, of the form \eqref{eq:Z2FixedPoint}. So these coincide with the singular points of the $\vph{\=}1/25,\,1/9,\,1$
      manifolds and have the effect of turning the conifold singularities into hyperconifold singularities. The fixed point $(1,1,1,1,1)$ is also fixed by the $\IZ/5\IZ$ generator. The other fixed points of the $\IZ/5\IZ$ generator are the four points
\beq
X_j\= \z^{jk}~;~~~k\= 1,\,2,\,3,\,4~,
\notag\eeq
where $\z$ is a nontrivial fifth root of unity. For such a point we have $\sum_j X_j{\=}\sum_j 1/X_j{\=}0$,
so these lie in the singular surface of the $\vph{\=}\infty$ manifold.
\newpage
\section{The Periods of $X_\vph$}\label{sec:periods}
\vskip-10pt
\subsection{The Picard-Fuchs equation}\label{subsec:PFequation}
\vskip-10pt
A method for finding the Picard Fuchs differential equation and the periods that satisfy it is given in~\cite{LocalZetaFunctionsI}. The differential operator for the family $X_\vph$ is
\beq
\cL\ = S_4\vth^4+S_3\vth^3+S_2\vth^2+S_1\vth+S_0
\notag\eeq
where $\vth \= \vph\,\dd / \dd\vph$ and
\beq\begin{split}
S_4&\= (\varphi -1) (9 \varphi -1) (25 \varphi -1)\\[3pt]
S_3&\= 2 \varphi  \left(675 \varphi ^2-518 \varphi +35\right)\\[3pt]
S_2&\= \varphi  \left(2925 \varphi ^2-1580 \varphi +63\right)\\[3pt]
S_1&\= 4 \varphi  \left(675 \varphi ^2-272 \varphi +7\right)\\[3pt]
S_0&\= 5 \varphi(180 \varphi ^2-57 \varphi +1).
\end{split}\notag\eeq

The operator $\cL$ appears as operator number 34 in the AESZ list \cite{almkvist2005tables} and has Riemann symbol

\beq
\cP\left\{
\begin{tabular}{ccccc}
0& $\frac{1}{25}$ & $\frac{1}{9}$ &~1~& \hskip-4pt $\infty$ \hskip-4pt{}\\[2pt]
\hline
0&0&0&0&1\\
0&1&1&1&1\\
0&1&1&1&2\\
0&2&2&2&2
\end{tabular}
\hskip5pt\vph~\right\}.
\notag\eeq

One can see that there is a point of maximal unipotent monodromy (the large complex structure point) at $\varphi=0$ and hyperconifold singularities when $\varphi\in\{\frac{1}{25},\frac{1}{9},1\}$.

\subsection{The Periods}
\vskip-10pt
One may use the method of Frobenius to solve the differential equation around $\varphi=0$ and find a basis of solutions that we shall term the {\it arithmetic Frobenius basis}
\beq\begin{split}
\varpi_0 &\= f_0(\vph)\\[3pt]
\varpi_1 &\= f_0(\vph)\,\log(\vph) + f_1(\vph)\\[3pt]
\varpi_2 &\= f_0(\vph)\,\log^2(\vph)+2f_1(\vph)\,\log(\vph)+f_2(\vph)\\[3pt] 
\varpi_3 &\= f_0(\vph)\,\log^3(\vph)+3f_1(\vph)\,\log^2(\vph)+3f_2(\vph)\,\log(\vph)+f_3(\vph)
\end{split}
\label{eq:ArithBasis}\eeq
where the $f_j$ are power series with $f_0(0)=1$ and $f_j(0)=0$ for $j\geq1$. 
\begin{figure}[!t]
\begin{center}
\framebox[\textwidth]{\begin{minipage}[c]{0.98\textwidth}
\vspace{15pt}
\centering\includegraphics[width=0.93\textwidth]{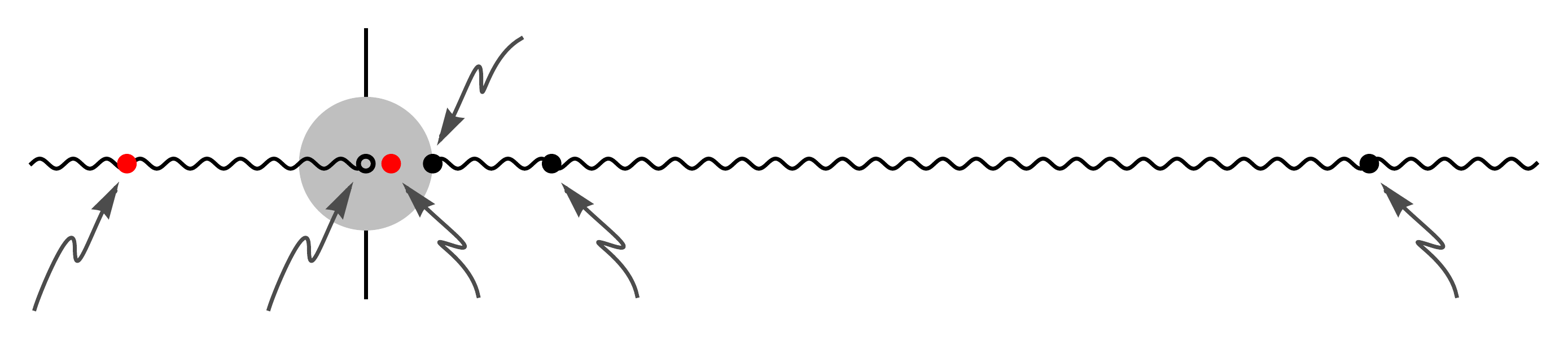}
\vspace{20pt}
\end{minipage}}
\vskip0pt 
\place{0.15}{0.25}{$\vph=-\frac17$}
\place{1.05}{0.23}{$\vph=0$}
\place{1.9}{0.20}{$\vph=\vph_{-}$}
\place{2.05}{1.55}{$\vph=\frac{1}{25}$}
\place{2.7}{0.35}{$\vph=\frac19$}
\place{5.6}{0.25}{$\vph=1$}
\vskip5pt
\capt{6.0in}{fig:CutPhiPlane}{The functions $f_j$, and so the periods, are defined initially in a disk of radius $\frac{1}{25}$. There is a branch cut on the negative real axis owing to our convention for the definition of the the logarithm. The branch cut that runs out along the positive real axis from $\vph{\;=\;}\frac{1}{25}$ is due to the singularities of the functions $f_j$. The two red dots indicate the attractor points at $\vph{\;=}-\frac17$ and $\vph{\;=\;}\vph_{-}$. The large complex structure limit is at $\vph{\;=\;}0$, and is marked by a hollow dot, and the black dots indicate (hyper) conifold points. The attractor point at $\vph{\;=\;}\vph_{+}$ and the conifold point at $\vph{\;=\;}\infty$ are not~shown.}
\end{center}
\end{figure}

As a practical matter, the coefficients of the functions $f_j$ are best calculated via recurrence relations. These are given in \cite{LocalZetaFunctionsI}, but, in any event, are easy to derive. We cannot however refrain from pointing out that there is an interesting closed form for the coefficients $a_n$, $n{\=}0,1,\ldots$, of the fundamental period $\vp_0$, that was found by Verrill.
\beq
a_n \= \sum_{p+q+r+s+t=n} \left( \frac{n!}{p! q! r! s! t!}\right)^2~.
\notag\eeq

If we want the periods to be single valued we can cut the $\vph$ plane along the negative real axis and along the positive real axis for $1/25<\vph<\infty$, as shown in the \fref{fig:CutPhiPlane}.

There are other bases of periods that will also concern us. The first of these which we can call the {\it complex Frobenius basis}, and whose utility will become evident shortly, is simply given~by
\beq
\widehat{\vp}_j(\vph)\= \frac{\vp_j(\vph)}{(2\p \ii)^j} ~.
\notag\eeq

We will continue to modify the basis $\vp_j$, but we pause to relate the $\vp_j$ to an integral basis and to set notation for the charge vector for the case that $X$ is used to reduce IIB string theory to a 4D black hole spacetime.

The prepotential $\cF$ transforms in a complicated way under simplectic changes of the basis forms $\{\a_a,\b^b\}$. It is believed however that, when there is a point of large complex structure, there is a choice of basis such that the prepotential takes the form
\beq
\cF \;= - \frac{1}{3!} Y_{abc} \frac{z^a z^b z^c}{z^0} + \ldots~,
\notag\eeq
where the elipsis indicates a power series in the exponentially small terms $\ee^{2\p\ii z^i/z^0}$, and the indices $a,b,c$ run over the values $0,1,\ldots,h^{11}(\widetilde{X})$.  With $\widetilde{X}$ denoting the mirror manifold. By choice of basis, the quantities $Y_{abc}$ are related to invariants of $\widetilde{X}$ . Let $i,j,k$ run over the values $1,\ldots,h^{11}(\check{X})$, omitting zero, then there is a choice of symplectic basis such that
\beq\begin{split}
Y_{ijk}&\= \int_{\widetilde{X}}\! e_i e_j e_k \\[4pt]
Y_{0jk}&\;\in\;\left\{0,\smallfrac12\right\}\\[4pt]
Y_{00k}&\;= -\frac{1}{12}\int_{\widetilde{X}}\! c_2\, e_k\\[4pt]
Y_{000}&\;= -3\frac{\z(3)}{(2\p\ii)^3}\,\chi\big(\widetilde{X}\big)
\end{split}\label{eq:GenCouplings}\eeq
where the $e_k$ are a basis for $H^2\big(\widetilde{X}\big)$. It is perhaps intuitive that the coefficients $Y_{0jk}$ should be given by the integral of $c_1 e_j e_k$ and so vanish. However, this is not quite true. The components can, by choice of basis, be made to take either the value 0 or $\frac12$. For the case of one parameter, the rule is simple and depends on whether $Y_{111}$ is even or odd. If $Y_{111}$ is even, then $Y_{011}$ can be taken to vanish, and if $Y_{111}$ is odd, it can be taken to be $1/2$. The history of the identification of these terms is a long one. The relation between the large complex structure of the prepotential and the intersection numbers $Y_{ijk}$ may be found in \cite{candelas1991moduli}. The identification of $Y_{000}$ appears in \cite{candelas1991pair}. The identification of the role of the coefficients $Y_{0jk}$ and $Y_{00k}$ may be found in \cite{Hosono:1994ax}. The advance that sets these observations in context is the Gamma class \cite{Halverson:2013qca}.

The utility of the prepotential $\cF$ is that we may express the components of  the holomorphic three-form with respect to a symplectic cohomology basis in terms of this
\beq
\O \= z^a\a_a - \cF_b(z)\b^b ~;~~~\cF_b=\pd{\cF}{z^b}~,
\label{eq:IntegralPeriods}\eeq
where $\a_a,\b^b\in H^3(X,\IZ)$ is the symplectic basis introduced in \sref{sec:AttractorMechanism}.

Returning to the Yukawa couplings, the specialization of \eqref{eq:GenCouplings} to our manifold is
\beq\begin{split}
Y_{111}&\;= \+ 12\k \\[3pt]
Y_{011}&\;= \+ 0 \\[3pt]
Y_{001}&\;= -\k \\
Y_{000}&\;= -24\k\frac{\z(3)}{(2\p\ii)^3}~.
\end{split}\notag\eeq

We form a vector from the integral periods
\beq
\P\=\left(\begin{matrix}\pd{\cF}{z^0}\\[5pt] \pd{\cF}{z^1}\\[3pt]
 z^0\\ z^1 \end{matrix}\right)
\label{eq:PiVector}\eeq
and by considering asymptotic forms in the large complex structure limit $\vph\to 0$, with the identification
$z_1/z_0 \sim \frac{1}{2\p\ii}\log\vph$, we deduce the relation between $\P$ and the vector $\widehat{\vp}$
formed from the periods $\widehat{\vp}_j$
\beq
\P\= \hat{\r}\,\widehat{\vp}~,
\notag\eeq
with
\beq
\hat{\rho}\= 
\left( \begin{array}{ccrr@{\hskip13pt}}
- \frac{1}{3} Y_{000} & -\frac{1}{2} Y_{001} & 0 &\+{\hskip13pt}\frac{1}{6} Y_{111}{\hskip-13pt}\\[3pt]
 -\frac{1}{2} Y_{001} & - \hphantom{\frac{1}{2}}Y_{011} & -\frac{1}{2} Y_{111}{\hskip-13pt} & 0\\[3pt]
 1 & 0 & 0 & 0 \\[3pt]
 0 & 1 & 0 & 0
\end{array} \right) 
\=
\left(\begin{array}{cccc}
8\k \frac{\z(3)}{(2\p\ii)^3} & ~\frac{1}{2}\k~ & 0 & ~2\k~ \\[5pt]
\frac{1}{2}\k & 0 & ~-6\k~ & 0\\[5
pt]
1 & 0 & 0 & 0\\[3pt]
0 & 1 & 0 & 0
\end{array}\right)~.
\notag\eeq

Let us now introduce another basis $\widetilde{\widehat{\vp}}_j$ which we shall term the {\it modified complex Frobenius basis} or, in the slightly shorter form, the {\it modified complex basis}. This basis differs from $\widehat{\vp}_j$ only when $j{\=}3$
\beq
\widetilde{\widehat\vp}_j \= 
\begin{cases} 
\widehat{\vp}_j &;~\text{for}~j=0,1,2\\[5pt]
\widehat{\vp}_3 - 2\frac{\raisebox{2pt}{$Y_{000}$}}{\raisebox{-3pt}{$Y_{111}$}}\, \widehat{\vp}_0 &;~\text{for}~ j=3~.
\end{cases}
\notag\eeq
This basis is related to the integral basis $\P$ by a matrix $\tilde{\hat\r}$ 
\beq
\P \= \tilde{\hat\r}\,\widetilde{\widehat\vp}
\notag\eeq
with
\beq
\tilde{\hat\rho}\= \left( \begin{array}{ccrr@{\hskip13pt}}
         0                       & -\frac{1}{2} Y_{001} & 0 &\+{\hskip13pt}\frac{1}{6} Y_{111}{\hskip-13pt}\\[3pt]
 -\frac{1}{2} Y_{001} & - \hphantom{\frac{1}{2}}Y_{011} & -\frac{1}{2} Y_{111}{\hskip-13pt} & 0\\[3pt]
 1 & 0 & 0 & 0 \\[3pt]
 0 & 1 & 0 & 0
\end{array} \right) 
\notag\eeq
which differs from $\hat{\r}$ only in so far as the irrational element $Y_{000}$ has been removed. 

Finally, we will require also a {\it modified arithmetic Frobenius basis}, or for short a {\it modified arithmetic basis}, which we shall denote by $\widetilde{\vp}_j$ where
\beq
\widetilde{\vp}_j \=
\begin{cases}
\vp_j &;~\text{for}~j=0,1,2\\[5pt]
\vp_3 - 2 (2\p\ii)^3\, \frac{\raisebox{2pt}{$Y_{000}$}}{\raisebox{-3pt}{$Y_{111}$}}\, \vp_0 &;~\text{for}~ j=3~.
\end{cases}
\notag\eeq
For the manifold we are considering we have simply
\beq
\widetilde{\vp}_3\=\vp_3 - 4\z(3)\,\vp_0~.
\notag\eeq

To justify our notation: we could regard $\widehat{*}$ as the operation that divides $\vp_j$ by $(2\p\ii)^j$, and by $\widetilde{*}$ the operation that adjusts the last component. Then the four versions of the Frobenius basis that we have introduced are 
\beq
\vp_j~,~~~\widehat{\vp}_j~,~~\widetilde{\vp}_j ~~\text{and}~~
\widetilde{\widehat\vp}_j~.
\notag\eeq

\subsection{The Periods on the real axis}
\vskip-10pt
The functions $f_j(\vph)$ are defined as series with real, in fact rational, coefficients and so are real for real values of $\vph$, that lie within the disks in which the series converge. For $\vph$ real and $\vph{\;>\;}\frac{1}{25}$ the values of the $f_j$, defined by analytic continuation, will in general be complex owing to the singularity at $\vph{\=}\frac{1}{25}$. For $\vph$ real and negative the $f_j$ are real but the periods
$\vp_j$ are complex owing to the presence of the logarithms. We cut the plane as in \fref{fig:CutPhiPlane} and understand the value of the periods, for $\vph$ real, to be the limit of approaching the real axis from above. 

For the modified arithmetic basis, let us define real and imaginary parts for the periods by the relation
\beq
\widetilde{\vp}_j(\vph+\ii\e) \= \x_j(\vph) + \ii \eta_j(\vph)~.
\notag\eeq
Thus, for example, for our manifold $\x_3$ is the real part of $\vp_3 - 4\z(3)\,\vp_0$. The operator $\cL$ is real, for real $\vph$, and the periods $\x_j$ form a basis of solutions on any interval $I$ of the real axis, that does not contain a singular point. Since the imaginary parts $\eta_j$ also satisfy the differential equation, there is a constant matrix $T_I$ such that
\beq
\eta(\vph)\= T_I\,\x(\vph)~;~~\vph\in I~.
\notag\eeq
For the interval $(0,\frac{1}{25})$ the $\widetilde{\vp}_j$ are real, so for this interval the $\eta_j$ and the corresponding $T$ vanish. By the Schwarz Reflection Principle the values of the periods just below a cut are
$\vp_j(\vph-\ii\e){\= }\x_j(\vph) - \ii \eta_j(\vph)$, so the real part $\x_j$ is, in fact, the average of the values just above and just below the cut.

The attractor point at $\vph{\;=}-\frac17$ lies in the interval $I{\=}(-\infty,0)$. Here the $T$ matrix is easily calculated from \eqref{eq:ArithBasis} using the fact that $\log\vph{\=}\log|\vph| + \ii\p$. We~find
\beq
T_{(-\infty,0)}\=
\left(\begin{matrix}
 0 & 0 & 0 & ~0 \\
 \p & 0 & 0 & ~0 \\
 0 & 2\p & 0 & ~0 \\
 2\p^3 & 0 & 3\p & ~0 
 \end{matrix}\right)~.
\notag\eeq
For $I{\=}\big(0,\smallfrac{1}{25}\big)$ we have already observed that $T_I{\=}0$. While for the interval $(1,\infty)$, that contains the attractor point $\vph_{+}$, we find by numerical calculation that
\beq
T_{(1,\infty)} \=
\left(
\begin{array}{cccc}
 0 & -\frac{45}{28 \p}\+ & 0 & -\frac{15}{28 \p^3}\+ \\[3pt]
 \frac{\pi }{4} & 0 & -\frac{3}{4 \p}\+ & 0 \\[3pt]
 0 & \frac{11 \pi }{28} & 0 & -\frac{15}{28 \pi}\+ \\[3pt]
 -\frac{\pi ^3\!}{4}\+ & 0 & \frac{3 \p}{4} & 0 \\[3pt]
\end{array}
\right)~.
\notag\eeq

The point that is being made is that the imaginary parts of the periods are readily calculated in terms of the real parts.

Now let $m_I$ denote the matrix
\beq
m_I \= \tilde{\hat{\r}}\n(\mathbbl{1} + \ii T_I)~,
\notag\eeq
where $\n$ is the diagonal matrix with entries $(2\p\ii)^j$, $j{\=}0,...,3$. The utility of this matrix is that we have
\beq
\P  \=  \tilde{\hat{\r}}\n (\x +\ii \eta) \=  \tilde{\hat{\r}}\n (\mathbbl{1} + \ii T_I)\, \x \= m_I\,\x~.
\notag\eeq
Let us further define matrices $\s_I$ and $\m_I$ by the relations
\beq
m_I^T\S m_I \= \s_I~~~\text{and}~~~m_I^\dag \S m_I \= \m_I~.
\notag\eeq
We record these matrices for the cases that we will need in the following table.

\begin{table}[H]
\renewcommand{\arraystretch}{1.1}
\begin{center}
\begin{tabular}{|>{$\displaystyle}c<{$}||>{$\displaystyle}r<{$}|>{$\displaystyle}r<{$}|}
\hline
\vrule height0.7cm depth0.4cm width0pt I & \s_I\hfil{} & \m_I\hfil{} \\
\hline\hline
\vrule height2.0cm depth1.7cm width0pt (-\infty,\,0) 
& -\frac{2\k}{(2\p\ii)^3}
\left(\begin{array}{@{\hskip3pt}c@{\hskip7pt}c@{\hskip10pt}r@{\hskip23pt}c}
0 & 6\p^2 & 0 & 1\\
- 6\p^2\hskip3pt{} & 0 & -3 & 0 \\
0 & 3 & 0 & 0 \\
-1\+ & 0 & 0 & 0
\end{array}\hskip3pt\right)
&-\frac{2\k}{(2\p\ii)^3}
\left(\begin{array}{@{\hskip3pt}c@{\hskip10pt}c@{\hskip18pt}r@{\hskip23pt}c}
0 & 6\p^2 & 0 & 1\\
6\p^2\hskip3pt{} & 0 & 3 & 0 \\
0 & 3 & 0 & 0 \\
1 & 0 & 0 & 0
\end{array}\hskip3pt\right) \\
\hline
\vrule height2.0cm depth1.7cm width0pt \left(0, \,\frac{1}{25}\right) 
& -\frac{2\k}{(2\p\ii)^3}
\left(\begin{array}{@{\hskip3pt}c@{\hskip18pt}c@{\hskip15pt}c@{\hskip15pt}c}
0 & 0 & 0 & 1\\
0 & 0 & -3\+ & 0 \\
0 & 3 & 0 & 0 \\
-1\+ & 0 & 0 & 0
\end{array}\hskip3pt\right)
& -\frac{2\k}{(2\p\ii)^3}
\left(\begin{array}{@{\hskip10pt}c@{\hskip23pt}c@{\hskip25pt}c@{\hskip22pt}c}
0 & 0 & 0 & 1\\
0 & 0 & 3 & 0 \\
0 & 3 & 0 & 0 \\
1 & 0 & 0 & 0
\end{array}\hskip3pt\right) \\
\hline
\vrule height2.0cm depth1.7cm width0pt (1,\,\infty)
& -\frac{\k}{28 (2\p\ii)^3}
\left(\begin{array}{@{\hskip-2pt}c@{\hskip5pt}c@{\hskip5pt}c@{\hskip8pt}c}
 0 & 39\p^2 & 0 & 41 \\
-39\p^2 & 0 & -51\+ &  0 \\
 0 & 51 & 0 & \frac{45}{\p^2} \\
- 41\+ & 0 & - \frac{45}{\p^2}\+ & 0 
\end{array}\right)
&-\frac{\k}{28 (2\p\ii)^3}
\left(\begin{array}{@{\hskip0pt}c@{\hskip10pt}c@{\hskip10pt}c@{\hskip15pt}c}
 0 & 39\p^2 & 0 & 41 \\
39\p^2 & 0 & 51 &  0 \\
 0 & 51 & 0 & \frac{45}{\p^2} \\
41 & 0 & \frac{45}{\p^2} & 0 \end{array}\right) \\
\hline
\end{tabular}
\capt{5.5in}{tab:SigmaAndMuMatrices}{The matrices $\s_I$ and $\m_I$ for the intervals of the real axis that contain the three attractor points.}
\end{center}
\end{table}
\subsection{Monodromy around the singular points}
\vskip-10pt
The Picard-Fuchs operator has singular points when $\vph\in\{ 0,\frac{1}{25},\frac{1}{9}, 1, \infty\}$. These singularities of the operator coincide with the values of $\vph$ for which $X_\vph$ is singular. Under monodromy about a singular point $\vph{\=}\ph$ the integral period vector undergoes a monodromy $\P\to M_\ph \P$. The monodromy matrices are the following:

\begin{align*}
M_0\=&\left(\begin{array}{@{\hskip10pt}r@{\hskip13pt}r@{\hskip10pt}r@{\hskip5pt}r}
1 &-1 & 3\k & 6\k\!\\
0 & 1 & -6\k & -12\k\!\\
0 & 0 & 1~ & 0~\\
0 & 0 & 1~ & 1~
\end{array}\right)&
M_{\frac{1}{25}}\=&\left(\begin{array}{@{\hskip10pt}c@{\hskip20pt}c@{\hskip25pt}c@{\hskip25pt}l}
1 & 0 & 0 & 0~\\
0 & 1 & 0 & 0\\
\!\!\! -\frac{10}{\k}~ & 0 & 1 & 0\\
0 & 0 & 0 & 1
\end{array}\right)\\[0.7cm]
M_{\frac{1}{9}}\=&\left(\begin{array}{@{\hskip-3pt}c@{\hskip8pt}c@{\hskip9pt}c@{\hskip22pt}l}
 -9 & -2\+ & 2\k & 0\hskip5pt{}\\[2pt]
\+ 0 & 1 & 0 & 0\\[2pt]
 -\frac{50}{\k} & -\frac{10}{\k}\+ & 11 & 0\\[3pt]
 -\frac{10}{\k} & -\frac{2}{\k}\+ & 2 & 1
\end{array}\right)&
M_1\=&\left(\begin{array}{@{\hskip0pt}r@{\hskip10pt}r@{\hskip10pt}c@{\hskip6pt}c}
-39 & -16 &\+ 16\k & -24\k\;{}\\[2pt]
 60 & 25 & -24\k &\+ 36\k\;{}\\[2pt]
-\frac{100}{\k} & -\frac{40}{\k} &\+ 41 & -60\;{}\\[3pt]
-\frac{40}{\k} & -\frac{16}{\k} &\+ 16 & -23\;{}
\end{array}\hskip-10pt\right)\\[0.7cm]
\multicolumn{4}{c}{$
M_\infty\=
\left(\begin{array}{@{\hskip0pt}rrcc}
 31 & 17 & -19 \k  &\+ 42 \k  \\[2pt]
-60 &-35 &\+ 42 \k  & -96 \k  \\[2pt]
 \frac{60}{\k} & \frac{30}{\k} &-29 & 60 \\[3pt]
 \frac{30}{\k} &\frac{16}{\k} & -17 & 37 
\end{array}\right)
$}\hskip15pt{}
\end{align*}
\vskip10pt
In these matrices, $\k{\=}1$ for the $\IZ/10\IZ$ quotient and $\k{\=}2$ for the $\IZ/5\IZ$ quotient. For the case that no quotient is taken, we have $\k{\=}10$. This case is not a one parameter family and indeed the monodromy matrices $M_\frac19$ and $M_1$ are not integral for this value of $\k$.

\begin{table}[H]
\renewcommand{\arraystretch}{1.3}
\begin{center}
\begin{tabular}{|lcc|}
\hline
\hfill ~Monodromy~ & ~~$c_\text{cnf}$~~ & $w^T$\\[2pt]
\hline\hline
\hskip25pt$M_{\frac{1}{25}}$ & $\frac{10}{\k}$ & $(0,0,1,0)$ \\[3pt]\hline
\hskip25pt$M_{\frac{1}{9}}$ & $\frac{2}{\k}$ & $(\k,0,5,1)$ \\[3pt]\hline
\hskip25pt$M_1$ & $\frac{4}{\k}$ & ~$(2\k,-3\k,5,2)$~ \\[3pt]\hline
\end{tabular}
\capt{5in}{tab:ConifoldVectors}{The coefficients and $w$-vectors for the three conifold points.}
\end{center}
\end{table}

The three monodromy matrices corresponding to the conifolds at $\vph{\=}1/25,1/9,1$, have the~form
\beq
M\= \mathbbl{1} - c_\text{cnf}\, w(\S w)^T~,
\label{eq:PLformOfMonodromy}\eeq
with $c_\text{cnf}$ a coefficient and $w$ a vector with integral components that corresponds to the vanishing cycle, the cycle that shrinks to zero at the conifold point. We will meet the coefficients $c_\text{cnf}$ again when we come to discuss the genus one corrections to the prepotential. These coefficients and the corresponding vectors are shown in \tref{tab:ConifoldVectors}.

The monodromy matrix $M_0$ is readily calculated by hand calculation. The three monodromies corresponding to the conifold points are calculated by integration of the Picard Fuchs equation along loops that encircle the conifold points. This technique can be applied also to the calculation of the monodromy matrix $M_\infty$, but it is easier to note that a contour, as in \fref{fig:TrivialContour}, that winds once about each of the singular points can be deformed to a point and this allows us to relate $M_\infty$ to the other matrices.
\beq
M_\infty\= \left( M_0 M_\frac{1}{25} M_\frac{1}{9} M_1\right)^{-1}~.
\notag\eeq
This matrix differs from the identity by a matrix of rank two and can be brought to a Jordan form with two $2{\times}2$ blocks. Let $J$ and $S$ denote the matrices
\beq
J\=
\begin{pmatrix}
1&1&0&0\\
0&1&0&0\\
0&0&1&1\\
0&0&0&1
\end{pmatrix}~;\qquad
S\=
\left(\hskip-5pt\begin{array}{rrrr}
\+ 2\k  & -\frac{\k}{2} &-\frac{\k}{2} & \frac{4 \k}{15} \\
   -6\k  & \k  & 2\k & -\frac{\k}{2} \\
 0 & 0 & 1 & 0 \\
 1 & 0 & 0 & 0 \\
\end{array}\right)~.
\notag\eeq
Then
\beq
M_\infty\= S J S^{-1}~.
\notag\eeq
\vskip20pt
\begin{figure}[H]
\begin{center}
\framebox[\textwidth][c]{
\begin{minipage}[c]{0.9\textwidth}
\vspace{15pt}
\centering
\includegraphics[width=0.9\textwidth]{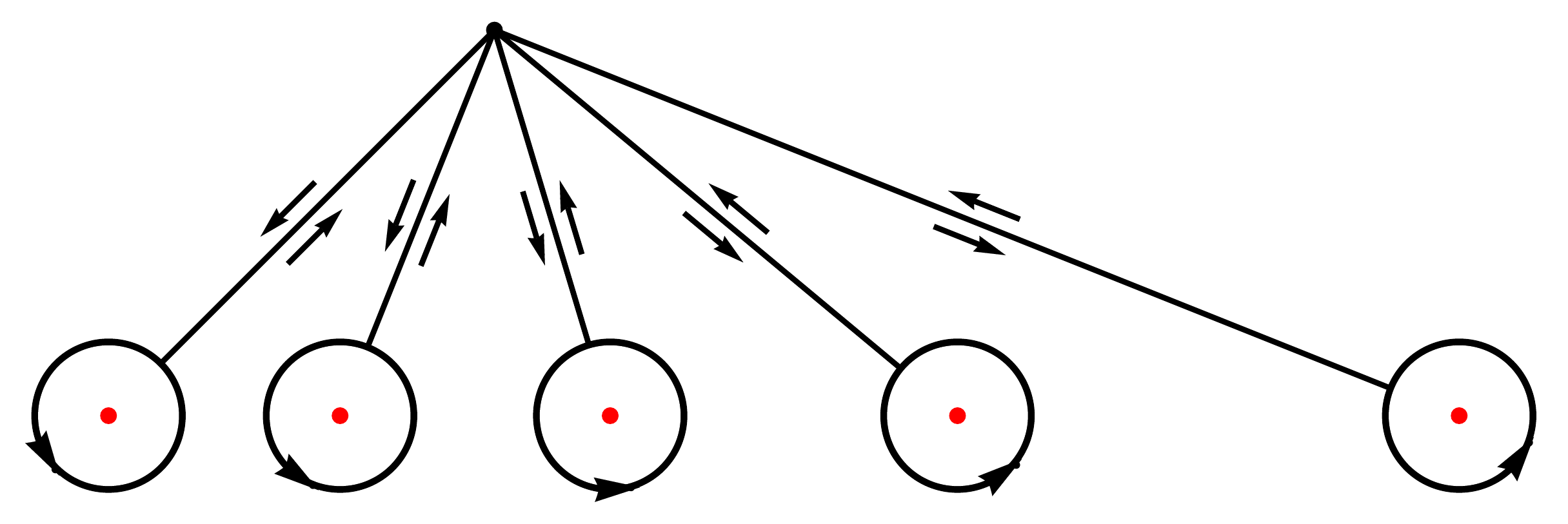}
\vspace*{15pt}
\end{minipage}}
\vskip0pt 
\place{0.8}{0.2}{$\vph{\;=\;}0$}
\place{1.55}{0.2}{$\vph{\;=\;}1/25$}
\place{2.45}{0.2}{$\vph{\;=\;}1/9$}
\place{3.65}{0.2}{$\vph{\;=\;}1$}
\place{5.3}{0.2}{$\vph{\;=\;}\infty$}
\place{2.25}{2.05}{$P$}
\vskip10pt
\capt{5.5in}{fig:TrivialContour}{A sketch of a contour, that can be deformed to a point, which shows that the product of all the monodromy matrices, taken in order, is the identity. In the figure, $P$ is a basepoint for the monodromies.}
\end{center}
\end{figure}
\newpage
\section{The Periods and their Derivatives at the Attractor Points}\label{sec:periodsandderivatives}
\vskip-10pt
\subsection{$L$-functions}
\vskip-10pt
Given a modular form $f$ we can define the associated $L$-function in terms of the Mellin transform
\beq
L(s)\= \frac{(2\p)^s}{\G(s)}\,\int_0^\infty\! \dd y\, f(\ii y) y^{s-1}~.
\notag\eeq
The growth of the coefficients in the $q$ series for $f$ is such that the integral above converges for $\text{Re}(s){\;>\;}1$. The ambiguity associated with the sign of the imaginary part of $f$ leads to a corresponding ambiguity in the choice of sign of the imaginary part of the $L$-function.

Let $\gamma(s)=(2\pi)^s\Gamma(s)$ which appears in the reflection formula for a weight $w$ modular form. We say that an integer $s_0$ is a {\em critical point} if neither $\gamma(s)$ nor $\gamma(w-s)$ has a pole there. In other words, the critical points for a weight $w$-modular form are $\{1,2,...,w-1\}$.

By Deligne's conjecture \cite{Deligne79Lvalues}, it is expected that the critical $L$-values are related to the periods of the modular Calabi-Yau manifold, so we may search numerically for relations between the periods and critical $L$-function values. See \cite{2017arXiv170909751C} for an example of this on a rigid Calabi-Yau manifold which is modular by \cite{GuveaYui}.

Since the reflection formula relates $L(s)$ to $L(w-k)$, the critical values can be taken to be $L_4(1)$ and $L_4(2)$ for weight 4 $L$-functions because the $L(3)$ can be expressed in terms of $L(1)$. Similarly, a weight two $L$ function has critical value $L(1)$.

\subsection{{$\varphi{\;=}-1/7$}}
\vskip-10pt
We have seen that the factorisation of $R(T)$ when $\vph{\;=}-\frac17$ is related to the group $\G_0(14)$ and the weight 2 and weight 4 modular forms with LMFDB designations {\bf 14.2.a.a} and {\bf 14.4.a.a}. We expect to find two linear relations between the periods, suitably understood, with rational coefficients. The caveat `suitably understood' refers to a `transcendentality degree' such that $\p$ has transcendentality degree 1 and $\z(3)$ has transcendentality degree 3. We will specify this more fully as we proceed. In this counting the periods $\vp_j$ and $\widetilde{\vp}_j$ have transcendentality degree $j$, so the two linear relations with rational coefficients that we find are most simply stated between the quantities $\vp_{\tilde{\jmath}}/\p^j$. The use of the modified arithmetic periods allows us to write relations without the explicit appearance of $\z(3)$.

The expected relations do exist, but more is true: the values of the periods at the attractor points are simply related to the values of of the $L$-functions, associated to the modular forms, at their critical points. The relations are most easily stated for the real parts $\x_j$ of the modified arithmetic basis. The critical points of the $L$-functions are $s{\=}1$ and $s{\=}2$ for the weight 4 function that we will denote by $L_4(s)$ and $s{\=}1$ for the weight two function that we will denote by $L_2(s)$. These have the values
\begin{align*}
L_4(1)\=&0.67496319716994177129269568273091339919322842904407\ldots\\
L_4(2)\=&0.91930674266912115653914356907939249680895763199044\ldots\\
\intertext{and}
L_2(1)\=&0.33022365934448053902826194612283487754045234078189\ldots~.
\end{align*}
The accuracy given is sufficient to check the simpler relations that follow, however, unless otherwise stated, our numerical calculations are performed with an accuracy of at least 1000 figures.

The relations between the periods and the $L$- functions are
\begin{align*}
\x_0&\;= \hskip6pt\frac{7}{\p^2} L_4(2) & \x_1&\= -\frac{5}{2}\, L_4(1) \\[8pt] 
\x_2&\;= -\frac{7}{3}\, L_4(2)    & \x_3&\=\! \frac{11\p^2}{2}L_4(1)~.
\end{align*}
So we find the linear relations
\beq
\x_0 + \frac{3}{\p^2}\,\x_2 \=0~~~\text{and}~~~\frac{11}{\p}\,\x_1 + \frac{5}{\p^3}\,\x_3\=0~.
\label{eq:linrelsone}\eeq
Notice that $L_2(1)$ does not appear in these relations and also that the $L_4(k)$, $k{\=}1,2$, have transcendentality degree $k$.

Let us try to see what we can say about the derivatives of the $\x_j$ at $\vph{\;=}-\frac17$. It is here that $L_2(1)$ appears along with the $L_4$-values. The relations are most simply stated for the covariant derivatives
\beq
D\x_j\= \x'_j + K'\x_j
\notag\eeq
where ${}'$ denotes the derivative with respect to $\vph$. The derivative of the \K potential can be calculated, from \eqref{eq:KahlerPotential}, in terms of the periods and their derivatives. Perhaps surprisingly, this quantity turns out to be rational, in fact $K'(-\frac17){\=}-\frac{35}{8}$. We quickly find the relations
\beq\begin{split}
19D\x_0 + \frac{15}{\p^2} D\x_2 &\=0\\[5pt]
5D\x_1 +\frac{3}{\p^2} D\x_3&\=0~.
\end{split}\notag\eeq

We can relate $D\x_0$ and $D\x_2$ directly to $L_2(1)$, but $D\x_1$ and $D\x_3$ depend also on a new irrational number $v^\perp$:
\begin{align*}
D\x_0&\;= - \frac{15{\cdot}7^2}{2^4 \p^2}L_2(1) & D\x_1&\;=\+\frac{3{\cdot}7^2}{2^5\p}\frac{L_2(1)}{v^\perp} \\[5pt]
D\x_2&\;= \+ \frac{19{\cdot}7^2}{2^4} L_2(1) & D\x_3&\;= - \frac{15{\cdot}7^2\,\p}{2^5}\frac{L_2(1)}{v^\perp}~.
\end{align*}
where
\beq
v^\perp\= 0.37369955695472976699767292752499463211766555651682...~.
\notag\eeq

To understand the role of $v^\perp$ let us revert to the integral basis $\P$. With a certain prescience, we also define a complex number
\beq
\t^\perp\= \frac12 + \ii v^\perp
\notag\eeq
In virtue of our results so far we find
\beq
D\P\left(-\smallfrac{1}{7}\right)\= \frac{3{\cdot}7^2}{2^5\p^2}\frac{\ii L_2(1)}{v^\perp}
\left\{\;
\left(\!\!\begin{array}{r}-5\kappa\\10\kappa\\-5\\-3\end{array}\!\right) - 
\t^\perp \left(\!\!\begin{array}{r}-7\kappa\\14\kappa\\-10\\-5 \end{array}\!\right)
\;\right\}
\label{eq:H21lattice}\eeq
The vector $D\P$ is the vector of periods of $D\O$, which, owing to the properties of special geometry, lies in $H^{2,1}(X)=\L^\perp\otimes \mathbb{C}$.  The relation above identifies $\t^\perp$ with the parameter of~$\L^\perp$. 

The $j$-invariant of this lattice is rational and given by
\beq
j(\t^\perp)\=\bigg(\frac{215}{28}\bigg)^3.
\notag\eeq

LMFDB contains only one elliptic curve defined over $\IQ$ with this $j$-invariant and with the form {\bf 14.2.a.a} as its associated weight 2 eigenform. This curve can be defined by the equation
\begin{equation}
\label{eq: equation of elliptic curve at -1/7}
y^2+x y+y \= x^3+4 x-6~.
\end{equation}
and is indeed the modular curve $X_0(14)$ itself.\footnote{As an aside, we note that a rank two attractor $X_{\varphi_{*}}$ with $h^{2,1}=1$ can be used to define a flux compactification with  internal manifold (an orientifold of) $X_{\varphi_{*}}\times T^2$ \cite{Moore:2004fg}. In this scenario, the G-flux of M-theory is given by
\begin{equation*}
G=\frac{1}{\tau-\overline{\tau}}\bigg\{\big(F-\overline{\tau}H\big)\wedge \dd z-\big(F-\tau H\big)\wedge \dd\bar{z}\bigg\}
\end{equation*}
where $F,H\in H^3(X_{\varphi_{*}},\mathbb{Z})$ and $\tau$ is the complex structure parameter of $T^2$. They satisfy
\begin{equation*}
F-\tau H\in H^{2,1}\big(X_{\varphi_{*}}\big).
\end{equation*}
We know that $H^{2,1}\big(X_{\varphi_{*}}\big)$ is generated by $D_{\varphi}\Omega$ and, by comparison, we see that the integral vectors in Equation~\ref{eq:H21lattice} can be identified with the periods of $F$ and $H$ and that $T^2$ can be identified with the elliptic curve defined by Equation~\ref{eq: equation of elliptic curve at -1/7}.}

We gather the periods and their covariant derivatives in \tref{tab:xivals}.

\begin{table}[H]
\renewcommand{\arraystretch}{2.3}
\begin{center}
\begin{tabular}{|>{$\displaystyle}c<{$}||>{$\displaystyle}c<{$}|>{$\displaystyle}c<{$}
|>{$\displaystyle}c<{$}|>{$\displaystyle}c<{$}|>{$}c<{$}|}
\hline
&\hskip10pt j=0 \hskip10pt{}&\hskip10pt j=1 \hskip10pt{}& \hskip10pt j=2 \hskip10pt{}
&j=3 \\[7pt]
\hline\hline
\x_j & \frac7{\p^2}\,L_4(2) & -\frac52\,L_4(1) &-\frac73\, L_4(2) & \frac{11\p^2}{2}\, L_4(1)\\[7pt]
\hline
D\x_j & -\frac{15{\cdot}7^2}{2^4\p^2}\,L_2(1) & \frac{3{\cdot}7^2}{2^5\pi} \frac{L_2(1)}{ v^\perp} & \frac{19{\cdot}7^2}{2^4}\,L_2(1) & -\frac{15{\cdot}7^2\pi}{2^5}\frac{L_2(1)}{ v^\perp} \\[7pt]
\hline
D^2\x_j & \frac{5{\cdot}7^3}{2^{8} \pi}\,\frac{ v^\perp}{L_2(1)} & \frac{7^3}{2^9}\,\frac{1}{L_2(1)} &
-\frac{19{\cdot}7^3\pi}{3{\cdot}2^8}\frac{ v^\perp}{L_2(1)} &\! -\frac{5{\cdot}7^3 \p^2}{2^9}\,\frac{1}{L_2(1)}\!\! \\[7pt]
\hline
y g^{\vph\bar\vph}\ee^K D\left(\frac{g_{\vph\bar\vph}\,\ee^{-K}}{y}\,D^2 \x_j\right)\!\!\!
&- \frac{3{\cdot}7^5}{2^{12}\p^2 L_4(1)} & -\frac{15{\cdot}7^4}{2^{13}L_4(2)} & \frac{7^5}{2^{12}L_4(1)}
& \frac{33{\cdot}7^4 \p^2}{2^{13}L_4(2)}\\[7pt]
\hline
\end{tabular}
\vskip10pt
\capt{5.5in}{tab:xivals}{A table showing the values for the $\x_j$ and their covariant derivatives when 
$\vph{\;=}-\frac17$. For the first two rows, the transcendentality degree for the $j$'th entry is $j$, while for the third and fourth rows, it is $3{-}j$.}
\end{center}
\end{table}

We give also a table of the values of quantities that enter in to the calculation of the covariant derivatives of \tref{tab:xivals}.

\begin{table}[H]
\renewcommand{\arraystretch}{2.3}
\begin{center}
\begin{tabular}{| >{$\displaystyle}c<{$} | >{$\displaystyle}c<{$} | >{$\displaystyle}c<{$} |
 >{$\displaystyle}c<{$} | >{$\displaystyle}c<{$} | >{$\displaystyle}c<{$} |}
\hline
\hskip10pt \ee^{-K} \hskip10pt{} & \hskip10pt K' \hskip10pt{}& \hskip10pt K'' \hskip10pt{} &
g_{\vph\bar\vph} & y & \G' + \G^2\\[7pt]
\hline\hline
\frac{7^2\k}{2\p^3} L_4(1)L_4(2)
& -\frac{5{\cdot}7}{2^3}
& -\frac{5{\cdot}7^3}{2^6}
& \frac{3^2 7^3 L_2(1)^2}{2^8 \p v^{\perp} L_4(1) L_4(2)} 
& - \frac{3{\cdot}7^6\, \k}{2^{10}(2\p\ii)^3}
& \frac{7}{2^7} ( 412\G {-} 1197 )\!{} \\[7pt]
\hline
\end{tabular}
\vskip10pt
\capt{5.5in}{tab:Kandgvals}{The values of quantities that enter into the calculation of the covariant derivatives of \tref{tab:xivals}. In this table, $\G$ denotes the Christoffel symbol $\G_{\vph\vph}^\vph$.}
\end{center}
\end{table}

We can continue with a computation of the second\footnote{It is best to restore the coordinate indices that have been suppressed on the derivatives when performing the calculation, in order to remember to include the Christoffel symbols $\G^\vph_{\vph\vph}$ that arise in the higher derivatives.} and third covariant derivatives of the~$\x_j$.

It is a pleasure to check identities such as 
\beq
\int D\O\wedge\O\=0~;~~~\int D^2\O\wedge\O\=0~;~~~\int\O'''\!\!\wedge\O\=y
~~~\text{and}~~~\int D^2\O\wedge D\O\;=-y~,
\notag\eeq
which translate into
\beq
(D\x)^T\! \s \x \= 0~,~~~(D^2\x)^T\!\s\x\=0~;~~~(\x''')^T\!\s\x\= y
~~~\text{and}~~~(D^2\x)^T\!\s D\x\;=-y~.
\notag\eeq
In the third of these identities, the third derivative $\x'''$ may be replaced by the third derivative of \tref{tab:xivals} without affecting the result.

Let us return to the integral period $\P$, whose components have not yet been stated explicitly 
\beq
\Pi\left(-\smallfrac{1}{7}\right)\= 
 \ii\frac{L_4(1)}{4\pi}
\left(\begin{array}{@{\hskip-1pt}r@{\hskip3pt}}
 8\k \\[3pt] -30\k \\[3pt] 0\hphantom{\k}  \\[3pt] 5\hphantom{\k} \end{array}\right) +
\frac{7}{2}\frac{L_4(2)}{\pi^2}
\begin{pmatrix} \; 0\;{} \\[3pt] 0 \\[3pt] 2  \\[3pt] 1 \end{pmatrix}~.
\label{eq:H30lattice}\eeq
The two integral vectors define a lattice but there is a finer lattice since the difference of the two vectors divides
\beq
( 8\k , -30\k , 0 , 5 ) - ( 0 , 0 , 2 , 1 ) \= 2\, (4\k, -15\k, -1, \,2)
\notag\eeq
One could define a lattice parameter $\t$ for either the coarser, or the finer, lattice but in both cases the invariant $j(\t)$ seems to be transcendental.

The vectors $\P$ and $D\P$ given in \eqref{eq:H21lattice} and \eqref{eq:H30lattice} reside in $\L$ and $\L^\perp$, respectively and allow us to identify the following bases for the lattices
\vskip10pt
\begin{table}[H]
\begin{center}
\renewcommand{\arraystretch}{2.0}
\begin{tabular}{| c | c |}
\hline
\null\hskip3.1cm$\L$\hskip3.1cm\null & $\L^\perp$\\[3pt]
\hline
\vrule height15pt width0pt ~~$(4\k,-15\k,-5,\,0)$, $(0,\,0,\,2,\,1)$~~ 
& ~~$(3\kappa,-6\kappa,\,0,\,1)$, $(\kappa,-2\kappa,-5,-1)$~~  \\ 
\hline 
\end{tabular}
\vskip5pt
\capt{6in}{tab:LambdaAndLambdaperpAtPhistar}{Generators for the lattices $\L$ and $\L^\perp$ for the attractor point at $\vph{\;=}-1/7$.}
\end{center}
\end{table}
\vskip-10pt
The basis for $\L$ is the finer basis discussed above, while the basis for $\L^\perp$ is a basis equivalent to that defined by \eqref{eq:H21lattice}. We observe that $\L\oplus\L^\perp$ has index $7^2\k^2$ within
     $H^3(X,\IZ)$.
     
Being orthogonal to $\L$, with respect to the symplectic product we see that $\L$ is the charge lattice and we have a two parameter family of charge vectors
\beq
Q_{k\ell}\= k\,(4\k,-15\k,-5,\,0) + \ell\,(0,\,0,\,2,\,1)~.
\notag\eeq
Equation~\ref{eq:areaformula} can now be used to find that the black hole with charge $Q_{k\ell}$ will have horizon area given by
\beq
\label{eq: horizon area at -1/7}
\frac{A(-1/7)}{4\p} \= \frac{ (5k-2\ell)^2}{8}\bigg(\frac{\p L_4(1)}{L_4(2)}\bigg)  + \frac{49 k^2}{2}\bigg(\frac{\p L_4(1)}{L_4(2)}\bigg)^{-1}~.
\eeq

We can rewrite \eqref{eq:H30lattice} in terms of the basis vectors of the finer lattice and, in this way, we see that, up to an $\text{SL}(2,\IZ)$ transformation, the lattice has parameter
\beq
\t \= -\frac12 + \ii v_* \quad\text{with}\quad v_*\= 7\,\frac{L_4(2)}{\p L_4(1)}~.
\notag\eeq
The area of the black hole can be rewritten in a simpler form in terms of $v_*$
\beq
A(-1/7) \= 14\p \left\{ k^2 v_* + \left( \ell - \frac{5k}{2} \right)^2 \frac{1}{v_*} \right\}~.
\notag\eeq

The parameter $\t$ is a ratio of periods and the periods are, as we have seen, $\IQ$-linear in the two quantities $\p L_4(1)$ and $L_4(2)$. So it is inevitable that $\t$ should be a fractional linear function $(a v_* + b)/(c v_* + d)$ of the ratio we have called $v_*$. For the $\t$ we have chosen, this is just a linear function, but an $\text{SL}(2,\IZ)$ transform of this would yield a a fractional linear function, in general. The special geometry coordinate $t$ is also a ratio of periods, so has this same general form. In fact we see from \eqref{eq:H30lattice} that
\beq
t_* \= \frac12 + \frac{5\ii}{4 v_*}~,
\label{eq:tstar}\eeq
where we have written $t(-1/7){\=}t_*$. 

This brings us to the three `Attractor Conjectures' formulated in \SS8 of \cite{Moore:1998pn}. Conjecture 2 amounts to the assumption, to which we subscibe, that the attractor points are algebraic in the parameter. Conjecture 1, however, asserts that the period vector $\P$, evaluated at the attractor point, is, projectively, a vector of algebraic numbers. Thus $t_*$, and so $v_*$ would have to be algebraic. While there is no proof that $v_*$ is transcendental, it is generally believed that the critical $L$-values are algebraically independent. If this is so, then Conjecture~1 is contradicted by \eqref{eq:tstar}.
Conjecture 3 concerns a conjectured extension of  Kronecker's Jugendtraum and depends for its formulation on
the periods being projectively algebraic.
\subsection{The rudiments of arithmetic in \smash{$\IQ(\sqrt{17})$}}
\vskip-10pt
As a preparation for discussing the attractor points $33\pm 8\sqrt{17}$, let us pause briefly to recall some elementary facts pertaining to the field $\IQ(\sqrt{17})$, which is the field of numbers of the form
\beq
t \= r + s\sqrt{17}~;~~~r,s\in\IQ~.
\label{eq:NumberQsqrt17}\eeq
The conjugate of $t$, denoted by $\bar{t}$ is the number
\beq
\bar{t}\=  r - s\sqrt{17}~.
\notag\eeq
For the avoidance of doubt: in this subsection, the quantity $t$ bears no relation to the coordinate of special geometry.

An integer in a field $\IK$ is a number $x{\;\in\;}\IK$ that is a root of an irreducible monic polynomial with coefficients in $\IZ$. Thus, for example, the rational integers, as well as numbers such as $\sqrt{17}$ and 
     $(1+\sqrt{17})/2$, are integers of $\IQ(\sqrt{17})$, since they satisfy the respective equations
\beq\begin{split}
x-m       &\=0~;~~~m\in\IZ~,\\
x^2 -17 &\=0~,\\
x^2 - x - 4&\=0~.
\end{split}\notag\eeq

If $x$ is an integer, then so is $-x$, and one can show that the sum and product of two integers is again an integer. 
It follows from the foregoing that the integers of $\IQ(\sqrt{17})$ are of the form
\beq
a + b\sqrt{17}~; a,b\in\IZ~~~\text{and}~~~\frac{a + b\sqrt{17}}{2}~\text{, if $a$ and $b$ are both odd integers.}
\notag\eeq
A number $t{\;\in\;}\IQ(\sqrt{17})$ of the form \eqref{eq:NumberQsqrt17} has a \emph{norm} $\cN(t)$
\beq
\cN(t) \= t\bar{t} \= r^2 - 17s^2~.
\notag\eeq
The term norm is universally used in this context, even though it is somewhat a misnomer, since $\cN(t)$ is not necessarily positive. It has however the property that $\cN(yz){\=}\cN(y)\cN(z)$, for all $y,z{\;\in\;}\IQ(\sqrt{17})$. Moreover, $\cN(x){\;\in\;}\IZ$ if $x$ is an integer of the field.

An integer, whose inverse is also an integer, is a \emph{unit} and the set of all units form a group. A unit necessarily has norm $\pm1$. For $\IQ(\sqrt{17})$ the unit group is infinite and is generated by $4{\;+\;}\sqrt{17}$ and we have 
\beq
\cN(4\pm\sqrt{17})\=-1~.
\notag\eeq
The conjugate satisfies $4{\;-\;}\sqrt{17}{\;=}-(4{\;+\;}\sqrt{17})^{-1}$ and so also generates. 

The attractor points $33{\;\pm\;}8\sqrt{17}$ are units, so are powers of the generator. In fact
\beq
33\pm8\sqrt{17} \= (4\pm\sqrt{17})^2~.
\notag\eeq

The existence of units complicates the process of factorizing integers. In general, for a field $\IQ(\sqrt{d})$, the factorisation of integers, even leaving aside multiplication by units, is not unique. However, for $\IQ(\sqrt{17})$, it is unique, up to multiplication by units. Given an integer $x$, that is not a unit, we can ask if it can be factored into a product $x{\=}yz$ of integers, neither of which is a unit. If $x$ cannot be factored, in this way, then $x$ is a \emph{prime} of the field. Since $\cN(x){\=}\cN(y)\cN(z)$ the integer $x$ can only factor if
     $\cN(x)$ factors as a rational integer. In particular, if $\cN(x)$ is a rational prime, then $x$ is a prime of the field. Note that some of the rational primes factor and so are not primes of the field. For example
\beq
2\;=-\left(\frac{3+\sqrt{17}}{2}\right)\left(\frac{3-\sqrt{17}}{2}\right)~~~\text{and}~~~17\=\left(\sqrt{17}\right)^2~.
\notag\eeq
We will often factorise integers in the following, in order both to save space, particularly in tables, and to show that otherwise inscrutable numbers are the products of a small number of primes with small norm. The numbers
     $4{\;\pm\;}\sqrt{17}$ and $(3{\;\pm\;}\sqrt{17})/2$, the latter being the prime with the smallest absolute value of the norm, so somewhat analogous to 2, are ubiquitous in expressions, so we will often write
\beq
4\pm\sqrt{17}\=\e_{\pm}~~~\text{and}~~~\frac{3\pm\sqrt{17}}{2}\=\d_{\pm}~.
\notag\eeq
As an illustration of the utility of this consider a relation that we will meet shortly
\beq
 j(\tau_{+}^\perp)\= \frac{1}{2^4}\e_{+}^4 \d_{-}^2 \big(2-\sqrt{17}\big)^3 \big(14-5\sqrt{17}\big)^3~.
\notag\eeq
If expanded, the right hand side becomes the somewhat more inscrutable number
\beq
\frac{1}{32} \left( 3832069 +915957\sqrt{17} \right)~.
\notag\eeq
\subsection{$\varphi_{\pm}{\=}33\pm8\sqrt{17}$}
\vskip-10pt
The relevant $L$-functions at $\varphi=33\pm8\sqrt{17}$ have LMFDB designations {\bf 34.2.b.a} and {\bf 34.4.b.a}. As in the previous section, we denote the corresponding weight-$j$ $L$-function by $L_j(s)$. These functions are complex but we can concentrate on the real parts since the imaginary parts are simply related to these. 
We set
\beq
L_j(s)\= \l_j(s)+ \ii\m_j(s)
\notag\eeq
and note that the real parts take the following values at their critical points
\begin{align*}
\lambda_4(1)&\=0.61300748403501690756896255581360559790853555213198\ldots\\
\lambda_4(2)&\=0.72053904959503349611018739597922735350251006854978\ldots\\
\intertext{and}
\lambda_2(1)&\=0.51696098116017249777442349444758176009873137273013\ldots~.\\
\end{align*}

At the critical values, the imaginary parts of the $L$-functions are determined in terms of the real parts up to a sign. This choice follows from the choice of sign in the square root in the Fourier expansion of the weight two form {\bf 34.2.b.a} and the weight four forms {\bf 34.4.33.a}. With the choices in \eqref{eq: Fourier expansion of 34.2.b.a} and~\eqref{eq: Fourier expansion of 34.4.b.a}, we find that
\beq
\m_4(1) \=\left(\frac{1+\sqrt{17}}{4}\right)^3 \l_4(1)~,\qquad
\m_4(2)\;= -\left(\frac{1-\sqrt{17}}{4}\right) \,\l_4(2)
\notag\eeq
and
\beq
\m_2(1)\;=- \left(\frac{3-\sqrt{17}}{2\sqrt{2}}\right)\,\l_2(1)~.
\notag\eeq
The coefficients in the first two relations are numbers in $\IQ(\sqrt{17})$ but the coefficient in the third relation is a number in the quartic extension $\IQ(\sqrt{17},\sqrt{2})$.

Just as was the case for $\vph{\;=}-\frac17$, we can determine the period matrix at $\vph{\;=\;}\vph_{\pm}$ in terms of $L$-function values and a single new modular parameter. We have gathered the values of the real parts of the periods and their derivatives into two tables. The tables contain the irrational numbers $v_{\!\pm}$, where
\beq\begin{split}
v_{+}^{\perp}&\= 1.9696894453517505490479716982864516913834531417517\ldots\\
v_{-}^{\perp}&\= 1.0153884942216545916762729868825409864938877880731\ldots~.\\
\end{split}\notag\eeq
Surprisingly we find that these numbers are simply related
\beq
v_{+}^{\perp}v_{-}^{\perp}\= 2~.
\notag\eeq
To understand the significance of these numbers we now set $\t_{\pm}^{\perp}{\=} \ii v_{\pm}^{\perp}$ and examine the the lattices defined by the covariant derivatives of the integral periods. We find that
\beq\begin{split}
D\P(\vph_{+}) &\;=\phantom{\d} -\frac{3}{2^5\p^2}\,\e_{-}^4\, \l_2(1)
\left\{\; 
\left(\begin{array}{@{\hskip-5pt}r@{\hskip3pt}} 9\k\\\ -16\k\\\ 20\\ 9 \end{array}\right) - \frac{1}{\t_{+}^{\perp}}
\left(\begin{array}{@{\hskip-5pt}r@{\hskip1pt}}15\k\\\ -36\k\\\ 15\\ 11\end{array}\right) 
\;\right\}  \\[15pt]
D\P(\vph_{-}) &\;= - \frac{3}{2^6 \p^2}\, \e_{+}^4 \d_{-} \,\l_2(1)
\left\{\; 
\left( \begin{array}{@{\hskip2pt}r@{\hskip5pt}} 0\\ -2\k\\\ 5\\ 0 \end{array} \right)
\,+\; \t_{-}^{\perp}\left( \begin{array}{@{\hskip8pt}r@{\hskip6pt}} 3\k\\\ 0\\ 0\\ 1 \end{array}\right) \;\right\}~.
\end{split}\notag\eeq
and we see that the $\t_{\pm}^{\perp}$ are the parameters of the lattices. We also find that these parameters have algebraic $j$-invariants. We find
\beq
 j(\tau_{+}^{\perp})\= \frac{1}{2^4}\e_{+}^4 \d_{-}^2 \big(2-\sqrt{17}\big)^3 \big(14-5\sqrt{17}\big)^3~,
\notag\eeq
with $j(\t_{-}^{\perp})$ satisfying the conjugate relation. Seeking elliptic curves, in LMFDB, with these $j$-invariants and with $f_{\bf 34.2.b.a}$ as the associated modular form, brings us to the curves $\cE_{\pm}$ listed as 4.1--a8.
\beq
\cE_{+}\,:\qquad 
y^2 + xy +\d_{+}\,y\= x^3 + \e_{+}\d_{-}^2\, x^2 - \d_{-}\,x -  \e_{+}\d_{-}^2~,
\label{eq:EllipticCurveAtPhiPM}\eeq
and $\cE_{-}$ is the curve with conjugated coefficients. We cannot resist reproducing a sketch of these curves in \fref{fig:ellipticcurves}.
\begin{figure}[H]
\centering
\includegraphics[width=3in]{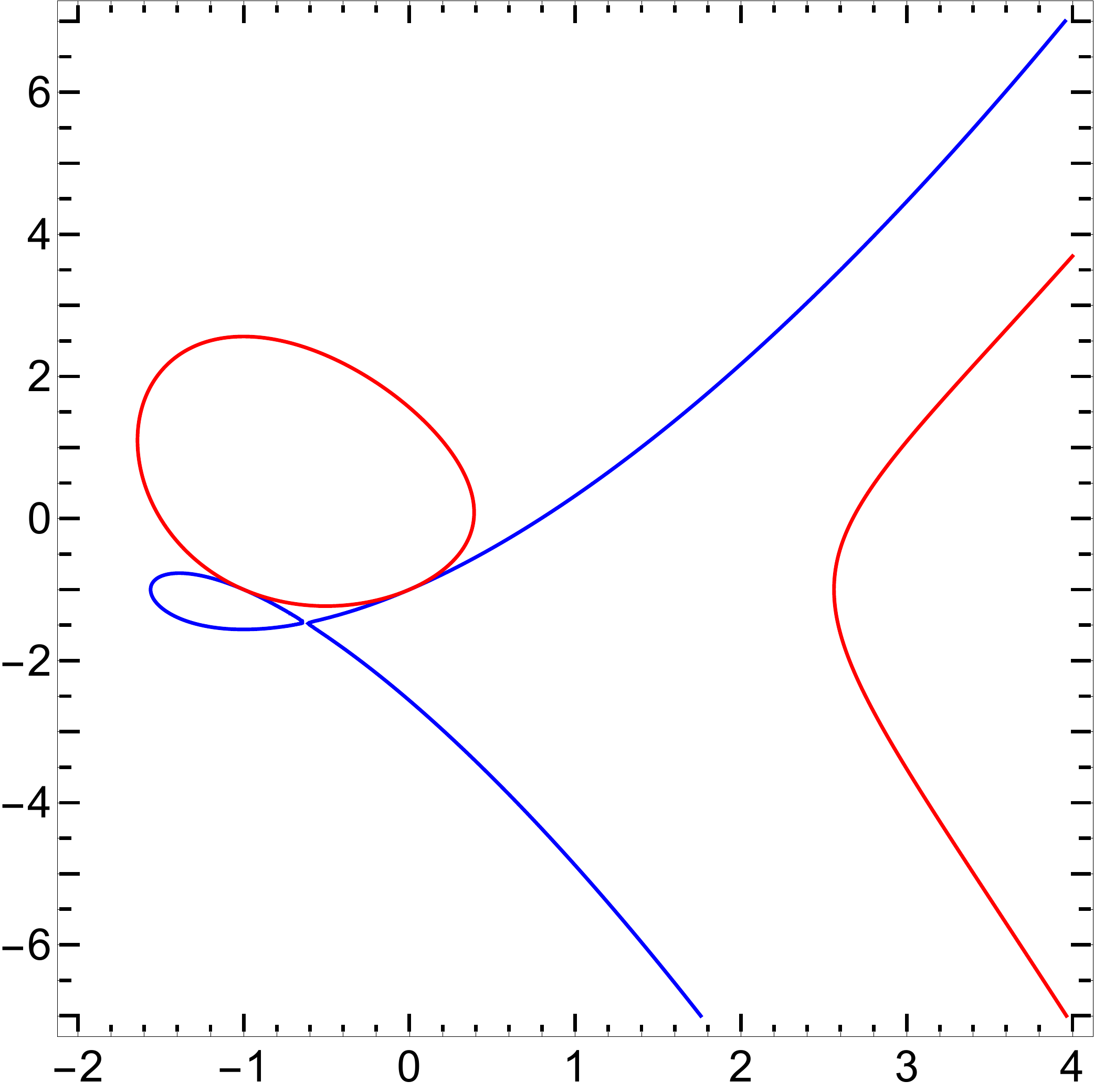}
\capt{6in}{fig:ellipticcurves}{The elliptic curves $\cE_{\pm}$. The curve $\cE_{+}$ is shown in blue while $\cE_{-}$ is shown in red. Despite appearances, the curve $\cE_{+}$ is smooth. Over the reals, this curve has two components and there is a gap where, at this scale, the curve appears to have self intersection.}
\end{figure}

We record the values of the periods and their derivatives in \tref{tab:xivalsplusminus}. Aside from the fact that the coefficients are in $\IQ(\sqrt{17})$ and the relevant $L$-function values are complex, the periods and their covariant derivatives at $\varphi_{\pm}$ look very similar to those at $\varphi=-\frac{1}{7}$. 

There are many interesting and mysterious relations in this table. These include relations analogous to those of
\eqref{eq:linrelsone}
\begin{align*}
25\, \x_0(\vph_{+}) - \frac{21}{\p^2}\,\x_2(\vph_{+})&\=0~; &
13\, \x_1(\vph_{+}) + \frac{5}{\p^2}\d_{-}\,\x_3(\vph_{+})&\=0\\[8pt]
5\,\x_0(\vph_{-}) - \frac{3}{\p^2}\,\x_2(\vph_{-}) &\=0~; &
9\,\x_1(\vph_{-}) - \frac{5}{\p^2}\d_{-}\, \x_3(\vph_{-})
&\=0
\end{align*}
and also relations such as the following
\begin{align*}
\frac{\x_j(\vph_{+})}{\x_j(\vph_{-})} &\;= \+\e_{-}^3\hphantom{\d_{+}}\left(\, 7, -1, -5, -\frac{13}{9}\right)\\
\intertext{and}
\frac{D\x_j(\vph_{+})}{D\x_j(\vph_{-})} &\;= 
-\e_{-}^8 \d_{+}\left( \,4,\; \frac{11}{2},\, \frac{52}{9},\;\frac{19}{10}\right)~,
\end{align*}

For $\quad j\=0,..,3$.
\begin{landscape}
\begin{table}[!p]
\renewcommand{\arraystretch}{1.4}
\setlength{\tabcolsep}{10pt}
\setlength{\extrarowheight}{4pt}
\begin{center}
\begin{tabular}{|>{$\displaystyle}c<{$}||>{$\displaystyle} c<{$}|>{$\displaystyle}c<{$}
|>{$\displaystyle} c<{$}|>{$\displaystyle}c<{$}|>{$\displaystyle}c<{$}|}
\hline
\multicolumn{5}{|c|}{$\varphi\;=\;33+8\sqrt{17}$}\\[5pt]
\hline
& j=0 & j=1 & j=2 & j=3\\[5pt]
\hline\hline
\x_j 
& -\frac{7\sqrt{17}}{2^3\pi^2}\e_{-}^2\d_{+}\l_4(2) 
& \frac{5}{2^4\sqrt{17}}\d_{-}^3\l_4(1) 
& \frac{5^2\sqrt{17}}{2^3{\cdot}3}\e_{-}^2\d_{+}\l_4(2) 
& \frac{13\pi^2}{2^4\sqrt{17}}\d_{-}^3 \l_4(1)\\[7pt]
\hline
D\x_j 
& -\frac{15}{2^3\pi^2}\e_{-}^4\l_2(1) 
& \frac{33}{2^5\pi}\e_{-}^4\l_2(1) v^{\perp}_{-} 
& \frac{13}{2^3}\e_{-}^4\l_2(1)
& -\frac{3{\cdot}19\,\pi}{2^5}\e_{-}^4\l_2(1) v^{\perp}_{-} \\[7pt]
\hline
D^2\x_j 
& -\frac{5}{2^7{\cdot}17\pi}\,\frac{\e_{-}^6\d_{-}}{\l_2(1) v^{\perp}_{-}} 
& -\frac{11}{2^9{\cdot}17}\,\frac{\e_{-}^6\d_{-}}{\l_2(1)} 
& \frac{13\p}{2^7{\cdot}3{\cdot}17}\frac{\e_{-}^6\d_{-}}{\l_2(1) v^{\perp}_{-}} 
& \frac{19\pi^2}{2^9{\cdot}17}\,\frac{\e_{-}^6\d_{-}}{\l_2(1)}\\[7pt]
\hline
g^{\vph\bar\vph} y\, \ee^K D\left( \frac{g_{\vph\bar\vph}\,\ee^{-K}}{y}\,D^2 \x_j\!\right)
& \frac{3{\cdot}7}{2^{12}\sqrt{17}\pi^2}\,\frac{\e_{-}^{10}\d_{+}^2}{\l_4(1)}
&-\frac{3{\cdot}5}{2^{12} 17^{3/2}}\,\frac{\e_{-}^8\d_{-}^2}{\l_4(2)} 
&-\frac{5^2}{2^{12}\sqrt{17}}\,\frac{\e_{-}^{10}\d_{+}^2}{\l_4(1)}
&-\frac{3{\cdot}13\pi^2}{2^{12}17^{3/2}}\,\frac{\e_{-}^8\d_{-}^2}{\l_4(2)}\\[7pt]
\hline
\noalign{\vskip0.3cm}
\hline
\multicolumn{5}{|c|}{$\varphi\;=\;33 - 8\sqrt{17}$}\\[5pt]
\hline
&j\;=\;0 &\; j=1 \;& \; j=2 \; & j\;=\;3\\[5pt]
\hline\hline
\x_j 
& \frac{\sqrt{17}}{2^3\pi^2}\e_{+}\d_{+} \l_4(2) 
& \frac{5}{2^4\sqrt{17}}\e_{+}^3 \d_{-}^3 \l_4(1) 
& \frac{5\sqrt{17}}{2^3{\cdot}3}\e_{+}\d_{+} \l_4(2) 
& \frac{9\pi^2}{2^4\sqrt{17}}\e_{+}^3 \d_{-}^3 \lambda_4(1)\\[7pt]
\hline
D\x_j 
& -\frac{15}{2^6\pi^2}\e_{+}^4\d_{-}\lambda_2(1) 
& \frac{3}{2^5\pi}\e_{+}^4\d_{-}\lambda_2(1)  v^{\perp}_{\!-}
& \frac{9}{2^6}\e_{+}^4\d_{-}\lambda_2(1)
& -\frac{15\pi}{2^5}\e_{+}^4\d_{-}\lambda_2(1)  v^{\perp}_{\!-} \\[7pt]
\hline
D^2\x_j 
&\frac{5}{2^{10}{\cdot}17\pi} \frac{\e_{+}^6 \d_{+}^ 2}{ \lambda_2(1) v^{\perp}_{\!-}}
&\frac{1}{2^9{\cdot}17}\frac{\e_{+}^6 \d_{+}^ 2}{\lambda_2(1)} 
&-\frac{3\pi}{2^{10}{\cdot}17}  \frac{\e_{+}^6 \d_{+}^ 2}{\lambda_2(1) v^{\perp}_{\!-}} 
&-\frac{5\pi^2}{2^9{\cdot}17} \frac{\e_{+}^6 \d_{+}^ 2}{\lambda_2(1)}  \\[7pt]
\hline
g^{\vph\bar\vph} y\, \ee^K D\left( \frac{g_{\vph\bar\vph}\,\ee^{-K}}{y}\,D^2 \x_j\!\right)
& \frac{3}{2^{13}\sqrt{17}\pi^2}\frac{\e_{+}^7\d_{+}^4}{\l_4(1)}
& \frac{15}{2^{11}17^{3/2}}  \frac{\e_{+}^9}{\lambda_4(2)}
& \frac{5}{2^{13}\sqrt{17}}\frac{\e_{+}^7\d_{+}^4}{\l_4(1)} 
& \frac{3^3\pi^2}{2^{11}17^{3/2}} \frac{\e_{+}^9}{\l_4(2)}\\[7pt]
\hline
\end{tabular} 
\vskip4pt
\capt{5.25in}{tab:xivalsplusminus}{The values of the $\x_j$ and their first three covariant derivatives.}
\end{center} 
\end{table}
\end{landscape}

%
%
%

A few geometric quantities that can be calculated exactly from the periods are collected in the following table.
\begin{table}[H]
\renewcommand{\arraystretch}{2.0}
\begin{center}
\begin{tabular}{|>{$} c<{$} | >{$\displaystyle}c<{$} | >{$\displaystyle}c<{$} | >{$\displaystyle}c<{$} |}
\hline
{} & \hskip10pt \ee^{-K} \hskip10pt{} & \hskip10pt K' \hskip10pt{} & \hskip10pt K'' \hskip10pt{} \\[5pt]
\hline\hline
~\vph_{+}
&\frac{17\k}{2^6\p^3}\e_{-}^2\d_{-}^2 \l_4(1)\l_4(2)
&\+\frac{5}{2^3\sqrt{17}}\e_{-}^2 (2 + \sqrt{17})
& -\frac{5}{2^6.17}\e_{-}^4 (135 + 16\sqrt{17})\\[7pt]
\hline
~\vph_{-}
&\frac{17\k}{2^6\p^3}\e_{+}^4\d_{-}^2 \l_4(1)\l_4(2)
&-\frac{5}{2^3\sqrt{17}}\e_{+}^2 (2 - \sqrt{17})
& -\frac{5}{2^6.17}\e_{+}^4 (135 - 16\sqrt{17})\\[7pt]
\hline
\noalign{\vskip1.0cm}
\hline
{}& g_{\vph\bar\vph} & y & \G' + \G^2\\[5pt]
\hline\hline
~\vph_{+}
&\frac{9}{2^6\p}\e_{-}^6 \d_{+}^2 \frac{\l_2(1)^2 v^{\perp}_{\!-}}{\l_4(1)\l_4(2)}
&\frac{3 \k}{2^{11}(2\p\ii)^3} \e_{-}^{10}\d_{-}
&\begin{aligned}&{\scriptstyle \+\frac{\e_{-}\d_{-}}{2^6\sqrt{17}}(2-\sqrt{17})(8 - \sqrt{17})(21+ 2\sqrt{17})\,\G_{+}\,+}\\
&\hskip30pt {\scriptstyle \frac{\e_{-}^3 \d_{-}}{2^8\cdot 17}(9-4\sqrt{17})(206+21\sqrt{17})}
\end{aligned}\\[10pt]
\hline
~\vph_{-}
&\frac{9}{2^5\p}\e_{+}^4 \frac{\l_2(1)^2 v^{\perp}_{\!-}}{\l_4(1)\l_4(2)}
&\frac{3 \k}{2^{11}(2\p\ii)^3} \e_{+}^{10}\d_{+}
&\begin{aligned}&{\scriptstyle -\frac{\e_{+}\d_{+}}{2^6\sqrt{17}}(2+\sqrt{17})(8 + \sqrt{17})(21 - 2\sqrt{17})\,\G_{-}\,+}\\
&\hskip30pt {\scriptstyle \frac{\e_{+}^3 \d_{+}}{2^8\cdot 17}(9+4\sqrt{17})(206-21\sqrt{17})}
\end{aligned}\\[10pt]
\hline
\end{tabular}
\vskip10pt
\capt{5.5in}{tab:Kandgvalsminus}{The values of quantities that enter into the calculation of the covariant derivatives of \tref{tab:xivalsplusminus}. In this table $\G_{\pm}$ denotes the Christoffel symbols 
      $\G_{\vph\vph}^\vph(\vph_{\pm})$.}
\end{center}
\end{table}

We identify generators for the lattices $\L_{\pm}$  and $\L^\perp_{\pm}$ by examining the vectors $\P$ and $D\P$, respectively. The $D\P(\vph_{\pm})$ have been give previously. For the $\P(\vph_{\pm})$ we have
\beq\begin{split}
\P(\vph_{+}) &\; = -\frac{\ii}{2^5\sqrt{17}\p}\, \phantom{\e_{+}^3}\d_{-}^3\, \l_4(1)\hskip2pt
\left(\hskip-7pt\begin{array}{c} - 4\k \\ \+30\k \\ \+30\phantom{\k} \\ \+5\phantom{\k} \end{array}\hskip-7pt\right)
- \frac{\sqrt{17}}{2^3\p^2}\, \e_{-}^2\d_{+}\, \l_4(2)
\left(\hskip-7pt\begin{array}{c} \+ 4\k \\ -9\k \\ \+ 7\phantom{\k} \\ \+ 4\phantom{\k} \end{array}
\hskip-5pt\right) \\[15pt]
\P(\vph_{-}) &\; =\+ \frac{\ii}{2^5\sqrt{17}\p} \, \e_{+}^3\d_{-}^3 \, \l_4(1)\hskip2pt
\left( \hskip-5pt\begin{array}{c} 2\k \\ 0 \\ 0 \\ -5\+ \end{array} \hskip-5pt \right) +
\frac{\sqrt{17}}{2^3\p^2}\, \e_{+}\d_{+}\, \l_4(2)
\left(\hskip0pt \begin{array}{c} 0 \\ 3\k \\ 1\\ 0 \end{array} \hskip-2pt\right)
\end{split}\label{eq:Pivecsplusminus}\eeq
Given these expressions, we identify the generators of $\L_\pm$ and $\L^\perp_\pm$
\vskip10pt
\begin{table}[H]
\renewcommand{\arraystretch}{2.0}
\begin{center}
\begin{tabular}{| c | c | c |}
\hline
{} & $\L_\pm$  & \hskip3.4cm $\L^\perp_\pm$\kern3.4cm{}\\[3pt]
\hline\hline
~$\vph_{+}$ & ~~$(4\k,-9\k,\,7,\,4)$,~ $(4\k,-30\k,-30,-5)$~~ & ~$(6\k,\,-20\k\,-5,\,2)$,~ $(3\k,\,4\k,\,25,\,7)$~~  \\[3pt]
\hline
~$\vph_{-}$ &  $(-2\k,\,0,\,0,\,5)$,~ $(0,\,3\k,\,1,\,0)$~~ & ~~$(0,-2\k,\,5,\,0)$,~  $(3\k,\,0,\,0,\,1)$~~  \\[3pt]  
\hline
\end{tabular}
\capt{6in}{tab:LambdaPlusAndLambdaMinus}{Generators for the lattices $\L_\pm$ and $\L_\pm^\perp$}
\end{center}
\end{table}
\vskip-10pt
The indices of the lattices $\L_{+}{\oplus}\L_{+}^\perp$ and  $\L_{-}{\oplus}\L_{-}^\perp$ in $H^3(X_\pm, \IZ)$ are, for both cases, $17^2\k^2$.

By taking combinations of generators with coefficients $k$ and $\ell$ as our charge vector, we can calculate the area of the horizon of the black hole 
\beq
\label{eq: horizon area at phipm}
\frac{A(\varphi_{\pm})}{4\p}\=\frac{k^2}{32}(9+\sqrt{17})\bigg(\frac{\pi\lambda_4(1)}{\lambda_4(2)}\bigg)+\frac{(17\ell)^2}{8}(9-\sqrt{17})\bigg(\frac{\pi\lambda_4(1)}{\lambda_4(2)}\bigg)^{-1}~.
\eeq

It is a surprising fact that the black holes associated with $\varphi_-$ and $\varphi_+$ have the same horizon areas. This is related to the fact that the expressions for the periods $\P(\vph_{\pm})$ in \eqref{eq:Pivecsplusminus} are remarkably similar. The coefficients of the generators for $\vph_{-}$ are related to those for $\P(\vph_{+})$ simply by multiplication by $-\e_{+}^3$. The parameters for these two lattices are therefore the same. Let us denote this parameter by $\t$ and write
\beq
\t \= \ii v~; ~~~\text{with}~~~v \= \frac{17}{4}(9-\sqrt{17})\,\frac{\l_4(2)}{\p\l_4(1)}
\=\frac{17}{2} \e_{-}^2 \d_{-}^4\,\frac{\l_4(2)}{\p\l_4(1)}~.
\notag\eeq
We find that the area can, analogously to the case of the attractor point at $\vph{\=}-1/7$, be written very succinctly in terms of $v$
\beq
A(\vph_{\pm}) \= 34\p\left(\frac{k^2} v + \ell^2 v \right)~.
\notag\eeq
\subsection{Identifying higher derivatives}
\vskip-10pt
In calculating the expressions in Tables~\ref{tab:xivals} and~\ref{tab:xivalsplusminus}, we have chosen to work with covariant derivatives instead of ordinary derivatives. We do this for two reasons: the first is that we obtain cleaner expressions. This is due to the fact that $\Omega$ takes values in $H^{3,0}$ and, owing to special geometry relations, $D_{\varphi}\Omega$ takes values purely in $H^{2,1}$, while $\partial_\vph\Omega$ takes values in $H^{3,0}\oplus H^{2,1}$. It follows that the periods of $\Omega$ can be expressed purely in terms of weight four $L$-values and the periods of $D_{\varphi}\Omega$ only depend on weight two $L$-values and the modulus of the relevant elliptic curve. Had we instead computed the periods of $\partial_\vph\Omega$, we would have found that they mix the weight two $L$-values with the weight four $L$-values. The second reason covers for our ignorance; had we calculated the partial derivatives, or even the covariant derivatives beyond those shown in the table, we would come across unidentified numbers. This happens first in evaluating the $\partial^2_{\varphi}\vp_j$. We can apportion the blame for this in various ways. We find that we need six numbers in order to compute all the covariant derivatives. Whereas, we have at our disposal only four. Namely, $L_4(1)$, $L_4(2)$, $L_2(1)$ and the modulus of the relevant elliptic curve. There are two numbers that we are unable to identify and, at $\varphi{\=}-\frac{1}{7}$, we can take these to be
\begin{align*}
\partial_{\bar\vph}{\partial_\vph}\hskip-3pt{}^2K(-\tfrac{1}{7}) &\;=\+13.3957566623799144847404045408028493504914256  \ldots\\
\partial^3_{\varphi}K(-\tfrac{1}{7}) &\;=-345.296197568387252384535830788469867726435775\ldots~
\intertext{Similarly, at $\vph_+$, we can take the unidentified numbers to be}
\partial_{\bar\vph}{\partial_\vph}\hskip-3pt{}^2K(\varphi_+) &\;=-2.11248092812853659831921795886813691685791340\ldots \times 10^{-8}\\
\partial^3_{\varphi}K(\varphi_+) &\;=\+6.41299157746065303963342880177316439551792591\ldots \times 10^{-6}~.
\intertext{Finally, at $\vph_-$, the unknown numbers can be taken to be}
\partial_{\bar\vph}{\partial_\vph}\hskip-3pt{}^2K(\varphi_-) &\=-9401.3272027230698289676141395408315362641649\ldots\\
\partial^3_{\varphi}K(\varphi_-) &\;=\+170631.685809372752493637298347668593555721135\ldots~.
\end{align*}
Given $\partial_{\bar\vph}{\partial_\vph}\hskip-3pt{}^2K$ and $\partial^3_{\varphi}K$ at a rank two attractor point, we can identify all the second and third derivatives at that point. All the higher derivatives are then fixed by invoking the Picard-Fuchs equation. We could then, for example, identify all the coefficients in an expansion of the periods about the rank two attractor points.

In the final stages of this work, we received communication from B\"onisch and Klemm~\cite{BonischAndKlemm} who inform us that they are able to express the second and third derivatives at $\vph{\;=}-\frac{1}{7}$, and so the unrecognized numbers above, in terms of periods and quasi-periods of the associated weight two and weight four forms.

\newpage
\section{Identities Involving the Instanton Numbers}\label{section: identities for instanton numbers}
\vskip-10pt
As foreseen in \cite{Moore:1998pn}, knowledge of the periods at the attractor points leads to interesting identities that involve the instanton numbers of the mirror \cym. In our case, we find identities that involve the instanton numbers, the critical values of the $L$-functions associated to each rank two attractor point and the modulus of the associated elliptic curve.

\subsection{Genus zero}
\vskip-10pt
The homogeneous genus zero prepotential $\cF_0$ and the corresponding inhomogeneous prepotential $F_0$  are given by
\beq
F_0\= \frac{\cF_0}{(z^0)^2}~;\qquad \cF_0\= \frac{1}{2} z^a\partial_a\cF_0~,
\label{eq:F0def}\eeq
where, in the second expression, $z^a$ and $\partial_{a}\cF_0$ are components of the vector of periods $\Pi$. In order to compute the genus zero instanton numbers, we must expand $F_0$ in terms of the complexified K\"ahler parameter $t$ of the mirror manifold $\widetilde{X}$
\beq
t(\varphi)\= \frac{z^1(\varphi)}{z^0(\varphi)}~.
\notag\eeq
For $\varphi{\;=}-1/7$, we have already seen in Eq.~\eqref{eq:tstar} that
\beq
t_{*}\= t(-\tfrac{1}{7})=\frac{1}{2}+\frac{5\ii}{28}\frac{\pi L_4(1)}{L_4(2)}
\notag\eeq
and we find also that
\beq
F_0(t_{*})\;= -\frac{\k}{10}(2t_* - 1)(15 t_*- 4)~.
\notag\eeq

We find similar relations at the other two attractor points,  $t_{\pm}{\=}t(\varphi_{\pm})$ 
\begin{equation}
\begin{alignedat}{2}
t_{-}&\=\frac{5\ii(9+\sqrt{17})}{16\cdot 17} \frac{\p\l_4(1)}{\l_4(2)}~;~~~~~~~~~~~~
                      & F_0(t_-)&\=\frac{13\k}{10}t_{-}\\[10pt] 
t_{+}&\= \frac{t_{-} - 4}{6t_{-} - 7}~;      
                      & F_0(t_+)&\=\frac{\k}{170}\left(-36 + 313\,t_{+} - 480\,t_{+}^2\right)\\
\end{alignedat}
\label{eq: F at phi plus minus}
\end{equation}

We can extract instanton numbers $n_k$ of genus zero and degree $k$ from the expansion of $F_0$ near the large complex structure point. For our situation the expansion is
\beq
F_0(t)\;= - 2\k t^3 + \frac12 \k t + 4\k \frac{\z(3)}{(2\p\ii)^3} - \cI(t)
\label{eq: prepotential at large t}\eeq
where $\cI$ is the instanton sum given by
\beq
\cI(t)\= \frac{1}{(2 \pi \ii)^3}\sum_{k=1}^{\infty} n_k\,\text{Li}_3(\ee^{2\pi \ii k t})~.
\label{eq:Idef}\eeq

The genus zero and genus one instanton numbers of small degree are listed in the following table.

\begin{table}[H]
\label{Table: instanton numbers}
\renewcommand{\arraystretch}{1.1}
\begin{center}
\begin{minipage}{\textwidth}
\centering
\begin{tabular}{| l | l |}
\hline
\vrule height 14pt depth5pt width0pt $k$ & \hfil $n_k$\\
\hline\hline
1&$ 12 \k $  \\
2&$ 24 \k $  \\
3&$ 112 \k $  \\
4&$ 624 \k $  \\
5&$ 4200 \k $  \\
6&$ 31408 \k $  \\
7&$ 258168 \k $  \\
8&$ 2269848 \k $  \\
9&$ 21011260 \k $  \\
10&$ 202527600 \k $  \\
11&$ 2017537884 \k $  \\
12&$ 20654747200 \k $  \\
13&$ 216372489804 \k $  \\
14&$ 2311525544064 \k $  \\
15&$ 25115533695300 \k $  \\
16&$ 276942939016224 \k $  \\
17&$ 3093639869100240 \k $  \\
18&$ 34957447938066952 \k $  \\
19&$ 399082284262216044 \k $  \\
20&$ 4598143339631725920 \k $  \\[2pt]
\hline
\end{tabular}
\hskip-4pt
\begin{tabular}{l |}
\hline
\vrule height 14pt depth5pt width0pt \hfil$d_k$\\
\hline\hline
$ 20-10 \k$  \\
$ 102-30 \k$  \\
$ 1180-438 \k$  \\
$ 12096-4428 \k$  \\
$ 133780-48938 \k$  \\
$ 1511730-550266 \k$  \\
$ 17647076-6407530 \k$  \\
$ 210201644-76161400 \k$  \\
$ 2545255572-920643442 \k$  \\
$ 31212421126-11273118446 \k$  \\
$ 386727907536-139494386712 \k$  \\
$ 4832555488984-1741106040676 \k$  \\
$ 60820504439296-21890039477888 \k$  \\
$ 770125991800110-276916193102934 \k$  \\
$ 9802710122549832-3521744606381596 \k$  \\
$ 125345358831091796-44996106417473728 \k$  \\
$ 1609189343845395964-577237489764357422 \k$  \\
$ 20732103878422556262-7431797271319182118 \k$  \\
$ 267947664660167267360-95989385991015664456 \k$  \\
$ 3472847998674908410256-1243366526895209656540 \k$  \\[2pt]
\hline
\end{tabular}
\end{minipage}
\capt{6in}{tab:InstantonNumbers}{The first few instanton numbers, $n_k$, for genus zero, and $d_k$, for genus one.}
\end{center}
\end{table}
\vskip-15pt
The attractor point at $\vph_- {\=}33-8\sqrt{17}$ lies within the radius of convergence of series expansion of the periods around $\vph{\=}0$ where the expansion in Eq.~\eqref{eq: prepotential at large t} is valid which means that by combining Equations~\eqref{eq: F at phi plus minus}, \eqref{eq: prepotential at large t} and \eqref{eq:Idef} we find a remarkable identity that involves the instanton numbers and special values of the weight four $L$-function associated with $\vph_-$.
\beq
\cI(t_{-}) \= \k\left( -2t_{-}^3 + \frac45 t_{-} + 4 \frac{\z(3)}{(2\p\ii)^3}\right)~.
\notag\eeq

The other two rank two attractors at $-\frac{1}{7}$ and $\varphi_+$ lie outside the radius of convergence of the instanton sum and the identities require a little more care. For $\vph{\;=}-1/7$ the partial sums of $\cI$ give rise to the plot \fref{fig: instanton partial sums}.  Although the sum diverges it responds well to the techniques of accelerated convergence \cite{Bender:1978aa}. The vertical axis in \fref{fig: instanton partial sums} is marked in steps of $10^{-7}$, so the simple expedient of computing the partial sum to say 100 terms and then taking half of the next term already gives the desired value to 8 figures. More sophisticated methods, such as an iterated Shanks transformation, or using a Pad\'{e} approximant, give better approximations. The Pad\'e approximant to $\cI$, for example, with numerator and denominator of order 400 in $q$, satisfies the expected identity to 435 figures. The point is that while $\cI(t)$ is defined by \eqref{eq:Idef} where the instanton sum converges, it is defined by \eqref{eq: prepotential at large t} and \eqref{eq:F0def} in terms of the periods, which are analytic throughout the cut plane. So any method of summation that returns the value of the analytic continuation will return a value that satisfies the identity. The attractor point $\vph_{+}$ is well outside the region where the instanton sum converges, yet the same Pad\'e approximants converge to the desired result albeit more slowly. The approximant with numerator and denominator of degree 400 now gives the desired value correct to 55 figures and this precision improves as we increase the number of terms in the approximant.

\begin{figure}[H]
\centering
\hspace*{-1.12cm}\includegraphics[keepaspectratio,height=7.5cm]{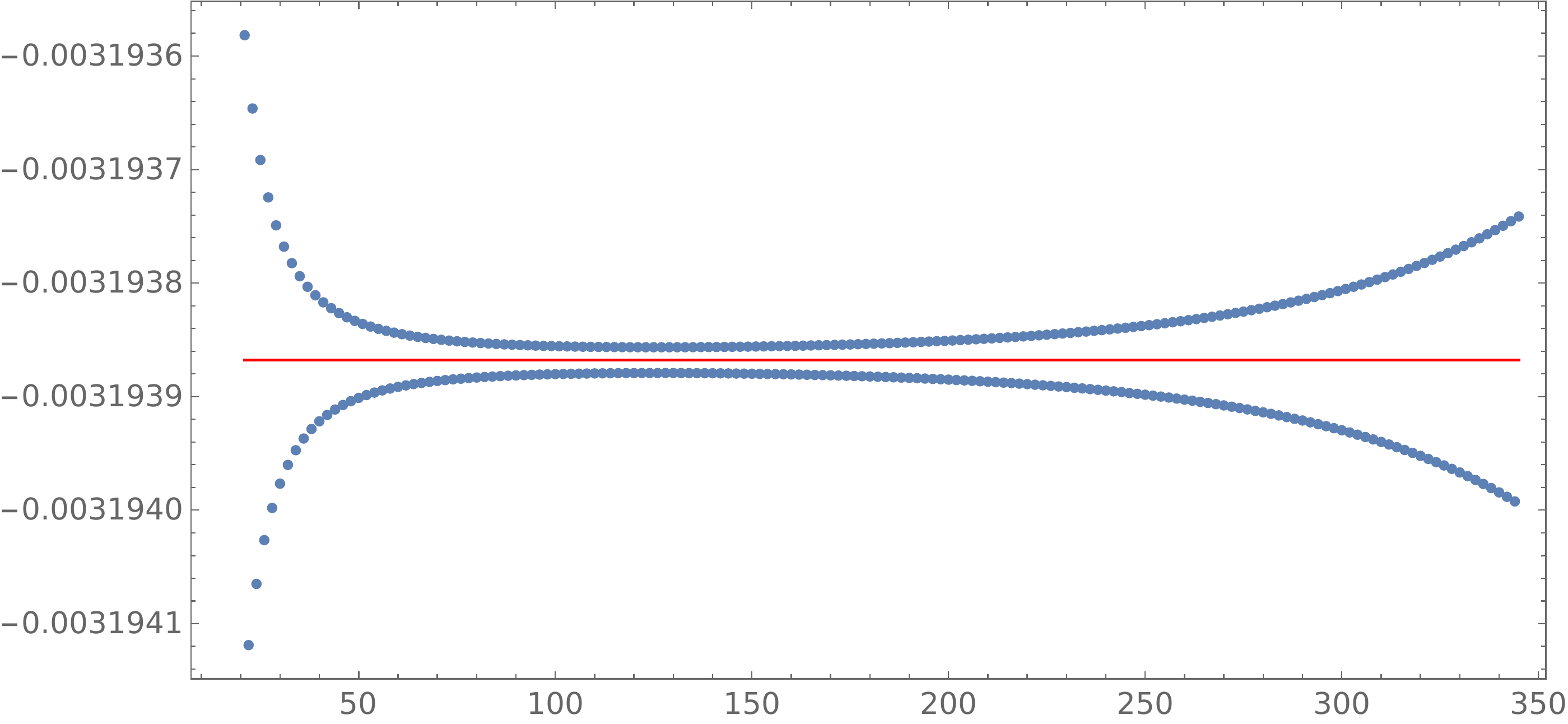}
\capt{5.3in}{fig: instanton partial sums}{A plot of the partial sums of\/ $\emph{Im}\,\cI(t_{*})$, for $\k{\;=\;1}$, up to order $q^n$ for $20\leq n \leq 350$. The expected value is shown in red.}
\end{figure}

Returning to the identities: note that in addition to computing the value of $F_0$ at the rank two attractor points, we may also compute the derivative of $F_0$ at these points and this leads to new identities. For example,
\beq
F_0'({t_{*}}) \= \k(3-6t_{*})~;\qquad
F_0'({t_-}) \= 3\k~; \qquad
F_0'({t_+}) \=  \frac{3\k}{17}(11 - 32 t_+)~.
\notag\eeq

Similarly, the second derivatives are given by
\begin{gather*}
F_0''(t_{*}) \= 
-\k\, \frac{ 30(1-2\t_{*}^\perp) t_{*} + 44\t_{*}^\perp - 25 }{5(1-2\t_{*}^\perp )t_{*} + 5\t_{*}^\perp - 3}~;\\[10pt]
F_0''(t_{-}) \= \frac{17\k}{5t_{-} - \t_{-}^\perp}~;\qquad
F_0''(t_{+})\= -\frac{\k}{17}\,\frac{ 480(3-4\t_{+}^\perp) t_{+} + 932\t_{+}^\perp - 1107 }{ 5(3-4\t_{+}^\perp)t_{+} + 9\t_{+}^\perp - 11 }~,
\end{gather*}
where $\t_{*}^\perp$, $\t_{-}^\perp$ and $\t_{+}^\perp$ are the parameters of the lattices $\L^\perp$ at the attractor points.

We find similar identities for the third derivatives. However, the expression become more complicated and, for example, we find that we cannot package the $L$-function values into~$t$. 

\subsection{Genus one}\label{sec:Genus1}
\vskip-10pt
The numbers of higher genus instantons can also be computed from a knowledge of the periods \cite{Bershadsky:1993ta,Bershadsky:1993cx}. Thus, we expect identities analogous to those in the previous subsection to be satisfied by the generating functions of higher genus instanton numbers. 

The genus one free energy can be computed from the holomorphic anomaly equation \cite{Bershadsky:1993ta} and, in the topological limit where $\overline{t}\rightarrow \ii \infty$ and $t$ is kept finite, we recover the generating function
\begin{equation*}
F_1(t)\=\log\bigg\{\vph^{-1-\k}\, \vp_0(\vph)^{4-\frac{2\k}{3}}\,\frac{d\vph}{dt}\,f_1(\vph)\bigg\}
\end{equation*}
where $f_1$ is the holomorphic ambiguity given by
\begin{equation}
f_1(\varphi)\=(1-25\varphi)^{a}(1-9\varphi)^{b}(1-\varphi)^{c}.
\label{eq:f1}
\end{equation}

$F_1$ has the the large volume expansion
\begin{equation}
F_1(t)\;= -2\pi \ii \k t -\sum_{k=1}^\infty \bigg\{2 d_k \log\bigg(\prod_{r=1}^\infty\big(1-q^{rk}\big)\bigg) + \frac{1}{6} n_k\,\log(1-q^k)\bigg\}+\text{const.}
\label{eq: F1 at large volume}
\end{equation}
from which the genus one instanton numbers can be extracted, once the holomorphic ambiguity is fixed.

The exponents appearing in the holomorphic ambiguity are determined by the singularity corresponding to each factor. For example, at a conifold point where an $S^3$ shrinks, the exponent is known to be $-\frac{1}{6}$. In cases where one considers the quotient of a conifold singularity by a finite group $G$, the exponent is given\cite{Gopakumar:1997dv,van38monodromy} by $-\frac{|G|}{6}$. From the analysis of \sref{section: A quotient of a Hulek--Verrill manifold}, we see that this happens at all three singularities. At $\varphi=\frac{1}{25}$ and $\varphi=\frac{1}{9}$, the exponent is determined by the order of the group that fixes the singularity when we take the quotient. The exponent at $\varphi=1$ is complicated by the fact that there are two singularities on the manifold that are each fixed by a group of order $\frac{2}{\kappa}$. We assume that the effect of the two singularities is to double the exponent and set
\begin{equation}
a\;=-\frac{10}{6\kappa}~;~~~~~~~~~b\;=-\frac{2}{6\kappa}~;~~~~~~~~~c\;=-\frac{4}{6\kappa}.
\label{eq:exponentsoff1}
\end{equation}
Alternatively, one may read off these exponents from the Picard-Lefschetz form of the monodromy matrices as in Eq.~\eqref{eq:PLformOfMonodromy} and \tref{tab:ConifoldVectors} of \sref{sec:periods}, see also \cite{van38monodromy}. Either way, this yields the integral instanton numbers listed in Table~\ref{tab:InstantonNumbers}.

As with $F_0$, we can evaluate $F_1$ at the attractor points. However, our expression for $F_1$  is not particularly enlightening due to the unknown constant in Equation~\ref{eq: F1 at large volume}. We can however write down identities involving the derivatives of $F_1$. 
\newpage
\section{Possible Geometrical Origin of the Splitting}\label{sec:SpeculationsOnGeometry}
\vskip-10pt
The calculations of this paper provide overwhelming evidence
for a splitting of $H^3(X_{\varphi})$ into a sum of two 2-dimensional
pieces for $\vph{\;=}-1/7$ and $\varphi{\=}33\pm 8\sqrt{17}$. Standard conjectures
(the Hodge conjecture and the Tate conjecture) imply that there exists a {\em geometrical 
explanation} which, once identified,  would lead to a rigorous
proof of our observations on the splitting of the Frobenius polynomials
and the expression of periods in terms of $L$-values.
For the sake of concreteness, we will concentrate on the variety $X:=X_{-1/7}$,
but the arguments are of a general nature and apply, mutatis mutandis,
also to $\varphi{\=}33 \pm 8\sqrt{17}$.

One of the simplest explanations for the splitting would be that $X$ has self-map $\iota$,
acting as $\mathbbl{1}$ on $H^{3,0}\oplus H^{0,3}$ and $\mathbbl {-1}$ on $H^{1,2}\oplus H^{2,1}$.
Such a transformation might arise from a self-correspondence of the family
$X_{\varphi}$, for which $\varphi=-1/7$ is a fixed point, but we have been unable 
to find such a map and the properties of the Picard-Fuchs equation make its
existence doubtful.  In \cite{Cynk2019} a very non-trivial
example of such a map (defined over $\IQ(\sqrt{2}))$ was exhibited for a 
certain Calabi-Yau threefold (defined over $\IQ$), which then led to a proof 
of Hilbert modularity for that particular variety. 

Cusp forms of weight two for $\Gamma_0(N)$ can be identified with 
holomorphic one-forms on the {modular curve} $X_0(N){\=}\overline{\IH/\Gamma_0(N)}$, which  is the moduli space of elliptic curves with a subgroup of order 
$N$. The union of these elliptic curves makes up the {elliptic modular surface} 
$\mathcal{E} \rightarrow X_0(N)$ and weight three modular forms for $\Gamma_0(N)$ can be identified with holomorphic two forms on $\mathcal{E}$. 
More generally, a weight $k$ cusp form for $\Gamma_0(N)$ gives a $(k-1)$-form on the {Kuga-Sato variety} $\mathcal{E}^{(k-2)}$, defined as
the $k-2$ fold fibre product of elliptic surface $\mathcal{E} \rightarrow X_0(N)$.

The Hodge conjecture would imply the existence of a correspondence 
between our variety $X$ and the Kuga-Sato threefold $\mathcal{E}^{(2)}$ for $\Gamma_0(14)$, such that the holomorphic three form of $X$ pulls back to the 
modular form $f_{\bf 14.4.a.a}$ of LMFDB. The correspondences are expected to 
exist  also for all rigid Calabi-Yau threefolds, but only in very few cases 
have these been found explicitly. For an overview of known cases we refer to 
the thesis of C.\ Meyer \cite{DissertationMeyer}. As critical $L$-values of the modular forms 
are tautologically periods of the corresponding three-form on $\mathcal{E}^{(2)}$, such correspondences also provide the rationale for Deligne's conjecture, referred to~in~\SS4. 

Note that in our case the modular curve $X_0(14)=:E$ is itself an elliptic curve.
The piece  \hbox{$H^{2,1}{\oplus }H^{1,2}$} of $H^3(X)$ corresponds to the weight two 
modular form, with LMFDB label {\bf 14.2.a.a}, via a Tate-twist. Now the Hodge 
conjecture, applied to $H^4(E \times X)$, predicts the existence of a 
surface $S$ inside $E \times X$, which can be seen as a family of curves 
in $X$, parametrized by $E$, which leads to a diagram
\[
\begin{array}{cccc}
E \times X ~\supset\hskip-10pt{}& S &\hskip-5pt\stackrel{\displaystyle q} \longrightarrow& X\\
                  & \phantom{p}\downarrow p& &\\
                  & \hskip-4pt E           & &
\end{array}
\] 
such that pulling back the $(2,1)$-form of $X$ via $q$ and integrating 
over the fibres of $p$ gives the holomorphic one-form {\bf 14.2.a.a} 
on $E$. Poincar\'e dually, taking the image under $q$ of the inverse 
image under $p$ of a cycle $\gamma \in H_1(E)$ produces elements 
$T(\gamma):=q_*p^!(\gamma) \in H_3(X)$, that maps $H_1(E)$ to the $H^{1,2}\oplus H^{2,1}$-part of $H_3(X)$.

Geometrically the simplest scenarios would be that $S$ is the union of
rational curves that are parametrized by $E$, so that $\pi:S \to E$ it is a 
ruled surface over $E$, embedded in $X$. Then clearly $H^3(S){\=}H^1(E)(-1)$ at 
the level  of Hodge structures, where the $(-1)$ denotes the Tate-twist, which 
makes from the weight one Hodge structure $H^1(E)$ a Hodge structure 
$H^1(E)(-1)$ of weight $3$, which on the level of arithmetic leads 
to the extra $p$ in the factor of~$R(T)$. 

As the normal bundle $N_C$ to a smooth rational curve $C$ in a Calabi-Yau 
threefold always has a degree $-2$, it follows from the fact 
that we have a one-parameter family of such curves that the normal bundle 
contains a trivial summand, and hence
\[ N_C=\cO_C \oplus \cO_C(-2) .\]
This means that each of the lines can be blown down to a point. When we
perform this contraction for all the rational curves of $S$, we obtain
a {\em singular Calabi-Yau threefold $Y$} that sits in a diagram
\[
\begin{array}{ccc}
S&\hookrightarrow\hskip-10pt{}&X\\
\downarrow&& \hphantom{p}\downarrow p\\
E & \hookrightarrow\hskip-10pt{}&Y\\
\end{array}
\] 
The elliptic curve can now be seen as the singular locus of $Y$;
the transverse type of singularity is a two-dimensional cone, i.e.
an $A_1$-singularity, which is resolved by a single blow-up and 
which restores the collapsed $\IP^1$'s. Although $Y$ is singular, 
its cohomology (with rational coefficients)
is just as that of a {smooth manifold}: Poincar\'e duality holds and
the Hodge structure remains pure. The reason is that the two dimensional
$A_1$ singularity is a {quotient singularity}; locally it is the 
quotient of $\IC^2$ by the $\IZ/2\IZ$ acting by identifying antipodal points.
As a result, the space  $Y$ also has only 
quotient singularities, so is what is sometimes called a $V$-manifold or 
a $\IQ$-homology manifold. Arithmetically, if we would count points on $Y$,
the Weil-conjectures would hold, and the factor $R(T)$ of $Y$ would 
be of degree two, and give rise to the weight four modular form.

The variety $X$ sits in a one-parameter family with fibres $X_{\varphi}$. 
Under this deformation, the surface $S$ completely disappears. In general,
if we have such a surface over a curve of genus~$g$, then after deformation
one generically ends up with $2g{\,-\,}2$ isolated rational curves with
normal bundle $\cO(-1) \oplus \cO(-1)$.

The singular variety $Y$ sits in a corresponding one-parameter family
$\mathcal{Y}$.
If we move away from the splitting point, each of the transverse 
cones is smoothed out and the variety $Y_{\varphi}$ becomes smooth.
If we denote the total space of the family $X_{\varphi}$ and $Y_{\varphi}$
by $X$ and $Y$ respectively, we get a diagram
\[
\begin{array}{ccccc}
X&\hookrightarrow&\mathcal{X} & \hookleftarrow & X_{\varphi}\\
\hphantom{p}\downarrow p&&\hphantom{p}\downarrow p&&\hphantom{p}\downarrow p \\
Y & \hookrightarrow&\mathcal{Y} &\hookleftarrow &Y_{\varphi}\\
\end{array}
\] 
For general $\varphi$ ($\neq -1/7$), the map at the right hand side is an
isomorphism, whereas on the left hand side we have the contraction
of the ruled surface $S$ onto its base $E$. Locally around each of the
singularities, we just have the phenomenon of {\em simultaneous resolution}
of the $A_1$-singularty, crossed with the elliptic curve $E$.

We can describe the change in cohomology between $Y$ and $Y_{\varphi}$
in terms of {\em vanishing cycles}. This general formalism also provides 
control on the level of Hodge structures. Without going into details, 
it can be shown that we obtain short exact sequences
\[ 0 \lra H^k(Y) \lra H^k_{\lim}(\mathcal{Y}) \lra \IH^k \lra 0 .\]
The middle term is a group  isomorphic to $H^k(Y_{\varphi})$, where $\varphi$ is
(infinitesimally) close to $-1/7$. It carries a so-called {\em limiting mixed Hodge structure}, that is described by the asymptotic limiting behaviour of the periods if  $\varphi$ tend to $-1/7$. In our case the limiting mixed Hodge structure is pure. The term $\IH^k$ decribes the vanishing cohomology. In this situation
it can be shown that
\[ \IH^k=H^{k-2}(E)(-1).\]
In particular for $k=3$ we find the sequence
\[ 
0 \lra H^3(Y) \lra H^3_{\lim}(\mathcal{Y}) \lra H^1(E)(-1) \lra 0~.
\]
We see that in fact we have an isomorphism of Hodge structures
\[ 
H^3_{\lim}(\mathcal{Y}) \= H^3(X)~.
\]

Dually to the group $\IH^3$, we have a rank two lattice of vanishing cycles
isomorphic to $H_1(E)$, which are now realised geometrically as union of
the vanishing two-spheres over a $1$-cycle $\gamma \in H_1(E)$. If we
use the isomorphism $X_{\varphi}{\=}Y_{\varphi}$, these cycles get mapped to
the cycles $T(\gamma) \in H_3(X)$, mentioned above.

There are several alternative scenarios that effectively 
produce similar phenomena. Rather then a single surface $S$, one
may have a chain $S_1,S_2,\ldots, S_r$ of surfaces that intersect in
copies of $E$. The whole chain could collapse, producing a singular
threefold $Y_*$ with an $A_r$-singularity. More generally, one may
consider collections of such surfaces intersection in an $ADE$-graph.
In all these cases the phenomenon of simultaneous resolution takes
place and one obtains $\IQ$-homology manifolds $Y$ and isomorphisms 
$X_{\varphi}{\=}Y_{\varphi}$. All this is studied by Katz, Morrison and 
Plesser in \cite{Katz:1996ht}.

In the paper of Hulek-Verrill \cite{hulek2005modularity} many splitting 
Calabi-Yau threefolds are identified by explicitly exhibiting certain 
non-trivial surfaces inside them. In these cases one is dealing with 
families of elliptic curves inside the threefold and something 
similar could happen in our exmples. A natural question seems to be:
is there a copy of the elliptic modular surface $\mathcal{E} \to E{\=}X_0(14)$ 
inside $X$? It is not clear to us what exactly we should be looking for; 
in fact one of the great problems with the Hodge conjecture 
is that it does not directly give geometrical information on the cycle 
that realises the splitting. 

In \cite{cynkfreitagsalvati-manni} there is a large  collection of 
Siegel-modular Calabi-Yau varieties with geometrically unexplained 
splits of $H^3$, that include the earlier examples from \cite{vanGeemen1995} 
and \cite{vangeemenvanstraten}.
The tables of the dissertation of Meyer~\cite{DissertationMeyer} also contain many examples of varieties
which split on an experimental level and in fact the split at $\varphi=-1/7$,
that we have studied here, was mentioned already on p.157 of the dissertation! 
The systematic study of these splits from the perspective of the attractor 
mechanism and special $L$-values seems a natural field of further inquiry.
\newpage
\section{Conclusion and Speculations}\label{section: conclusion and speculations}
\vskip-10pt
We have found examples of rank two attractor varieties by studying the arithmetic structure of the Calabi-Yau manifold AESZ34. More precisely, one expects that the Frobenius polynomial associated to the middle cohomology of a one-parameter family of Calabi-Yau manifolds will factor into two quadrics whenever the parameter solves a certain polynomial $G(\vph)$, with integer coefficients. A computer analysis of these factorisations found a linear and a quadratic factor of $G(\vph)$ and the associated roots of this polynomial revealed the examples in this paper.\footnote{In fact, many more examples of non-rigid modular Calabi-Yau threefolds can be found in \cite{DissertationMeyer} that should also correspond to attractors of rank two.}

In the remainder of this section, we speculate on the physical significance of our results and certain unanswered questions.

\subsection{Entropy and topological strings}
\vskip-10pt
A consequence of modularity is that certain physical quantities may be expressed in terms of critical $L$-function values, such as the area of the horizon of a black hole, as in Eqs.~\eqref{eq: horizon area at -1/7} and \eqref{eq: horizon area at phipm}. Since this is proportional to the entropy of a black hole, in the limit of large charges, it is natural to presume that the modular forms are playing a role in the counting of microstates. A direct count of these, for this class of black holes, should shed light on the precise role that these modular forms play. This enumeration of microstates, however, remains a difficult problem for $\mathcal{N}{\=}2$ black holes. 

Rigid Calabi-Yau manifolds are trivially rank two attractors and are known to be modular over $\IQ$ \cite{GuveaYui}, as a consequence of the proof of the Serre  conjecture. Moreover, one expects that the periods of rigid \cys are given by critical $L$-values of the associated weight four form. For example, expressions for the periods, similar to those presented here, can be found in \cite{2017arXiv170909751C}. So, it is expected that the area of the horizon of a black hole, associated to a rigid Calabi-Yau manifold, may also be expressed in terms of ratios of critical $L$-values. 

We have seen, in \sref{section: identities for instanton numbers}, that topological string free energies at genus zero and one, when evaluated at a rank two attractor point, may be expressed in terms of $L$-function values and the modulus of an elliptic curve. We expect that these relations have analogues for all genera. The computation of topological string free energies and the computation of black hole entropy are not independent. They are related, for example, by the well known conjecture of Ooguri, Strominger and Vafa \cite{Ooguri:2004zv} which states that one may compute the entropy of $\cN=2$ black holes, that arise in Type II compactifications, by computing topological string free energies. We summarise this triangle of ideas with the following diagram.

\vskip10pt
\begin{figure}[H]
\centering
\hspace*{17pt}
\begin{tikzpicture}[auto]
\node[rounded corners, ultra thick, minimum height=1cm, minimum width = 6.5 cm, draw] (entropy) {\textbf{Entropy of 4d Black Hole}};
\node[draw=none,fill=none, below=0cm of entropy] (belowentropy) {};
\node[draw=none,fill=none, right=0cm of entropy] (rightofentropy) {};

\node[rounded corners, ultra thick, minimum height=1cm, minimum width = 6.5 cm, draw, below = 3 cm of entropy] (instantons) {\textbf{GV Invariants of Mirror CY}};
\node[draw=none,fill=none, above=0cm of instantons] (aboveinstantons) {};
\node[draw=none,fill=none, right=0cm of instantons] (rightofinstantons) {};

\draw[<->, >=stealth, very thick] (belowentropy) -- node [text width=3.3cm,midway,left] {OSV Conjecture} (aboveinstantons)
node[rounded corners, ultra thick, minimum height=1cm,minimum width=4.7cm, draw, midway,right=5] (modularity) {\textbf{Modularity of CY}};

\node[draw=none,fill=none, above left=-0.5cm and 0cm of modularity] (leftofmodularity1) {};
\node[draw=none,fill=none, below left=-0.5cm and 0cm of modularity] (leftofmodularity2) {};

\draw[<->,  >=stealth,very thick] (rightofentropy) -- node [text width=6cm,midway,above right=-0.1cm and 0.4cm]{} (leftofmodularity1);
\draw[<->, >=stealth,very thick]  (rightofinstantons) -- node [text width=6cm,midway,below right=-0.1cm and 0.4cm] {Identities such as $F_0(t_-)=\frac{13\k}{10}t_-$ \\[3pt]  where $t_-=\frac{5\ii}{v}$\\[-10pt]}(leftofmodularity2);
\end{tikzpicture}
\vskip0pt
\place{3.75}{2.05}{
\parbox{6cm}{\small 
$S(Q_{kl})= \frac{17\pi}{2}\left\{\frac{k^2} v + \ell^2 v \right\}+...$\\[2pt] 
where $v=\frac{17(9-\sqrt{17})}{4}\frac{\lambda_4(2)}{\pi\lambda_4(1)}$}}
\end{figure}
\vskip10pt

\subsection{Massless states in $\L^\perp$}
\vskip-10pt
Rank two attractor varieties, in one parameter families of Calabi-Yau manifolds, come with two rank two lattices
\begin{equation*}
\Lambda\oplus\Lambda^{\perp}\subset H^3(X,\IZ) 
\end{equation*}
which we recall are such that
\begin{equation*}
\Lambda\otimes\IC=H^{3,0}\oplus H^{0,3}~~~~~~~\text{and}~~~~~~~\Lambda^{\perp}\otimes\IC=H^{2,1}\oplus H^{1,2}  .
\end{equation*}
In this paper, we have mostly focused on the charge lattice $\Lambda$ and put less emphasis on $\Lambda^{\perp}$, even though the elements of $\Lambda^{\perp}$ define central charges with the same critical point as those in $\Lambda$. The main difference is that the central charges corresponding to the points $\Lambda^{\perp}$ vanish at the attractor point and so lead to ``massless black holes"\footnote{Since $\Gamma\in\Lambda^\perp\subset H^{2,1}\oplus H^{1,2} \Longrightarrow Z(\Gamma,\varphi_{*})\propto\int_{X_{\varphi_*}}\Gamma\wedge\Omega=0$.}. There is an apparent paradox: if $\Lambda^{\perp}$ contains BPS states, then these would lead to a singularity in the moduli space at the attractor point, as in the case of a conifold point. Moreover, if there are infinitely many BPS states among the points of $\Lambda^{\perp}$, then this singularity at the attractor point will be very severe and, at least conjecturally, will be at infinite distance in moduli space.

A possible resolution of the paradox is that the putative singularities cancel out.\footnote{We are grateful to Albrecht Klemm for pointing this out and for discussions on the resulting field theory.}
It was shown in \cite{Vafa:1995ta} that, at a point in the moduli space where D-branes become massless, the genus one free energy develops a logarithmic singularity. This singularity is of the form
\begin{equation}
F_1\sim-\frac{1}{6}\sum_i (-1)^{s_i}~\text{log}\,m_i^2~,
\notag\end{equation}
where $m_i$ is the mass of a state that vanishes at the singularity and $s_i=0$ or $1$ for a hypermultiplet or a vectormultiplet respectively \cite{Katz:1996ht,Klemm:1996kv}. For example, at a conifold point one introduces a single hypermultiplet, which agrees with the exponents in the genus one holomorphic ambiguity in Eqs.~\eqref{eq:f1} and \eqref{eq:exponentsoff1}.

As discussed in \sref{sec:SpeculationsOnGeometry}, a plausible scenario, that explains the geometric origin of the weight two eigenform and $\Lambda^{\perp}$, is that there exists a $\IP^1$ bundle over an elliptic curve $E$ such that the total space $S$ is embedded in the attractor variety. The elliptic curves relevant for our examples are given by Eqs.~\eqref{eq: equation of elliptic curve at -1/7} and~\eqref{eq:EllipticCurveAtPhiPM} and $\Lambda^\perp$ is identified with the dual of the image of $H_3(S,\IZ)$. In this scenario, one obtains massless states by wrapping D3-branes on $\IP^1$ and either of the 1-cycles of the elliptic curve. We expect that one of these states contributes an adjoint hypermultiplet while the other contributes a vectormultiplet, so the singularity in~$F_1$~cancels.

A better understanding of the field theory at a rank two attractor point and the complete resolution of the above paradoxes requires a more involved analysis of the geometry to which we hope to return elsewhere.
\newpage
\appendix
\section{The Polyhedron and its Dual for the Singular Variety}
\vskip-10pt
We start by setting $X_5{\=}1$ in the Laurent polynomial \eqref{eq:PolynomialP} and listing the 21 monomials that the polynomial contains. These are 
\beq
1~;\qquad X_i~;\qquad  \frac{1}{X_j}~;\qquad \frac{X_i}{X_j}~,~i\neq j~;
\notag\eeq
where the indices take the values $i,j{\;\=\;}1,\ldots,4$. Writing these in a multi-index notation
\beq
X^v\= X_1^{v^1} X_2^{v^2} X_3^{v^3} X_4^{v^4}
\notag\eeq
we have a list of 21 vectors $v$ in $\IZ^4$. The convex hull of these points yields a four dimensional polyhedron $\D$. We run this data through a computer code which produces the data shown in \tref{tab:DeltaData} and \tref{tab:NablaData}. The code numbers the vertices of $\D$ in an arbitrary way.
      However, it is not possible to order the points of both $\D$ and $\nabla$ in a nice way and also have a nice form for the duality map. So we accept this ordering and at least have a simple duality~map.
      
\tref{tab:DeltaData} gives the data for $\D$. The first sub-table lists and numbers the vertices. There are 20 of these, so all the points of $\D$, apart from the interior point $\{0,0,0,0\}$, corresponding to the monomial $1$, are vertices. We see from the second sub-table that $\D$ has 30 three-faces. We will follow the usage of toric geometry and refer to the top dimensional faces as \emph{facets} in the following. The table gives the equations of each of these facets and lists the vertices of $\D$ that lie in each of these. The equations of the facets each have integral coefficients and the constant terms are all $1$. This, together with the fact that there is precisely one interior point, makes the polyhedron reflexive.

Given that the constant term of each equation is $1$, each facet is specified by listing the coefficients of the coordinates $x[j]$, $j=1,\ldots,4$ in the corresponding equation. These vectors are the vertices of the dual polyhedron $\nabla$, whose data is given in \tref{tab:NablaData}, with the dual vertices listed in the order corresponding to the facets of $\D$. Thus vertex 1 of $\nabla$ is $\{-1,0,0,0\}$, for example. The dual of $\nabla$ is again $\D$ so we see, for example, that the coefficients defining the first dual-face are $\{-1,0,0,1\}$ which is just vertex 1 of $\D$.
      It happens that the only lattice point of $\nabla$, apart from the vertices is the origin.
      
Let us return to considering $\D$ and \tref{tab:DeltaData}. We see that $\D$ has 10 facets that each have 4 vertices, so these are tetrahedra, and 20 facets that have 6 vertices, each of these is a prism with triangular section. We can hope to gain some understanding of the combinatorics of the polyhedra by seeing how the faces fit together. A first consideration is how the symmetries act on the polyhedra. Let us denote by $\ccA$ the $\IZ/5\IZ$ generator with the action
\beq
\ccA\,: X_i \to X_{i+1}~.
\notag\eeq
This acts on the monomials and so on the vertices $v_r$, of the polyhedron via the rule
\beq\begin{split}
\ccA_v\= \big\{
&v_1\to v_{12},\,v_2\to v_3,\,v_3\to v_{15},\,v_4\to v_5,\,v_5\to v_{14},\,v_6\to v_{11},\,v_7\to
   v_6,\,v_8\to v_{16},\\
&v_9\to v_8,\,v_{10}\to v_{13},\,v_{11}\to v_{18},\,v_{12}\to v_9,\,v_{13}\to v_2,\,v_{14}\to v_{17},\,v_{15}\to v_{10},\,v_{16}\to v_1,\\
&v_{17}\to v_{19},\,v_{18}\to v_{20},\,v_{19}\to v_4,\,v_{20}\to v_7\big\}~.
\notag\end{split}\eeq
\begin{table}[H]
\centering
\begin{minipage}{\textwidth}
\centering 
\begin{tabular}[t]{|rr@{\hskip1pt}r@{\hskip3pt}r@{\hskip3pt}r@{\hskip3pt}r@{\hskip3pt}r|}
\hline
\multicolumn{7}{|c|}{}\\[-14pt]
\multicolumn{7}{|c|}{Vertices}\\[2pt]
\hline\hline
\multicolumn{7}{|c|}{}\\[-12pt]
 1 & \{&-1,&0,&0,&1&\} \\
 2 & \{&-1,&0,&1,&0&\} \\
 3 & \{&0,&-1,&0,&1&\} \\
 4 & \{&0,&0,&-1,&1&\} \\
 5 & \{&0,&0,&0,&-1&\} \\
 6 & \{&0,&0,&0,&1&\} \\
 7 & \{&0,&0,&1,&-1&\} \\
 8 & \{&0,&1,&0,&-1&\} \\
 9 & \{&1,&0,&-1,&0&\} \\
 10 & \{&1,&0,&0,&-1&\} \\
 11 & \{&-1,&0,&0,&0&\} \\
 12 & \{&0,&-1,&0,&0&\} \\
 13 & \{&0,&1,&0,&0&\} \\
 14 & \{&1,&0,&0,&0&\} \\
 15 & \{&0,&0,&-1,&0&\} \\
 16 & \{&0,&0,&1,&0&\} \\
 17 & \{&-1,&1,&0,&0&\} \\
 18 & \{&1,&-1,&0,&0&\} \\
 19 & \{&0,&-1,&1,&0&\} \\
 20 & \{&0,&1,&-1,&0&\} \\[3pt]
 \hline
\end{tabular}
\hskip50pt
\begin{tabular}[t]{| r >{$}r<{$} >{$}l<{$} |}
\hline
\multicolumn{3}{|c|}{}\\[-14pt]
\multicolumn{3}{|c|}{Faces}\\[2pt]
\hline\hline
\multicolumn{3}{|c|}{}\\[-12pt]
 1 & -x[1]+1 & \{9,10,14,18\} \\
 2 & x[3]+x[4]+1 & \{5,8,9,10,15,20\} \\
 3 & -x[3]-x[4]+1 & \{1,2,3,6,16,19\} \\
 4 & -x[4]+1 & \{1,3,4,6\} \\
 5 & x[1]+x[4]+1 & \{2,5,7,8,11,17\} \\
 6 & -x[1]-x[4]+1 & \{3,4,6,9,14,18\} \\
 7 & x[4]+1 & \{5,7,8,10\} \\
 8 & x[1]+x[2]+1 & \{1,2,3,11,12,19\} \\
 9 & -x[2]-x[3]-x[4]+1 & \{1,2,6,13,16,17\} \\
 10 & x[1]+1 & \{1,2,11,17\} \\
 11 & x[1]+x[2]+x[3]+1 & \{1,3,4,11,12,15\} \\
 12 & -x[2]-x[4]+1 & \{1,4,6,13,17,20\} \\
 13 & x[1]+x[3]+1 & \{1,4,11,15,17,20\} \\
 14 & x[1]+x[2]+x[4]+1 & \{2,5,7,11,12,19\} \\
 15 & -x[2]-x[3]+1 & \{2,7,8,13,16,17\} \\
 16 & -x[3]+1 & \{2,7,16,19\} \\
 17 & x[2]+x[3]+1 & \{3,4,9,12,15,18\} \\
 18 & -x[1]-x[3]-x[4]+1 & \{3,6,14,16,18,19\} \\
 19 & x[2]+1 & \{3,12,18,19\} \\
 20 & -x[1]-x[2]-x[4]+1 & \{4,6,9,13,14,20\} \\
 21 & x[3]+1 & \{4,9,15,20\} \\
 22 & x[2]+x[4]+1 & \{5,7,10,12,18,19\} \\
 23 & x[1]+x[3]+x[4]+1 & \{5,8,11,15,17,20\} \\
 24 & x[2]+x[3]+x[4]+1 & \{5,9,10,12,15,18\} \\
 25 & x[1]+x[2]+x[3]+x[4]+1 & \{5,11,12,15\} \\
 26 & -x[1]-x[2]-x[3]-x[4]+1 & \{6,13,14,16\} \\
 27 & -x[1]-x[2]-x[3]+1 & \{7,8,10,13,14,16\} \\
 28 & -x[1]-x[3]+1 & \{7,10,14,16,18,19\} \\
 29 & -x[1]-x[2]+1 & \{8,9,10,13,14,20\} \\
 30 & -x[2]+1 & \{8,13,17,20\} \\[3pt]
 \hline
\end{tabular}
\end{minipage}
\vskip2cm
\capt{5.5in}{tab:DeltaData}{The data for the Newton polyhedron, $\D$. The first table lists the vertices of $\D$, while the second lists the three-faces. The lists on the right of the second table give the vertices of the corresponding face.}
\end{table}
\newpage
\begin{table}[H]
\centering
\begin{minipage}{\textwidth}
\centering
\begin{tabular}[t]{|rr@{\hskip1pt}r@{\hskip3pt}r@{\hskip3pt}r@{\hskip3pt}r@{\hskip3pt}r|}
\hline
\multicolumn{7}{|c|}{}\\[-14pt]
\multicolumn{7}{|c|}{Dual Vertices}\\[2pt]
\hline\hline
\multicolumn{7}{|c|}{}\\[-12pt]
 1 & \{&-1,&0,&0,&0&\} \\
 2 & \{&0,&0,&1,&1&\} \\
 3 & \{&0,&0,&-1,&-1&\} \\
 4 & \{&0,&0,&0,&-1&\} \\
 5 & \{&1,&0,&0,&1&\} \\
 6 & \{&-1,&0,&0,&-1&\} \\
 7 & \{&0,&0,&0,&1&\} \\
 8 & \{&1,&1,&0,&0&\} \\
 9 & \{&0,&-1,&-1,&-1&\} \\
 10 & \{&1,&0,&0,&0&\} \\
 11 & \{&1,&1,&1,&0&\} \\
 12 & \{&0,&-1,&0,&-1&\} \\
 13 & \{&1,&0,&1,&0&\} \\
 14 & \{&1,&1,&0,&1&\} \\
 15 & \{&0,&-1,&-1,&0&\} \\
 16 & \{&0,&0,&-1,&0&\} \\
 17 & \{&0,&1,&1,&0&\} \\
 18 & \{&-1,&0,&-1,&-1&\} \\
 19 & \{&0,&1,&0,&0&\} \\
 20 & \{&-1,&-1,&0,&-1&\} \\
 21 & \{&0,&0,&1,&0&\} \\
 22 & \{&0,&1,&0,&1&\} \\
 23 & \{&1,&0,&1,&1&\} \\
 24 & \{&0,&1,&1,&1&\} \\
 25 & \{&1,&1,&1,&1&\} \\
 26 & \{&-1,&-1,&-1,&-1&\} \\
 27 & \{&-1,&-1,&-1,&0&\} \\
 28 & \{&-1,&0,&-1,&0&\} \\
 29 & \{&-1,&-1,&0,&0&\} \\
 30 & \{&0,&-1,&0,&0&\} \\[3pt]
 \hline
\end{tabular}
\hskip1in
\begin{tabular}[t]{| r >{$}r<{$} >{$}l<{$} |}
\hline
\multicolumn{3}{|c|}{}\\[-14pt]
\multicolumn{3}{|c|}{Dual Faces}\\[2pt]
\hline\hline
\multicolumn{3}{|c|}{}\\[-12pt]
1 & -y[1]+y[4]+1 & \{3,4,8,9,10,11,12,13\} \\
 2 & -y[1]+y[3]+1 & \{3,5,8,9,10,14,15,16\} \\
 3 & -y[2]+y[4]+1 & \{3,4,6,8,11,17,18,19\} \\
 4 & -y[3]+y[4]+1 & \{4,6,11,12,13,17,20,21\} \\
 5 & -y[4]+1 & \{2,5,7,14,22,23,24,25\} \\
 6 & y[4]+1 & \{3,4,6,9,12,18,20,26\} \\
 7 & y[3]-y[4]+1 & \{5,7,14,15,16,22,27,28\} \\
 8 & y[2]-y[4]+1 & \{2,5,7,15,23,27,29,30\} \\
 9 & y[1]-y[3]+1 & \{1,2,6,17,20,21,24,29\} \\
 10 & y[1]-y[4]+1 & \{1,2,7,22,24,27,28,29\} \\
 11 & -y[1]+1 & \{5,8,10,11,13,14,23,25\} \\
 12 & -y[2]+1 & \{8,11,14,17,19,22,24,25\} \\
 13 & y[2]+1 & \{9,12,15,20,26,27,29,30\} \\
 14 & y[1]+1 & \{1,6,18,20,26,27,28,29\} \\
 15 & -y[3]+1 & \{2,11,13,17,21,23,24,25\} \\
 16 & y[3]+1 & \{3,9,15,16,18,26,27,28\} \\
 17 & -y[1]+y[2]+1 & \{5,9,10,12,13,15,23,30\} \\
 18 & y[1]-y[2]+1 & \{1,6,17,18,19,22,24,28\} \\
 19 & -y[2]+y[3]+1 & \{3,8,14,16,18,19,22,28\} \\
 20 & y[2]-y[3]+1 & \{2,12,13,20,21,23,29,30\} \\[3pt]
 \hline
\end{tabular}
\end{minipage}
\vskip2cm
\capt{5.5in}{tab:NablaData}{The data for the dual polyhedron, $\nabla$.}
\end{table}
\newpage
It is an agreeable fact that there is a $4{\times}4$ matrix $A$ that represents $\ccA$ as an action on the vertices, considered as four component column vectors
\beq
\ccA\,:\,v\to Av~;\qquad A\=\left(\hskip-3pt
\begin{array}{rrrr}
-1 & -1 & -1 & -1\\
 1 &  0 &  0  &  0\\
 0 &  1 &  0  &  0\\
 0 &  0 &  1  &  0\\
 \end{array}
 \right)
\notag\eeq

If $\ccB$ denotes the $\IZ/2\IZ$ generator with the action
\beq
\ccB\,: X_i \to \frac{1}{X_i}~,
\notag\eeq
then $\ccB$ permutes the vertices according to the rule
\beq\begin{split}
\ccB_v = \big\{
&v_1\to v_{10},\,v_2\to v_9,\,v_3\to v_8,\,v_4\to v_7,\,v_5\to v_6,\,v_6\to v_5,\,v_7\to
   v_4,\,v_8\to v_3,\\
&v_9\to v_2,\,v_{10}\to v_1,\,v_{11}\to v_{14},\,v_{12}\to v_{13},\,v_{13}\to
   v_{12},\,v_{14}\to v_{11},\,v_{15}\to v_{16},\,v_{16}\to v_{15},\\
&v_{17}\to v_{18},\,v_{18}\to
   v_{17},\,v_{19}\to v_{20},\,v_{20}\to v_{19}\big\}~.
\end{split}\notag\eeq
As a linear action on $v$ we have simply
\beq
\ccB\,:\,v\to - v~.
\notag\eeq

Since $\ccA$ and $\ccB$ act on the vertices of $\D$, they act also on the facets. The action of $\ccA$ is given~by
\beq\begin{split}
\ccA_f = \big\{
&f_1\to f_{30},\,f_2\to f_{27},\,f_3\to f_{11},\,f_4\to f_{25},\,f_5\to f_{18},\,f_6\to
   f_{23},\,f_7\to f_{26},\,f_8\to f_{17},\\
&f_9\to f_8,\,f_{10}\to f_{19},\,f_{11}\to f_{24},\,f_{12}\to f_{14},\,f_{13}\to f_{22},\,f_{14}\to f_6,\,f_{15}\to f_3,\,f_{16}\to f_4,\\
&f_{17}\to f_2,\,f_{18}\to f_{13},\,f_{19}\to f_{21},\,f_{20}\to f_5,\,f_{21}\to
   f_7,\,f_{22}\to f_{20},\,f_{23}\to f_{28},\\
&f_{24}\to f_{29},\,f_{25}\to f_1,\,f_{26}\to f_{10},\,f_{27}\to f_9,\,f_{28}\to f_{12},\,f_{29}\to f_{15},\,f_{30}\to f_{16}\big\}~.
\end{split}\label{eq:Arulef}\raisetag{17pt}\eeq
While the action of $\ccB$ is given~by
\beq\begin{split}
\ccB_f = \big\{
&f_1\to f_{10},\,f_2\to f_3,\,f_3\to f_2,\,f_4\to f_7,\,f_5\to f_6,\,f_6\to f_5,\,f_7\to f_4,\,f_8\to f_{29},
\,f_9\to f_{24},\\
&\,f_{10}\to f_1,\,f_{11}\to f_{27},\,f_{12}\to f_{22},\,f_{13}\to f_{28},\,f_{14}\to f_{20},\,f_{15}\to f_{17},\,f_{16}\to f_{21},\\
&f_{17}\to f_{15},\,f_{18}\to f_{23},\,f_{19}\to f_{30},\,f_{20}\to f_{14},\,f_{21}\to f_{16},\,f_{22}\to f_{12},\,f_{23}\to f_{18},\\
&f_{24}\to f_9,\,f_{25}\to f_{26},\,f_{26}\to f_{25},\,f_{27}\to f_{11},\,f_{28}\to f_{13},\,f_{29}\to f_8,\,f_{30}\to f_{19}\big\}~.
\end{split}\label{eq:Brulef}\raisetag{17pt}\eeq

Now we may think of the facets of $\D$ as the vertices of $\nabla$ and the vertices of $\D$ as the facets of $\nabla$, so the above rules determine how $\ccA$ and $\ccB$ act on $\nabla$. It is now an easy check that
\beq
\ccA\,: f\to \tilde{A}f ~;\qquad \tilde{A} \=
\left( \begin{array}{rrr@{\hskip5pt}r}
 0 & 0 & 0 & -1 \\
 1 & 0 & 0 & -1 \\
 0 & 1 & 0 & -1 \\
 0 & 0 & 1 & -1 \\
\end{array} \hskip-3pt\right)
\notag\eeq
and as a linear action for $\ccB$ we again simply have
\beq
\ccB\,:\,f\to - f~.
\notag\eeq

The following two figures give some insight into the combinatorics of the faces. In \fref{fig:FaceConnections}, the first pentagon corresponds to 15 facets which comprise 3 orbits of $\ccA$. The five vertices are facets which are tetrahedra. Each facet of a tetrahedron is joined to a triangular facet of a prism and the other triangular facet is joined to another tetrahedron. The lines of the pentagon correspond to the prisms and show these connections. The action of $\ccA$ on the pentagon corresponds to a $2\p/5$-rotation in the positive sense. The image, under $\ccB$, of the pentagon on the left, is the pentagon on the right, which rotates in the same way under $\ccA$.

\begin{figure}[H]
\centering
\includegraphics[width=0.4\textwidth]{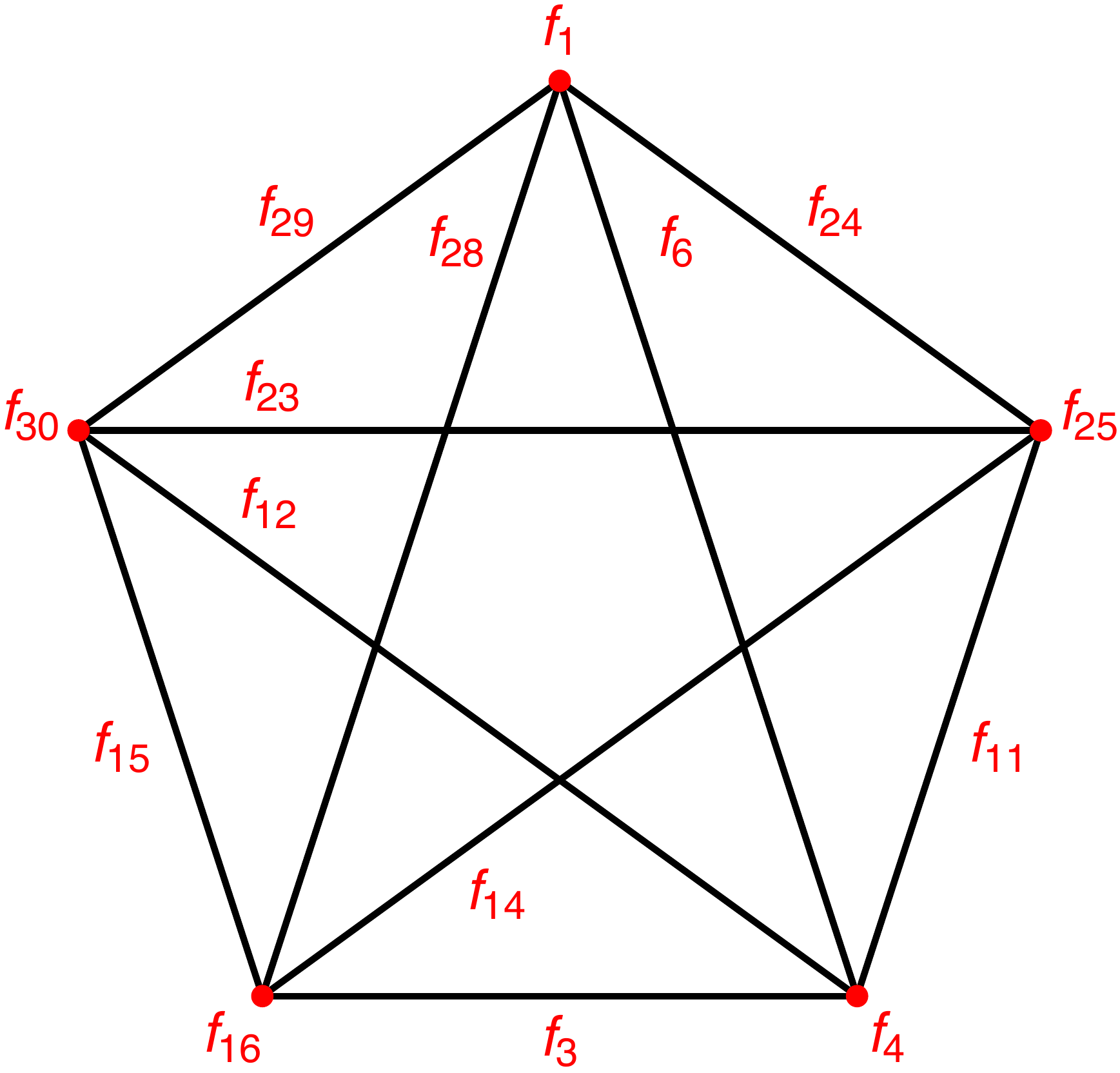}\hskip30pt
\includegraphics[width=0.4\textwidth]{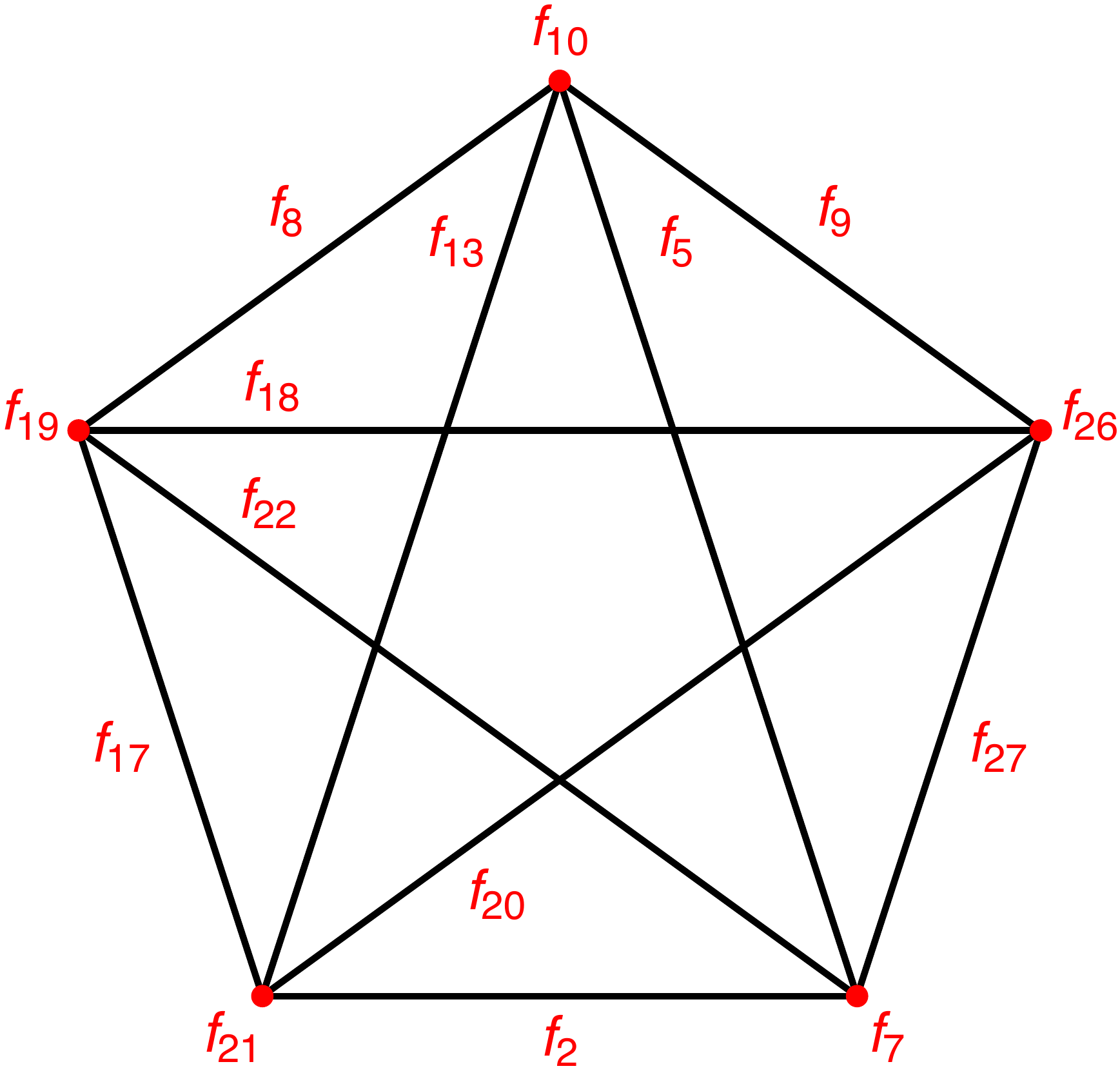}
\capt{5.4in}{fig:FaceConnections}{These pentagons display some of the connections between the facets. The vertices of the pentagons correspond to facets that are tetrahedra. Each prism has two triangular facets and each of these is joined to a tetrahedron. The lines of the pentagons correspond to these prisms and show how they link to the tetrahedra. The two pentagons are interchanged by $\ccB$, and $\ccA$ acts by rotation by~$2\p/5$.}
\end{figure}

In \fref{fig:tetrahedra}, we give a partial realisation of, say, the first pentagon in 3 dimensions. We start with a tetrahedron, say $f_1$, that is shown on the left in \fref{fig:tetrahedra} in red. To this are attached 4 prisms, three of which are visible in the figure and are coloured blue, green and yellow. Note that we refer to these solids as prisms and indeed they each have two triangular and three quadrilateral faces, but they are not regular prisms. The four prisms that have been attached to $f_1$ are $f_{29}$, $f_{28}$, $f_6$ and $f_{24}$. We make, in this way, a bigger tetrahedron. To each face of this big tetrahedron is attached another tetrahedron. These are $f_{30}$, $f_{16}$, $f_4$ and $f_{25}$. This corresponds to the figure on the right.
      
So far we have accounted for the four lines that emanate directly from $f_1$ in \fref{fig:FaceConnections}. There remain six lines, in the first pentagon and these correspond to six further prisms each of which connects two of the triangular faces visible in the figure on the right. These are not easily added to a three-dimensional figure.
      
\begin{figure}[H]
\centering
\includegraphics[width=\textwidth]{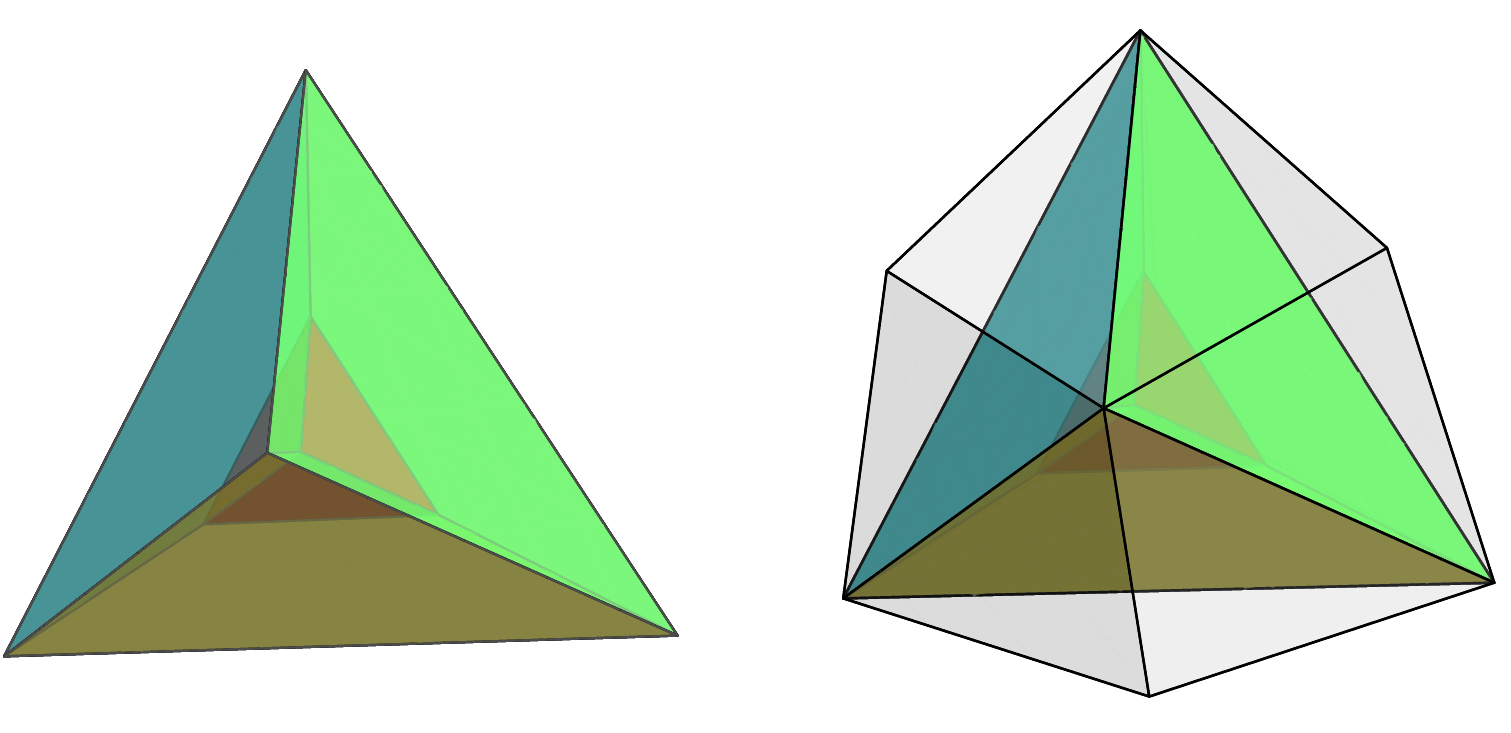}
\capt{4.8in}{fig:tetrahedra}{This figure shows a partial construction in 3 dimensions of the 4 dimensional situation depicted in \fref{fig:FaceConnections}.}
\end{figure}

We turn now to the combinatorics of the facets of $\nabla$. Two sketches follow in \fref{fig:orbit2} and \fref{fig:orbit1}. Each shows a $\IZ/10\IZ$ orbit, with generator $\ccA^2\ccB$ of the facets of $\nabla$, with facets of the same colour corresponding to orbits of $\ccA$. We refer to, and draw, the facets as cubes when they are in reality hexahedra. The facets are numbered in large boldface and the vertices, which are vertices of the facets and also of $\nabla$, are numbered in eight-point font. 

The figures are superficially different: in \fref{fig:orbit2}, for example, a dual facet $v$ meets
      $\ccA^2\ccB v$ in an edge, while in \fref{fig:orbit1} a dual facet $v$ meets $\ccA^2\ccB v$ in a facet of each. There are however additional identifications to made in these figures. In \fref{fig:orbit2}
      a facet $v$ meets a facet three steps on, so $(\ccA^2\ccB)^3 v{\=}\ccA\ccB v$ in a common facet. Thus $v_{14}$ meets $v_6$ in the common facet with vertices $\{f_6, f_{18}, f_{26}, f_{20}\}$ and $v_6$ meets $v_4$ in the facet $\{f_6, f_{20}, f_{12}, f_4\}$, for example, and all three of $v_{14}$, $v_6$ and $v_4$ meet in the common edge $\{f_6, f_{20}\}$. With these identifications, the two figures reveal the same reality. Note also that, despite appearances, two dual facets never meet in just a vertex. So in \fref{fig:orbit1} the dual facets $v_{12}$ and $v_{10}$, for example, appear to meet in just $f_{24}$,
      but in fact meet in the edge $\{f_{24}, f_{22}\}$, which has to be identified between the two cubes. This identification ensures that cubes of different colours do, in fact, meet in an edge as in \fref{fig:orbit2}.  
      
From the dual polyhedron we can read off a polynomial that defines $\nabla$ and is the analogue of \eqref{eq:PolynomialP}. Note that the vectors corresponding to the vertices of $\nabla$ in \tref{tab:NablaData}
      have the property that the components are all 0 or $\pm1$ and that the components, within a given vector, all have the same sign. Thus, introducing coordinates $Y_r$, $r=1,2,3,4$, the dual vertices, together with the interior point, correspond to the 31 Laurent monomials
\beq
1~,\quad Y_r~,\quad Y_r Y_s~,\quad Y_r Y_s Y_t~,\quad Y_1 Y_2 Y_3 Y_4~,
\quad \frac{1}{Y_r}~,\quad \frac{1}{Y_r Y_s}~,\quad \frac{1}{Y_r Y_s Y_t}~,\quad \frac{1}{Y_1 Y_2 Y_3 Y_4}~,
\notag\eeq
where, in each monomial, the indices take distinct values.

We need to combine the monomials into a Laurent polynomial that is invariant under the symmetries. The action of $\ccB$ on the coordinates is simply $Y_r\to 1/Y_r$. The action of $\ccA$ is slightly more involved. Consider an orbit of $\ccA$ that starts with $f_{10}$, say. We see from \eqref{eq:Arulef} that this induces the following action on the coordinates
\beq
Y_1 \to Y_2 \to Y_3 \to Y_4 \to \frac{1}{Y_1 Y_2 Y_3 Y_4} \to Y_1 \to\dots~.
\notag\eeq
We can simplify this rule by introducing a fifth coordinate
\beq
Y_5 \= \frac{1}{Y_1 Y_2 Y_3 Y_4}~,
\notag\eeq
so that $Y_1 Y_2 Y_3 Y_4 Y_5{\=}1$, then the rule is $Y_r\to Y_{r+1}$, with the indices understood mod 5.
The most general polynomial invariant under $\ccA$ is
\beq\begin{split}
\widetilde{P} \= A_0\; +\; &A_1 \sum Y_r + A_2 \sum Y_r Y_{r+1} + A_3 \sum Y_r Y_{r+2} 
+ A_4 \sum Y_r Y_{r+1} Y_{r+2} + \\[3pt] 
&A_5 \sum Y_r Y_{r+1} Y_{r+3} + A_6 \sum Y_r Y_{r+1} Y_{r+2} Y_{r+3}~.
\end{split}\notag\eeq
There is no need to separately include inverse powers of the $Y_r$, since these are already included through the $Y_5$'s. If we now require also that $\widetilde{P}$ should be invariant under $\ccB$ then we find that
\beq
A_6 \=  A_1~,\qquad A_5\= A_3~,\qquad A_4 \= A_2~.
\notag\eeq
The fixed points of the symmetries occur at certain discrete points of the embedding space. For example, the fixed points of $\ccA$ are where all the $Y_r$ are equal to the same fifth root of unity. For a generic choice of the free coefficients $A_0,\,A_1,\,A_2,\,A_3$ these points will not lie on the locus $\widetilde{P}{\=}0$.
\newpage
\begin{figure}[H]
\begin{center}
\includegraphics[height=0.8\textheight]{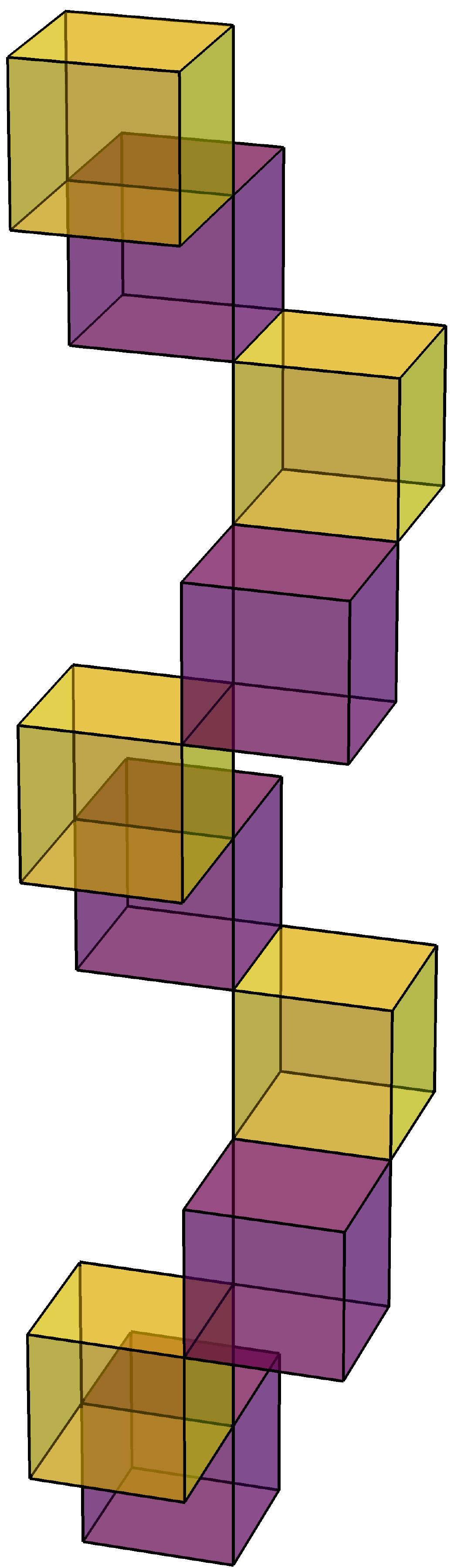}
\vskip13pt
\place{2.295}{7.1}{\bf\Large 14}
\place{3.1}{6.06}{\bf\Large 7}
\place{3.68}{5.68}{\bf\Large 17}
\place{3.6}{4.65}{\bf\Large 6}
\place{2.32}{4.07}{\bf\Large 19}
\place{3.0}{3.2}{\bf\Large 11}
\place{3.8}{2.8}{\bf\Large 4}
\place{3.45}{1.8}{\bf\Large 18}
\place{2.47}{0.89}{\bf\Large 5}
\place{2.95}{0.58}{\bf\Large 20}
\place{2.94}{7.3}{\eightrm 20}
\place{3.3}{7.5}{\eightrm 29}
\place{2.47}{7.6}{\eightrm 1}
\place{2.13}{7.33}{\eightrm 6}
\place{2.87}{6.51}{\eightrm 26}
\place{2.11}{6.39}{\eightrm 18}
\place{2.54}{6.63}{\eightbf 28}
\place{2.73}{7.02}{\eightbf 22}
\place{3.11}{6.73}{\eightrm 27}
\place{3.57}{6.88}{\eightrm 7}
\place{2.82}{6.27}{\eightbf 14}
\place{2.36}{5.86}{\eightrm 16}
\place{3.57}{6.22}{\eightrm 5}
\place{4.32}{6.05}{\eightrm 23}
\place{4.08}{5.77}{\eightrm 30}
\place{3.11}{5.77}{\eightrm 15}
\place{4.3}{5.3}{\eightrm 13}
\place{3.5}{5.3}{\eightbf 10}
\place{3.13}{5.15}{\eightrm 9}
\place{4.08}{5.0}{\eightrm 12}
\place{3.7}{4.9}{\eightrm 20}
\place{2.87}{4.9}{\eightrm 26}
\place{2.4}{4.58}{\eightrm 16}
\place{3.32}{4.48}{\eightbf 3}
\place{4.1}{4.33}{\eightrm 4}
\place{3.9}{4.05}{\eightrm 6}
\place{2.87}{4.22}{\eightrm 18}
\place{2.87}{4.22}{\eightrm 18}
\place{2.13}{4.27}{\eightrm 28}
\place{3.55}{3.9}{\eightrm 10}
\place{2.82}{3.97}{\eightbf 5}
\place{2.55}{3.67}{\eightbf 14}
\place{3.11}{3.63}{\eightrm 8}
\place{2.88}{3.5}{\eightrm 19}
\place{2.88}{3.5}{\eightrm 19}
\place{2.13}{3.42}{\eightrm 22}
\place{2.8}{3.3}{\eightbf 23}
\place{3.1}{2.9}{\eightrm 11}
\place{3.53}{3.38}{\eightrm 13}
\place{4.22}{3.28}{\eightrm 12}
\place{4.05}{2.85}{\eightrm 4}
\place{2.4}{2.99}{\eightrm 25}
\place{3.1}{2.3}{\eightrm 17}
\place{4.05}{2.2}{\eightrm 6}
\place{4.25}{2.52}{\eightrm 20}
\place{3.55}{2.7}{\eightbf 21}
\place{2.9}{2.0}{\eightrm 19}
\place{3.67}{2.0}{\eightrm 18}
\place{3.3}{1.7}{\eightbf 24}
\place{4.05}{1.55}{\eightrm 1}
\place{3.85}{1.2}{\eightrm 28}
\place{2.9}{1.25}{\eightrm 22}
\place{2.4}{1.8}{\eightrm 25}
\place{2.16}{1.45}{\eightrm 14}
\place{2.6}{1.02}{\eightbf 23}
\place{2.95}{0.75}{\eightrm 7}
\place{3.12}{0.94}{\eightrm 2}
\place{2.22}{0.7}{\eightrm 5}
\place{2.75}{1.5}{\eightbf 13}
\place{3.43}{1.42}{\eightbf 21}
\place{3.53}{0.69}{\eightrm 20}
\place{2.68}{0.84}{\eightbf 12}
\place{3.3}{0.27}{\eightrm 29}
\place{2.42}{0.40}{\eightrm 30}
\capt{6in}{fig:orbit2}{A $\IZ/10\IZ$ orbit of the dual facets. The generator $\ccA^2\ccB$ runs through these dual facets, in the order given, from top to bottom. The facets shown in yellow and purple are distinct $\ccA$ orbits. The dual facets are numbered in large boldface, according to \tref{tab:DeltaData} and the dual vertices are numbered in eight point font. The edge $\{f_{20}, f_{29}\}$ of $v_{20}$ is identified with the corresponding edge of $v_{14}$.}
\end{center}
\end{figure}
%

\newcommand{\hshearbox}[3]{\scalebox{0.866025}[#2]{\rotatebox{210}%
{\scalebox{1.73205}[-0.57735]{\rotatebox{60}{\scalebox{-1.1547}[#1]{#3}}}}}}

\newcommand{\shear}[1]{\hshearbox{0.4}{2}{#1}}
\begin{figure}[H]
\begin{center}
\includegraphics[width=0.8\textwidth]{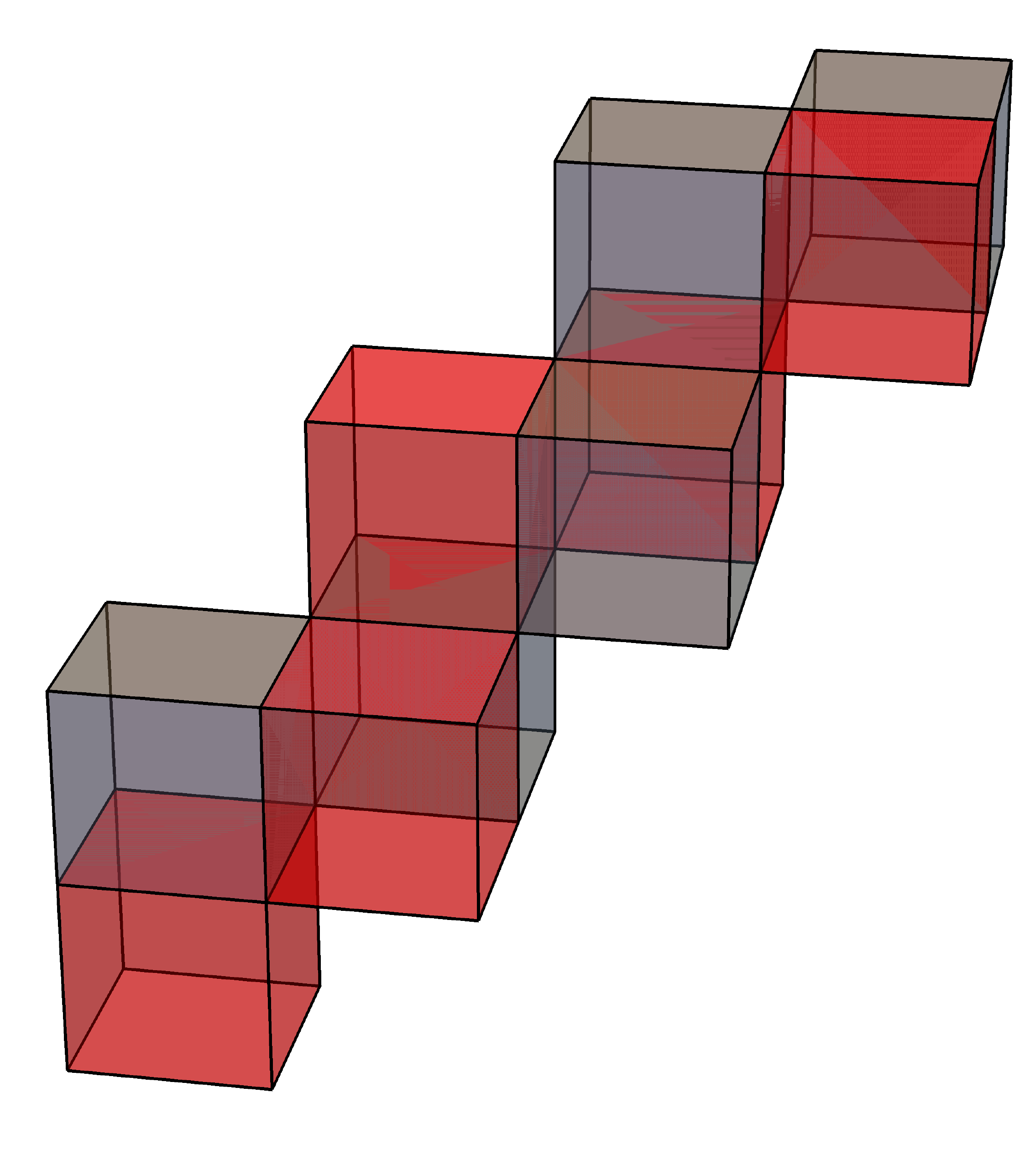}
\vskip13pt
\place{4.75}{6.18}{\shear{\Large\bf 13}}
\place{5.23}{5.2}{\Large\bf 16}
\place{4.27}{5.25}{\Large\bf 2}
\place{3.65}{4.83}{\shear{\Large\bf 1}}
\place{4.1}{3.87}{\Large\bf 3}
\place{2.92}{3.92}{\Large\bf 12}
\place{2.42}{3.7}{\shear{\Large\bf 15}}
\place{2.85}{2.5}{\Large\bf 9}
\place{1.6}{2.55}{\Large\bf 10}
\place{1.75}{1.6}{\Large\bf 8}
\place{2.0}{0.65}{\eightrm 30}
\place{0.8}{0.8}{\eightrm 15}
\place{1.3}{1.3}{\eightbf 5}
\place{2.32}{1.3}{\eightrm 23}
\place{0.73}{1.85}{\eightrm 27}
\place{1.28}{2.18}{\eightbf 7}
\place{2.30}{2.13}{\eightbf 2}
\place{3.33}{2.15}{\eightrm 21}
\place{3.1}{1.55}{\eightrm 20}
\place{2.1}{1.82}{\eightbf 29}
\place{2.48}{2.58}{\eightbf 23}
\place{3.5}{2.6}{\eightrm 13}
\place{0.72}{2.84}{\eightrm 28}
\place{1.08}{3.35}{\eightrm 22}
\place{2.05}{3.27}{\eightrm 24}
\place{2.97}{2.73}{\eightbf 6}
\place{1.88}{2.8}{\eightbf 1}
\place{3.09}{3.2}{\eightbf 17}
\place{4.38}{2.95}{\eightrm 6}
\place{4.55}{3.43}{\eightrm 4}
\place{3.5}{3.4}{\eightbf 11}
\place{2.28}{3.62}{\eightbf 25}
\place{3.67}{3.97}{\eightbf 13}
\place{4.65}{3.87}{\eightrm 12}
\place{2.02}{4.2}{\eightrm 22}
\place{2.27}{4.6}{\eightrm 14}
\place{3.32}{4.58}{\eightrm 8}
\place{4.38}{4.51}{\eightbf 3}
\place{4.22}{4.13}{\eightbf 18}
\place{3.13}{4.19}{\eightbf 19}
\place{5.6}{4.29}{\eightrm 18}
\place{3.67}{4.91}{\eightbf 10}
\place{4.68}{4.67}{\eightbf 9}
\place{4.79}{5.18}{\eightbf 12}
\place{5.8}{5.1}{\eightrm 20}
\place{5.71}{4.7}{\eightrm 26}
\place{3.28}{5.53}{\eightrm 14}
\place{4.38}{5.51}{\eightbf 16}
\place{5.43}{5.45}{\eightbf 28}
\place{5.52}{5.77}{\eightbf 27}
\place{5.82}{6.05}{\eightrm 29}
\place{4.6}{6.1}{\eightrm 30}
\place{4.5}{5.83}{\eightrm 15}
\place{3.52}{5.89}{\eightrm 5}
\vskip3.1cm
\capt{5in}{fig:orbit1}{The remaining dual facets form a second orbit of $\ccA^2\ccB$, descending from $v_{13}$, in this figure. The facet $\{f_{15}, f_{30}, f_{29}, f_{27}\}$ of $v_8$ is identified with the corresponding facet of $v_{13}$.}
\end{center}
\end{figure}
\newpage
\section{Are There Other Rank Two Attractor Points for AESZ34?}
\vskip-10pt
It is natural to ask if one can find further rank two attractor points in the moduli space of AESZ34. A satisfactory answer to this question is probably contingent on a good understanding of the geometry of AESZ34 and answering the questions raised in \sref{sec:SpeculationsOnGeometry}. In lieu of this, we make some observations about the interpretation of the data on factorisations, that we have, and the prospects for computer searches for other attractor points of rank two, in this moduli space. We study also the statistical distribution of the $a$ and $b$ coefficients and ask how many factorisations can be attributed to chance.

The reader is warned, from the outset, that we prove no theorems here and that statistical trends that appear compelling might be reversed by the acquisition of more data.

\subsection{Brute-force searches and the Chebotar{\"e}v Theorem}
\vskip-10pt
A brute-force search over degree $n$ polynomials
\beq
c_n\vph^n + c_{n-1}\vph^{n-1}+\cdots+c_0\=0
\notag\eeq
rapidly becomes onerous as the degree and the search space of the coefficients is increased. 
However, we can, to some extent, see whether it is likely, for a given degree, that there should be a polynomial as above, based on the frequency of factorisations in \fref{fig:AESZ34andQuintic}. We have already observed that the fact that there is at least one factorisation, for AESZ34, for each $p$ in the range $7{\;\leq\;}p{\;\leq\;}p_{502}$, where $p_j$ denotes the $j$'th prime, makes it highly likely that there should exist a linear equation $c_1\vph{\;+\;}c_0{\=}0$, corresponding to an attractor point. A~converse is that the fact that there is no factorisation for the mirror quintic threefold, for many $p$,  makes it very unlikely that there should exist a linear equation in that case. If such an equation were to exist, then $c_1$ would have to be divisible by all the primes for which there is no factorisation and so by the product of these, which is an integer with 1217 digits!
\vskip5pt
\begin{figure}[H]
\begin{center}
\includegraphics[width=\linewidth]{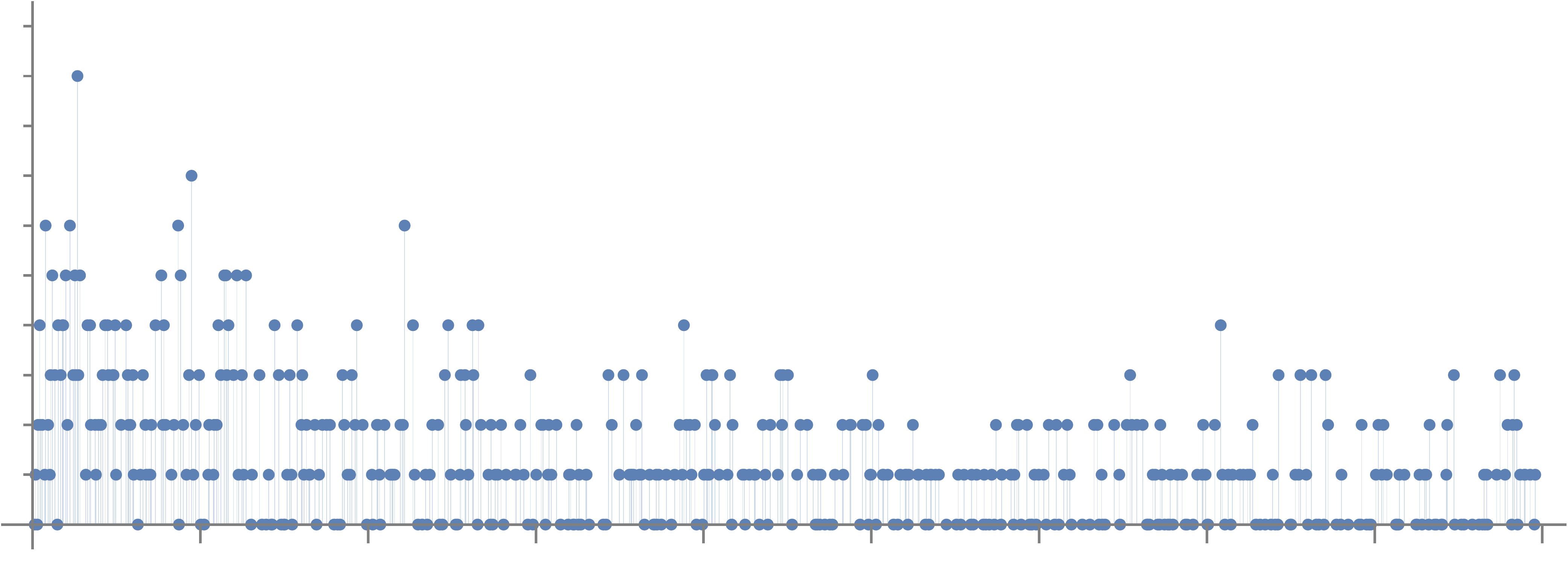}
\vskip0pt 
\place{-0.05}{0.54}{\scriptsize 1}
\place{-0.05}{0.75}{\scriptsize 2}
\place{-0.05}{0.96}{\scriptsize 3}
\place{-0.05}{1.16}{\scriptsize 4}
\place{-0.05}{1.37}{\scriptsize 5}
\place{-0.05}{1.57}{\scriptsize 6}
\place{-0.05}{1.78}{\scriptsize 7}
\place{-0.05}{1.99}{\scriptsize 8}
\place{-0.05}{2.20}{\scriptsize 9}
\place{-0.07}{2.41}{\scriptsize 10}
\place{0.73}{0.15}{\scriptsize 400}
\place{1.43}{0.15}{\scriptsize 800}
\place{2.10}{0.15}{\scriptsize 1200}
\place{2.80}{0.15}{\scriptsize 1600}
\place{3.49}{0.15}{\scriptsize 2000}
\place{4.19}{0.15}{\scriptsize 2400}
\place{4.89}{0.15}{\scriptsize 2800}
\place{5.59}{0.15}{\scriptsize 3200}
\place{6.29}{0.15}{\scriptsize 3600}
\place{6.6}{0.3}{$p$}
\vskip-8pt
\capt{6in}{fig:ReducedFacnums}{The residual factorisations for AESZ34 after the factorisations for $\vph{\; =}-1/7$ and $\vph{\;=\;}\vph_{\pm}$ have been removed.}
\end{center}
\end{figure}

Passing to quadratic equations: recall that a quadratic equation has two roots in $\IF_{\! p}$ if the discriminant $\D{\=}c_1^2{\;-\;}4c_0 c_2$ is a nonzero square mod $p$, none, if $\D$ is not a square, and one root if $p|\D$. For given $\D$, this last condition is satisfied for only finitely many $p$. For a large set of primes, a quadratic equation will have no roots or two roots, each with frequency that approaches $1/2$. So, if there is a quadratic factor to $G(\vph)$, we would expect $R(T)$ to factorise at least twice, with frequency at least 1/2. For the mirror quintic the plot of \fref{fig:AESZ34andQuintic} gives a frequency of $16/500$, and which is moreover decreasing as $p$ increases. So it seems very unlikely that there is a quadratic factor to $G(\vph)$, for this space. 
    
For equations of degree $n{\;>\;}2$, we can have recourse to a consequence of the Chebotar{\"e}v density theorem. This states that such an equation will have $n$ roots in $\IF_{\! p}$ with frequency $1/|\mathfrak{G}|$, where $\mathfrak{G}$ is the Galois group of the equation. Since this group is always a subgroup of $S_n$,
      the group of permutations of $n$ objects, we know that an equation of degree $n$ has $n$ roots in $\IF_{\! p}$
      with frequency at least $1/n!$. This would seem to rule out cubic and quartic equations for the mirror quintic,
      since three is the largest number of factorisations, in our data, and this occurs for only three primes. For higher $n$, we really need data for several times $n!$ primes to draw a conclusion.
\begin{figure}[!b]
\begin{center}
\includegraphics[width=\linewidth]{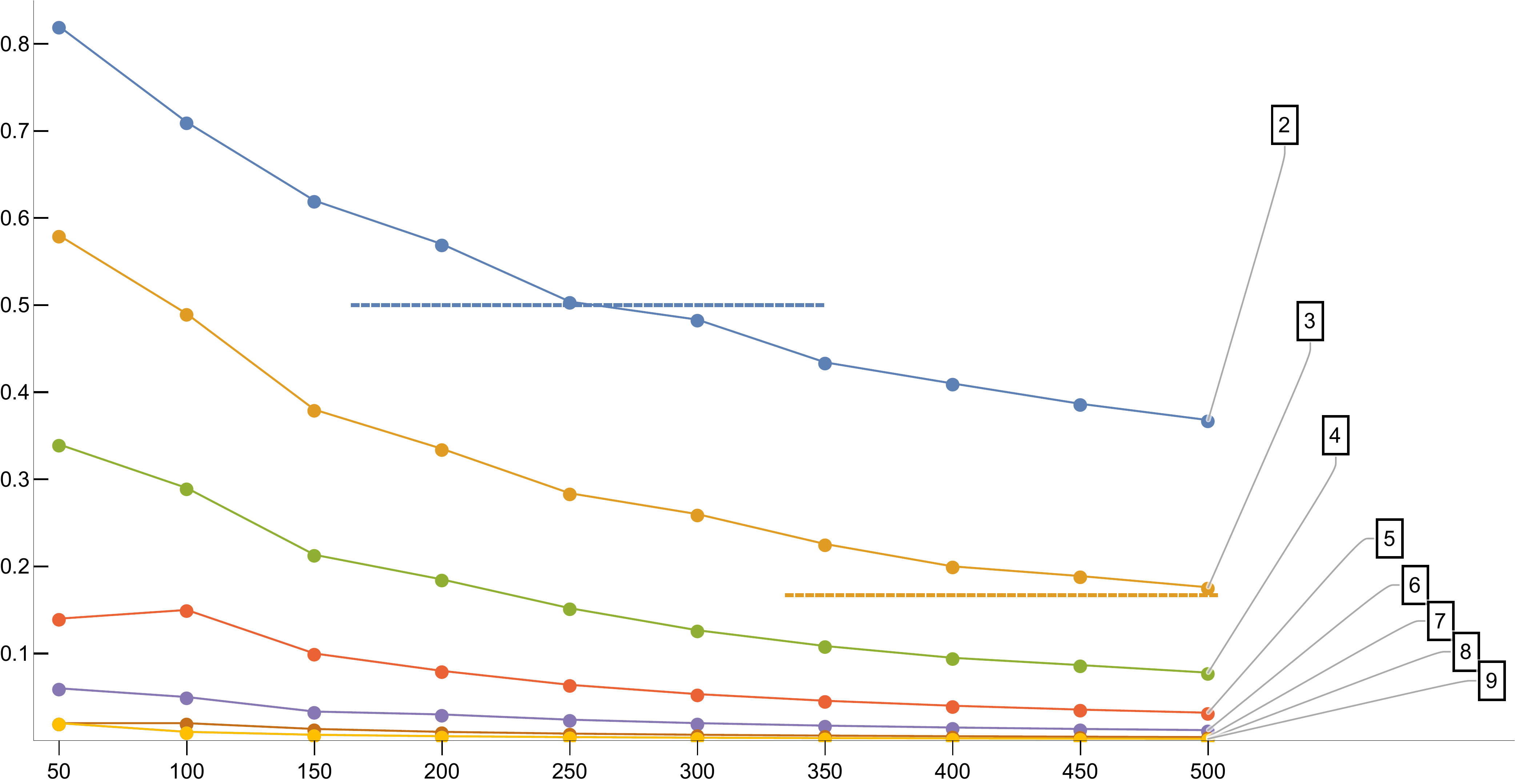}
\capt{6.5in}{fig:frequencyplot2}{The frequencies for which there are at least $n$ factorisations for the primes
$5{\;\leq\;}p{\;\leq\;}p_\text{max}$, with increasing $p_\text{max}$, in steps of 50 primes. The blue dashed line has height $1/2!$ and the yellow dashed line has height $1/3!\,$.}
\end{center}
\end{figure}

Let us see how these considerations may apply to AESZ34. \fref{fig:ReducedFacnums} shows the number of residual factorisations for AESZ34 after the factorisations for $\vph{\;=}-1/7$ and $\vph{\=}\vph_\pm$ have been removed. Note that, even so, there are many more factorisation than for the mirror quintic.
     We can plot the proportion of primes $5{\;\leq\;}p_\text{max}$ for which there are at least $n$ factorisations for $2{\;\leq\;}n{\;\leq\;}9$, and do this in bins of 50, that is for $p_\text{max}{\=}p_{52},p_{102},\ldots,p_{502}$. In this way, we can see how the frequencies evolve with $p_\text{max}$. The result is \fref{fig:frequencyplot2}. 
     
The blue horizontal line corresponds to $1/2$ and we see that the proportion of times for which there are at least two residual factorisations passes below this value, and appears to be decreasing, so it seems unlikely that there is a second quadratic equation.
     The yellow horizontal line corresponds to $1/3!$ and it seems that the proportion of times that there are at least 3 residual factorisations is about to pass below this line. 
     
The evolution of the frequencies has a long tail that is dominated by the large number of factorisations for small primes. If we eliminate the primes up to $p_\text{200}$, say, we are left with a distribution that still corresponds to 302 primes but is more uniform. 
\begin{figure}[H]
\begin{center}
\includegraphics[width=\linewidth]{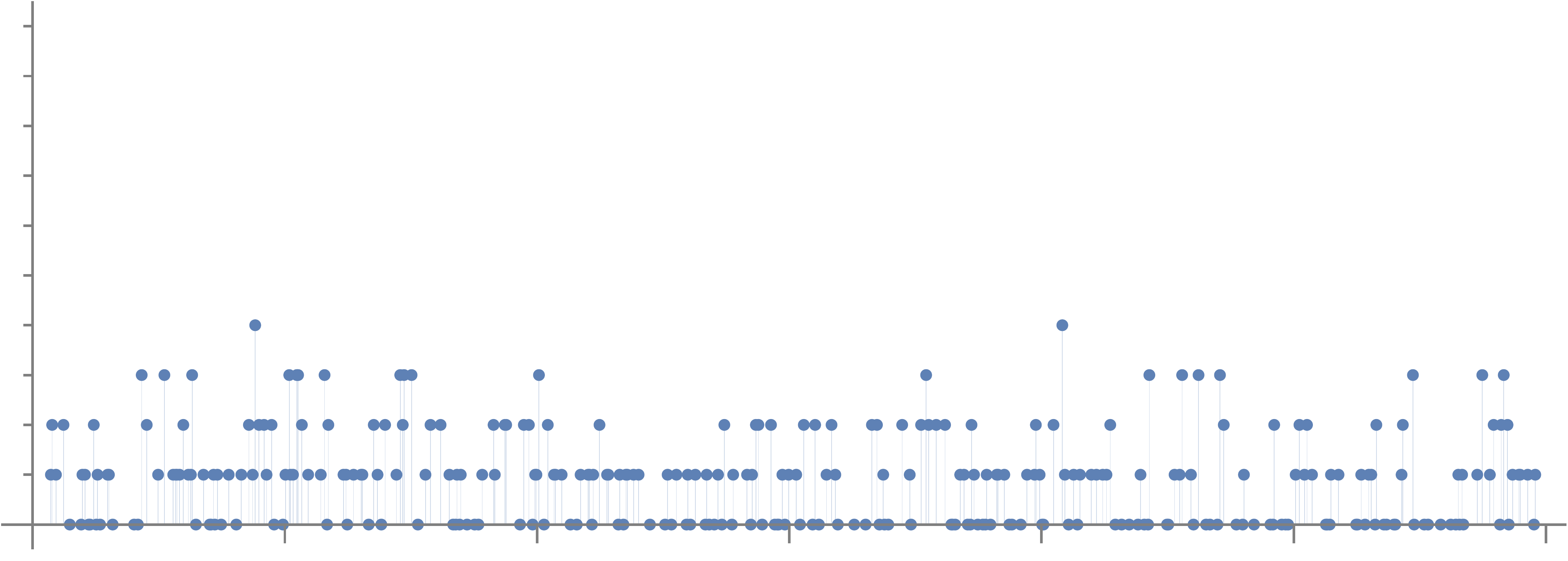}
\vskip0pt 
\place{-0.05}{0.54}{\scriptsize 1}
\place{-0.05}{0.75}{\scriptsize 2}
\place{-0.05}{0.96}{\scriptsize 3}
\place{-0.05}{1.16}{\scriptsize 4}
\place{-0.05}{1.37}{\scriptsize 5}
\place{-0.05}{1.57}{\scriptsize 6}
\place{-0.05}{1.78}{\scriptsize 7}
\place{-0.05}{1.99}{\scriptsize 8}
\place{-0.05}{2.20}{\scriptsize 9}
\place{-0.07}{2.41}{\scriptsize 10}
\place{0.00}{0.15}{\scriptsize 1200}
\place{1.05}{0.15}{\scriptsize 1600}
\place{2.10}{0.15}{\scriptsize 2000}
\place{3.15}{0.15}{\scriptsize 2400}
\place{4.20}{0.15}{\scriptsize 2800}
\place{5.25}{0.15}{\scriptsize 3200}
\place{6.29}{0.15}{\scriptsize 3600}
\place{6.6}{0.3}{$p$}
\vskip-10pt
\capt{4.7in}{fig:ReducedFacnums302}{The residual factorisations for AESZ34 after eliminating the primes up to $p_{200}{\;=\;}1223$.}
\end{center}
\end{figure}
\vskip-20pt
The number of times that there are at least two factorisations, in this plot, is $71/302{\=}0.235$, and for at least three factorisations it is $21/302{\=}0.0695$; in each case less than half the lower bound suggested by the Chebotar{\"e}v theorem. For at least four factorisations, the frequency is $2/302{\=}0.00662$ which is about
     $1/6$ of $1/4!$, the lower bound suggested by the \hbox{Chebotar{\"e}v} theorem. For degrees of 5 and above
     we cannot say more without more extensive~data.
     
\subsection{Random factorisations}
\vskip-10pt
We wish to ask now what frequency of factorisation is to be expected ``at random''. To this end, consider again the factorisation
\beq
1+ a\,T + b\,p T^2 + a\,p^3 T^3 +p^6T^4 \=(1-\a\,pT+p^3T^2)(1-\b\,T+p^3T^2)~.
\notag\eeq
We are interested in the cases that the coefficients $a,b,\a,\b$ are integers, but let us temporarily take them to be merely real. Over $\IR$, factorisation, as above, is always possible and we have the relations
\begin{align}
a &\;= -(p\a + \b)~,& b &\= 2p^2 + \a\b~,\label{eq:maptoS}\\
\intertext{which we rewrite as}
\tilde{a}&\;= -(\tilde{\a} + \tilde{\b})~,& \tilde{b}&\= 2 +\tilde{\a}\tilde{\b}~,\label{eq:maptoStilde} 
\end{align}
with
\beq
\tilde{a}\= \frac{a}{p^{3/2}}~,\qquad \tilde{b}\=\frac{b}{p^2}~,\qquad
\tilde{\a}\=\frac{\a}{p^{1/2}}~,\qquad\tilde{\b}\=\frac{\b}{p^{3/2}}~.
\notag\eeq
\begin{figure}[!b]
\framebox[\textwidth]{\begin{minipage}[c]{\textwidth}
\vspace{20pt}
\hspace*{-15pt}\includegraphics[height=7cm]{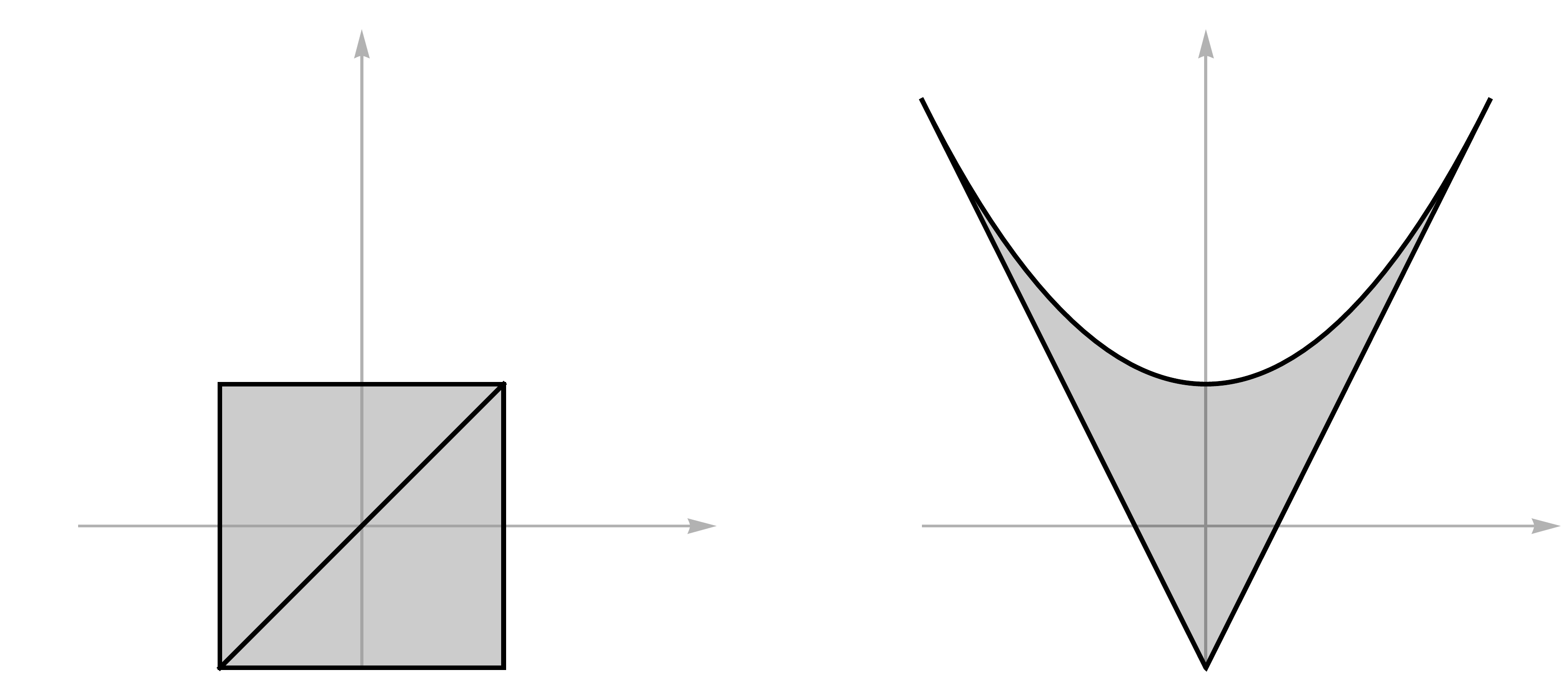}
\vspace{20pt}
\end{minipage}}
\vskip-10pt 
\place{0.36}{0.1}{$(-2,-2)$}
\place{1.58}{0.1}{$(2,-2)$}
\place{0.36}{1.6}{$(-2,2)$}
\place{1.62}{1.6}{$(2,2)$}
\place{4.32}{0.1}{$(0,-2)$}
\place{3.12}{2.73}{$(-4,6)$}
\place{5.52}{2.73}{$(4,6)$}
\place{2.68}{0.9}{$\a/p^{1/2}$}
\place{1.08}{3.0}{$\b/p^{3/2}$}
\place{6.01}{0.9}{$a/p^{3/2}$}
\place{4.43}{3.0}{$b/p^2$}
\place{4.35}{1.15}{$\tilde{\cS}$}
\place{1.45}{0.59}{$\tilde{\cS}_0$}
\place{0.9}{1.12}{$\tilde{\cS}_0'$}
\begin{center}
\capt{5.5in}{fig:coefficientregions}{The allowed regions for the $(a,b)$ and $(\a,\b)$ coefficients. The region on the left maps 2--1 onto the region on the right. Either of the two triangular regions shown can be taken to be a fundamental region for the $(\a,\b)$ coefficients.}
\end{center}
\end{figure}

Now, in the case that $R(T)$ arises from the $\z$-function, then, quite apart from the question of factorisation, the point $(\tilde{a},\tilde{b})$ is constrained by the Weil Conjectures to lie within a region, $\cS$, bounded\cite{DissertationSamol,LocalZetaFunctionsI} by the curves
\beq
\tilde{b}\= 2|\tilde{a}| - 2 \quad \text{and} \quad \tilde{b}\= \frac{\tilde{a}^2}{4} + 2~.
\notag\eeq
The preimage of $\cS$ in $(\tilde{\a},\tilde{\b})$-space is the square 
\beq
|\tilde{\a}| \;\leq\; 2~,\qquad |\tilde{\b}| \;\leq\; 2~.
\notag\eeq
Owing to the symmetry of \eqref{eq:maptoStilde} under interchange of $\tilde{\a}$ and $\tilde{\b}$, the map to $\cS$ is generically 2--1 with $(\tilde{\a},\tilde{\b})$ and $(\tilde{\b},\tilde{\a})$ mapping to the same point of $\cS$. We can divide the square into two triangles by the diagonal $\tilde{\a}{\=}\tilde{\b}$; either the lower triangle, $\cS_0$, or the upper triangle, $\cS_0'$, can be taken to be a fundamental region for parametrizing the points $(\tilde{a},\tilde{b})$. These regions are sketched in \fref{fig:coefficientregions}.

Now if $(\a,\b)$ are integers, then, as we see from \eqref{eq:maptoS}, so are $(a,b)$. The converse however is not true in general, which is just the statement that $R(T)$ factorises only ocaisionally over~$\IZ$. \fref{fig:IntegerPointsOfS0} sketches, for $p{\=}13$, how these integral $(\a,\b)$ points lie in $\tilde{\cS}$. Individual points are plotted but, at this scale, they run together to form lines. These lines, which are the images of horizontal and vertical lines in $\tilde{\cS}_0$, are tangent to the to the upper boundary of $\tilde{\cS}$.
\begin{figure}[H]
\begin{center}
\includegraphics[height=7cm]{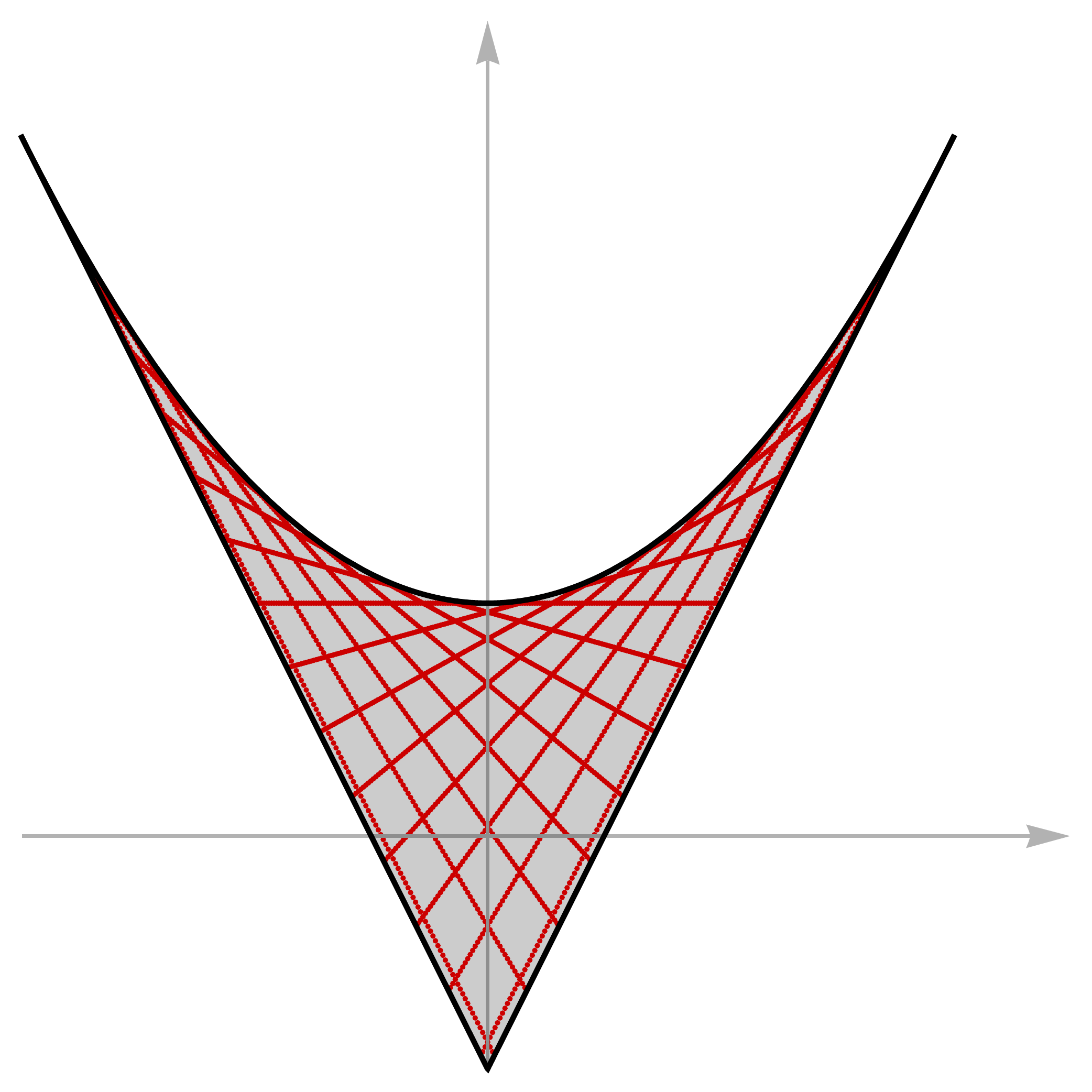}
\capt{5.5in}{fig:IntegerPointsOfS0}{The images of the points of $\tilde{\cS}_0$, that correspond to integral
$(\a,\b)$-points, in $\tilde{\cS}$, plotted for $p{\;=\;}13$. At this scale, the points run together to form lines.}
\end{center}\end{figure}
Let $\cS$ and $\cS_0$ denote the regions of $(a,b)$-space and $(\a,\b)$-space that correspond to $\tilde{\cS}$ and~$\tilde{\cS}_0$. The numbers of integral points in these regions is, for large $p$, closely approximated by their areas. We have
\beq
\int_\cS \dd a\, \dd b \= p^{7/2}\int_{\tilde{\cS}} \dd\tilde{a}\, \dd\tilde{b} \= \frac{32}{3}\,p^{7/2}
\qquad\text{and}\qquad
\int_{\cS_0} \dd\a\, \dd\b \= p^2\int_{\tilde{\cS_0}} \dd\tilde{\a}\, \dd\tilde{\b} \= 8p^2~.
\notag\eeq

Let us denote by $\r$, in this appendix, the modulus of the jacobian of the transformation between $(\a,\b)$ and $(a,b)$
\beq
\r \= \left|\pd{(\a,\b)}{(a,b)}\right| \= \frac{1}{|p\a - \b|}~.
\notag\eeq
\begin{figure}[!th]
\begin{center}
\framebox[\textwidth]{\begin{minipage}[c]{\textwidth}
\vspace{20pt}
\hspace{15pt}\includegraphics[height=7cm]{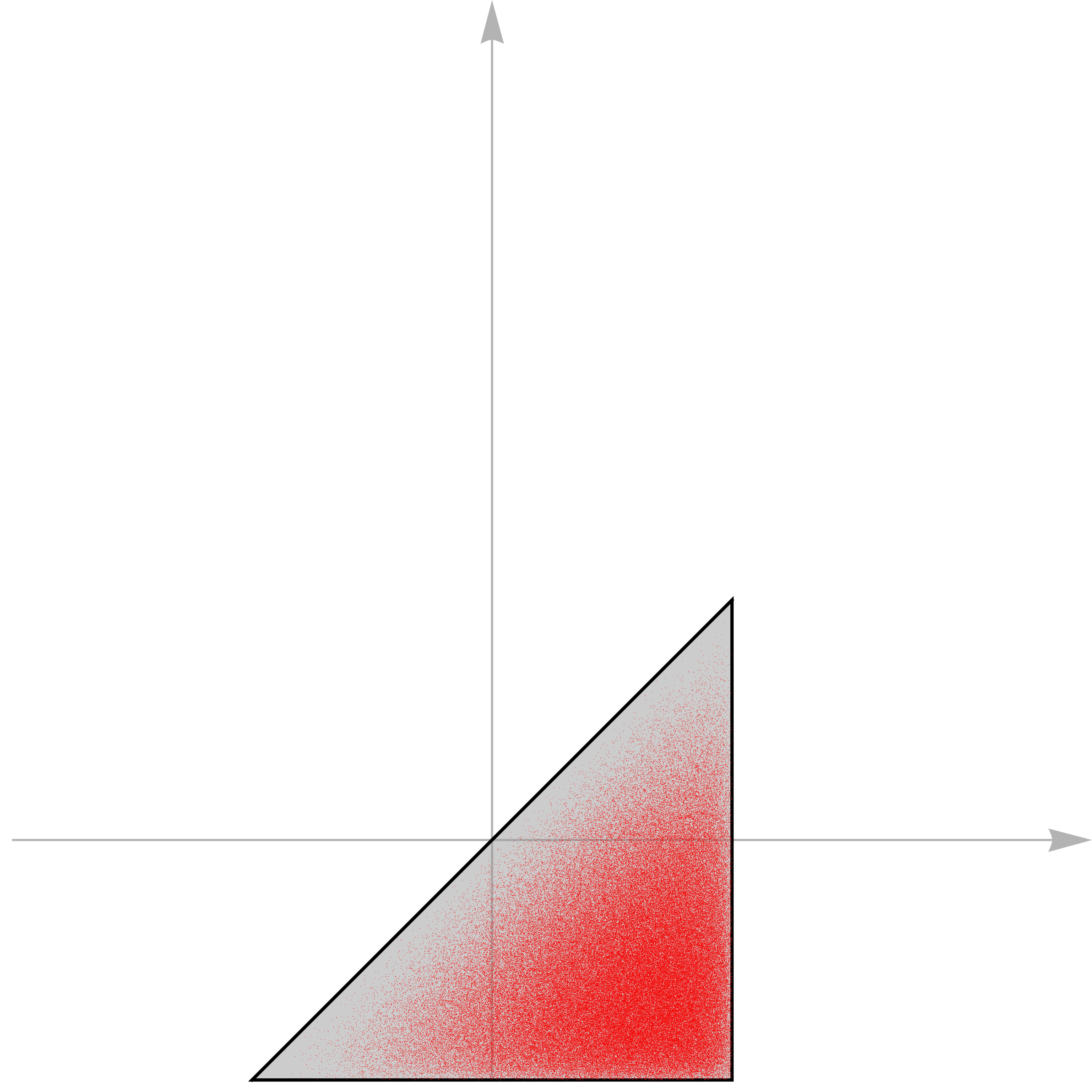}\hspace{1.2cm}
\includegraphics[height=7cm]{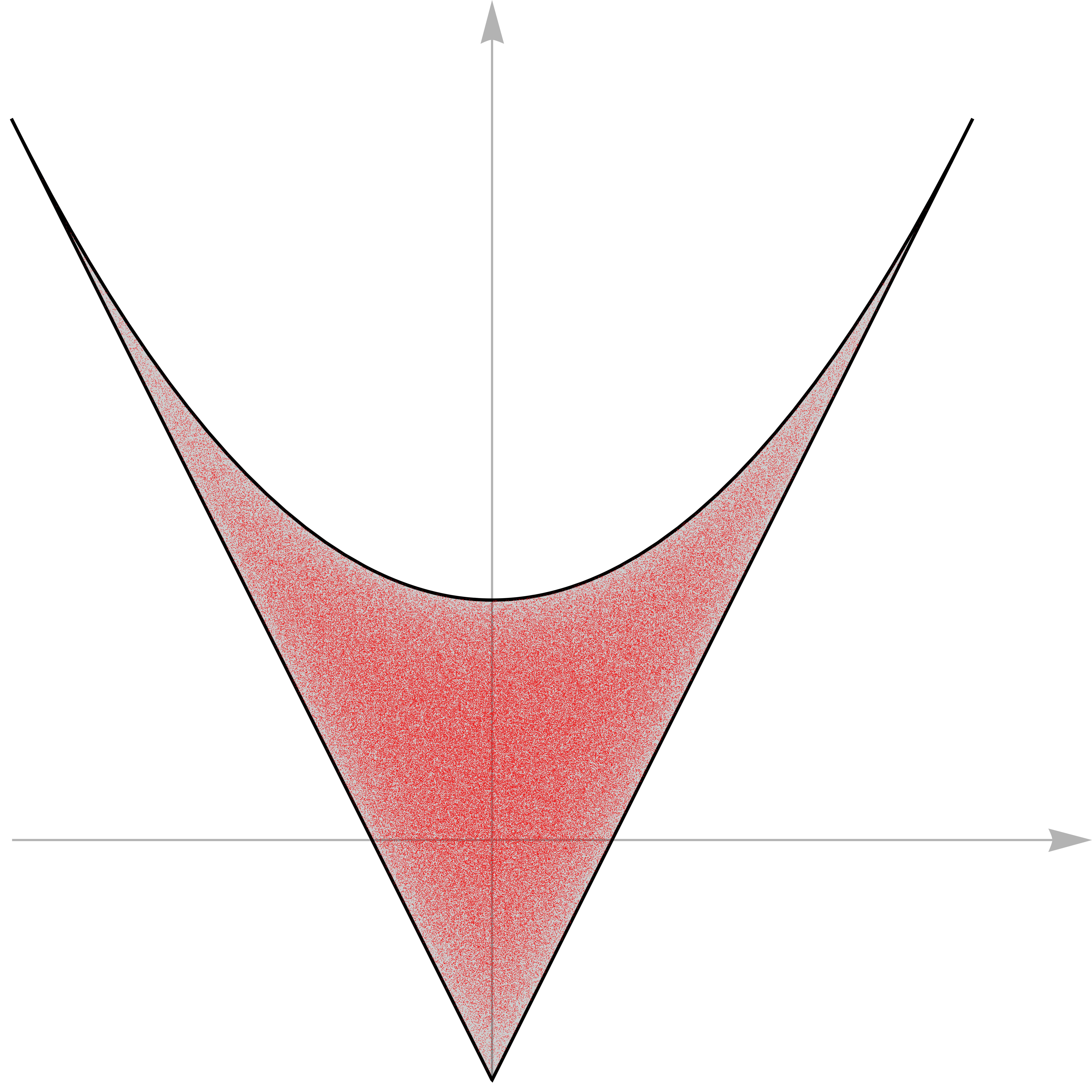}
\vspace{20pt}
\end{minipage}}
\capt{6.0in}{fig:CumulativePlots}{A cumulative plot for the 75 primes $p_{428}{\;\leq\;}p{\;\leq\;}p_{502}$ showing how the $(a,b)$-points lie in $\tilde\cS$ and how the preimages of these points lie in $\tilde{\cS}_0$. Note how the frequency declines near the diagonal boundary of $\tilde{\cS}_0$.}
\end{center}
\end{figure}

We have
\beq
\r\,\dd a \dd b \= \dd\a\dd\b~;
\notag\eeq
so, since the factorisations over $\IZ$ are distributed with density one with respect to $(\a,\b)$, they are distributed with density $\r$ with respect to $(a,b)$. Let us suppose now that the $(\a,\b)$ are distributed with a probability density function $h$, then the $(a,b)$ are distributed with frequency $\r h$ and the density of factorisations, for the $(a,b)$ coefficients, is then $\r$ times this, so $\r^2 h$. Let us write $\m/p$ for the `probability' that there is a factorisation for a given $p$ and~$\vph$. We have
\beq
\frac{\m}{p} \= \int_\cS \r^2 h\, \dd a \dd b \= p^2\int_{\tilde{\cS}_0} \r h\, \dd\tilde{\a}\dd\tilde{\b}~.
\label{eq:muoverp}\eeq

In \fref{fig:HistogramAndPdf} we plot a histogram for the frequency of points in $\tilde{\cS}_0$. This uses the data for all 500 primes $p_{3}{\;\leq\;}p{\;\leq\;}p_{502}$.

\begin{figure}[!t]
\begin{center}
\includegraphics[width=0.48\textwidth]{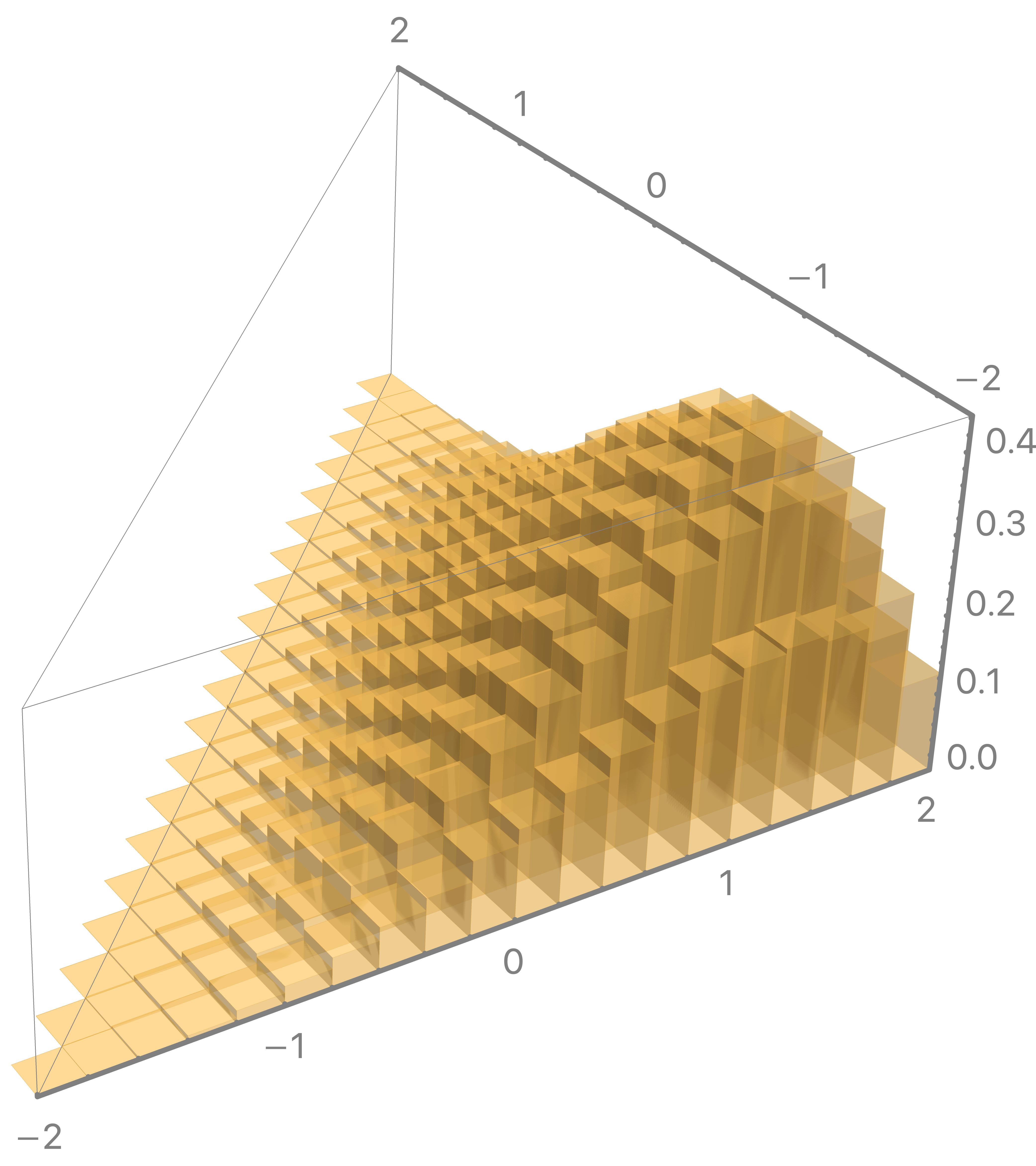}\hfill
\includegraphics[width=0.48\textwidth]{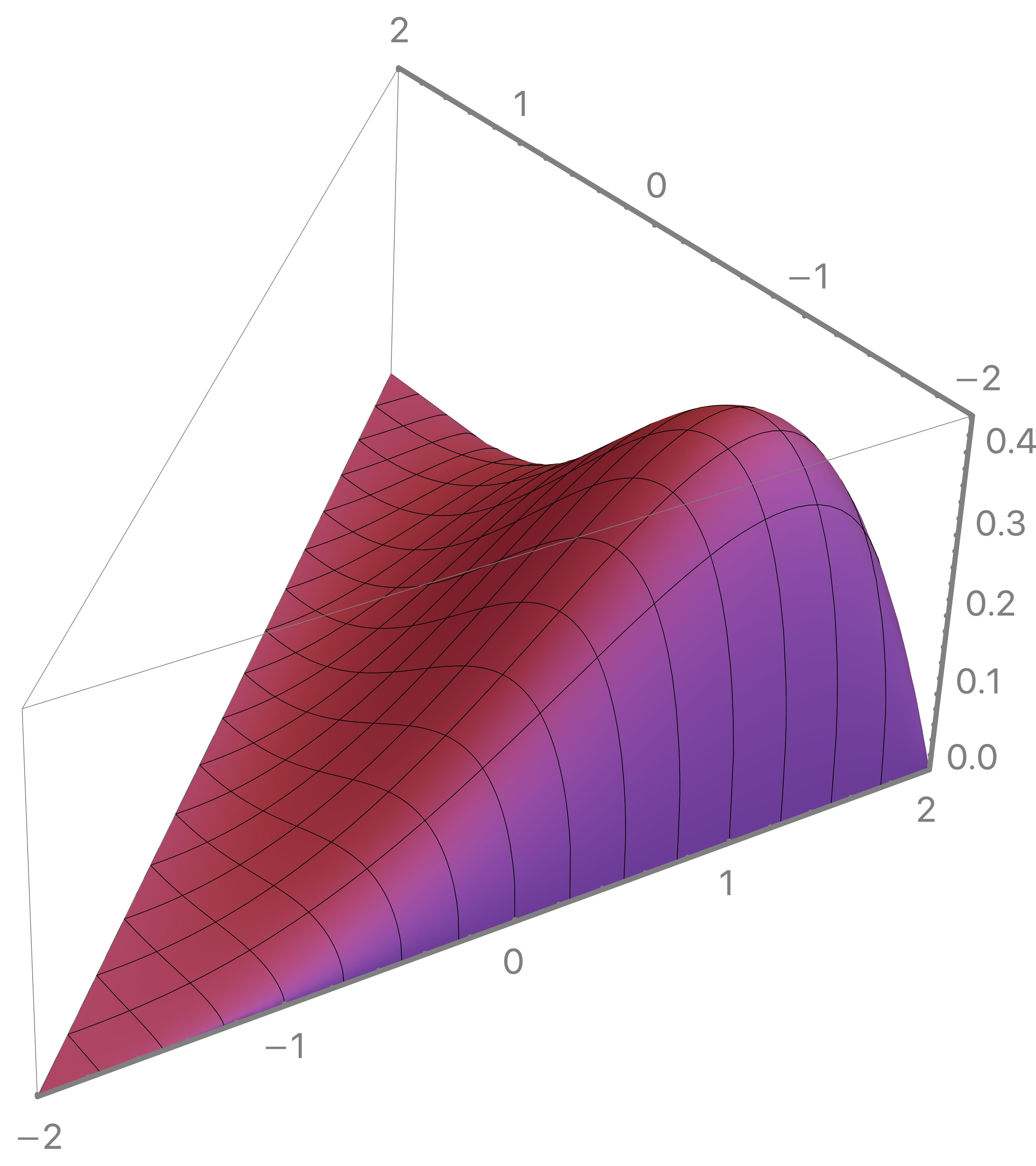}
\capt{5.5in}{fig:HistogramAndPdf}{On the left: a histogram of the frequency of $(\tilde\a,\tilde\b)$-points using data for the 500 primes, $p_{3}{\;\leq\;}p{\;\leq\;}p_{502}$. On the right: a plot of the function $p^2 h$ from \eqref{eq:hexpression}.}
\end{center}
\end{figure}
We write the probability as $\m/p$, since, for a given $p$ there are $p{\;-\;}1$ values of $\vph$ in our tables. We take $p$ to be large, in the following, so we will not distinguish between $p{\;-\;}1$ and $p$. The expected number of factorisations, for given $p$, is then $p$ times the probability above, so $\m$. 
     The probability that there are precisely $k$ factorisations, for a given $p$, assuming that $p$ is large compared to $k$, is then
\beq
\binom{p}{k}\,\left(\frac{\m}{p}\right)^k \left( 1 - \frac{\m}{p} \right)^{p-k} \;\sim\; 
\ee^{-\m}\,\frac{\m^k}{k!}~,
\notag\eeq
which characterizes a Poisson process with parameter $\m$. For such a process, the mean and variance are both $\m$.

It remains to estimate the frequency $h$ and so $\m$.

In order to discuss the form suggested by this histogram we make a further change of variables by writing
\beq
\tilde{\a} \;= -2 \cos\th_1 ~,\qquad \tilde{\b} \;= -2\cos\th_2~.
\notag\eeq
Then
\beq
h\,\dd\a\dd\b \= p^2 h \,\dd\tilde{\a}\dd\tilde{\b} \= 4p^2 h \sin\th_1 \sin\th_2\, \dd\th_1\dd\th_2
\label{eq:pdfs}\eeq
The quantity that is plotted in the histogram is $p^2 h$ and it is compelling to suppose that this quantity takes the form
\beq
p^2 h \= \frac{4}{\p^2} \sin\th_1 \sin\th_2\, (\cos\th_1 - \cos\th_2)^2~,
\label{eq:hexpression}\eeq
the constant corresponding to the need to normalize the total probability to unity. This function is plotted on the right in \fref{fig:HistogramAndPdf}. The correspondence seems remarkably close: we overlay the two plots in \fref{fig:CombinedHistogramAndPdf}

The corresponding frequency relative to the coordinates $(\th_1,\th_2)$ is given by the last term in \eqref{eq:pdfs}; let us denote this by $f$
\beq
f \= \frac{16}{\p^2} \sin^2\th_1 \sin^2\th_2\, (\cos\th_1 - \cos\th_2)^2~.
\notag\eeq
\begin{figure}[!th]
\begin{center}
\null\vskip-15pt
\includegraphics[width=0.7\textwidth]{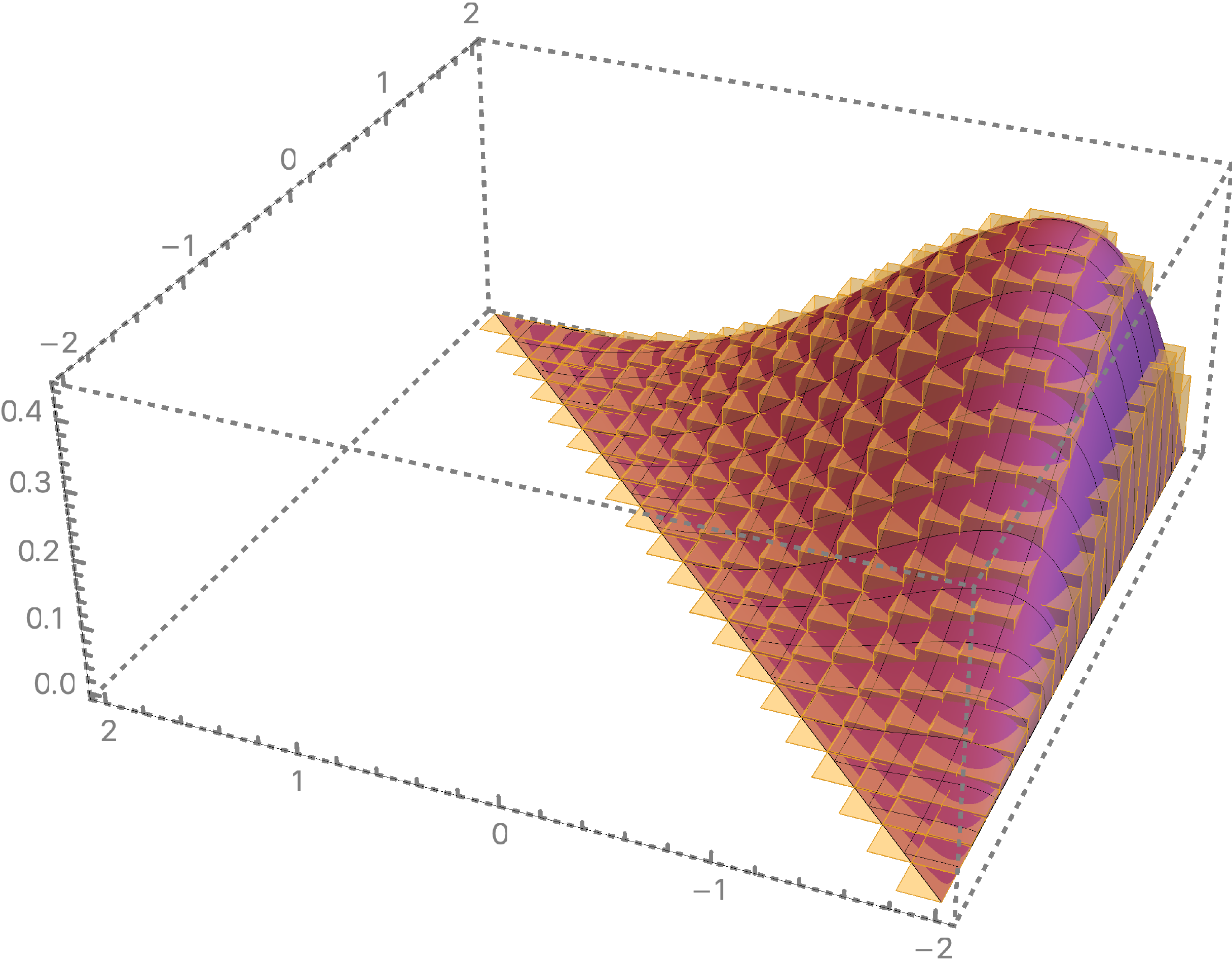}
\capt{5.5in}{fig:CombinedHistogramAndPdf}{An overlay of the histogram and function $p^2 h$ from the previous figure.}
\end{center}
\end{figure}

The factors of $\sin^2\th_1$ and $\sin^2\th_2$ are reminiscent of the Sato-Tate probability density function that
describes the distribution of the analogue of $\a$ for a large class of elliptic curves. The factor of 
$(\cos\th_1 {-} \cos\th_2)^2$ owes to the fact that the fundamental region $\tilde{\cS}_0$ is a triangle, rather than the full square and at least one power of $(\cos\th_1 {-} \cos\th_2)$ is required to cancel the singularity introduced by the factor of $\r$ on the right hand side of \eqref{eq:muoverp}.

The probability density function $f$ is intriguing because this corresponds to the eigenvalue distribution of $\text{USp}(4)$ matrices that are distributed randomly with respect to the Haar measure. This being so,
      and given the closeness of the fit of the, admittedly limited, statistical data, we conjecture that $f$ is the true density function. A probability density function corresponding to randomly distributed $\text{USp(2g)}$ matrices has appeared in relation to the distribution of the coefficients of the Frobenius polynomials for hyperelliptic curves of genus $g\leq 3$, see \cite{kedlaya2008}.
      
The following tables express the bivariate moments $\langle \tilde{\a}^m \tilde{\b}^n \rangle$ for $m{\,+\,}n{\;\leq 8}$. The first table gives the value of the moments calculated from the assumed probability density function, the second gives the values calculated from the data, the third gives the ratios of the corresponding entries in these terms and shows that these differ by at most 3 parts per thousand.  

\begin{table}[H]
\beq
\setlength{\extrarowheight}{1pt}
\begin{array}{c|@{\hskip3pt}c@{\hskip-11pt}c@{\hskip0pt}c@{\hskip-13pt}c@{\hskip-6pt}c@{\hskip-11pt}c@{\hskip0pt}c@{\hskip-11pt}c@{\hskip-2pt}c}
& 0 &\+ 1& 2 &\+ 3& 4 &\+5 & 6 &\+ 7 &  8 \\[2pt]
\hline
{} &\\[-16pt]
0&1 & -\frac{2^{14}}{3^2 5^2 7 \p^2} & \frac{3}{2} & -\frac{2^{16} 11}{3^3 5^2 \p^2} & \frac{7}{2} 
& -\frac{2^{18} 233}{3^4 5^2 7^2 11 \p^2} & \frac{19}{2} 
& -\frac{2^{20} 61{\cdot} 389}{3^4 5^2 7^2 11^2 13\p^2} & 28 \\[3pt]
1&\frac{2^{14}}{3^2 5^2 7 \p^2} & -1 & \frac{2^{16}}{3^3 5^2 7 \p^2} & -2 
& \frac{2^{18} 43}{3^3 5^2 7^2 11 \p^2} & -5 & \frac{2^{20} 101}{3^4 5^2 7^2 13 \p^2} & -14\\[3pt]
2&\frac{3}{2} & -\frac{2^{16}}{3^3 5^2 7 \p^2} & 2 & -\frac{2^{18} 13}{3^2 5^2 7^2 11 \p^2} 
& \frac{9}{2} & -\frac{2^{20} 137}{3^4 5^2 7{\cdot}11{\cdot}13 \p^2} & 12\\[3pt] 
3&\frac{2^{16} 11}{3^3 5^2 7^2 \p^2} & -2 & \frac{2^{18} 13}{3^2 5^2 7^2 11 \p^2} & -4 
& \frac{2^{20} 281}{3^3 5^2 7^2 11{\cdot}13\p^2} & -10\\[3pt]
4&\frac{7}{2} & -\frac{2^{18} 43}{3^3 5^2 7^2 11 \p^2} & \frac{9}{2} 
& -\frac{2^{20} 281}{3^3 5^2 7^2 11{\cdot}13 \p^2} & 10\\[3pt]
5&\frac{2^{18} 233}{3^4 5^2 7^2 11 \p^2} & -5 & \frac{2^{20} 137}{3^4 5^2 7{\cdot}11{\cdot}13\p^2} 
& -10\\[3pt] 
6&\frac{19}{2} & -\frac{2^{20} 101}{3^4 5^2 7^2 13 \p^2} & 12\\[3pt]  
7&\frac{2^{20} 61{\cdot} 389}{3^4 5^2 7^2 11^2 13 \p^2} & -14\\[3pt] 
8& 28\\
\noalign{\vskip14pt}
& 0 &\+ 1& 2 &\+ 3& 4 &\+5 & 6 &\+ 7 &  8 \\[2pt]
\hline
{} &\\[-17pt]
0&1.0000 & -1.0547 & 1.5010 & -2.2099 & 3.5019 & -5.6728 & 9.5032 & -16.156 & 28.003\\
1&1.0547 & -1.0018 & 1.4077 & -2.0037 & 3.1443 & -5.0085 & 8.3314 & -14.021\\
2&1.5008 & -1.4071 & 2.0016 & -2.8496 & 4.5018 & -7.1830 & 12.000\\
3&2.2103 & -2.0033 & 2.8503 & -4.0050 & 6.3170 & -10.008\\
4&3.5041 & -3.1455 & 4.5061 & -6.3221 & 10.011\\
5&5.6782 & -5.0114 & 7.1920 & -10.019\\
6&9.5170 & -8.3413 & 12.025\\
7&16.184 & -14.040\\
8&28.061\\
\noalign{\vskip14pt}
& 0 &\+ 1& 2 &\+ 3& 4 &\+5 & 6 &\+ 7 &  8 \\[2pt]
\hline
{} &\\[-17pt]
0&1.0000 & 0.9993 & 0.9993 & 0.9993 & 0.9995 & 0.9995 & 0.9997 & 0.9998 & 0.9999\\
1&0.9994 & 0.9982 & 0.9983 & 0.9982 & 0.9984 & 0.9983 & 0.9985 & 0.9985\\
2&0.9995 & 0.9987 & 0.9992 & 0.9992 & 0.9996 & 0.9997 & 1.0000\\
3&0.9991 & 0.9983 & 0.9989 & 0.9988 & 0.9992 & 0.9992\\
4&0.9988 & 0.9980 & 0.9986 & 0.9984 & 0.9989\\
5&0.9986 & 0.9977 & 0.9984 & 0.9981\\
6&0.9982 & 0.9973 & 0.9979\\
7&0.9981 & 0.9972\\
8&0.9978\\
\end{array}
\notag\eeq
\vskip-7pt
\capt{6.5in}{tab:AlphaBetaMoments}{The first table gives the the moments $\langle \tilde{\a}^m \tilde{\b}^n \rangle$ for $m{\,+\,}n{\,\leq 8}$, calculated from the distribution function \eqref{eq:hexpression}. The second table gives the same moments calculated from the data of \cite{LocalZetaFunctionsI}. The third table gives the result of dividing the entries of the first table by the corresponding entries of the second. For the cases shown, these ratios differ from unity by less than 3 parts per thousand.}
\end{table}

Returning to the estimation of $\m$: by using \eqref{eq:hexpression} we compute
\beq
\m \= \frac{256}{45\p^2\, p^{1/2}}~.
\notag\eeq
While this suggests an explanation for the tendency for the number of factorisations to fall off as $p$ increases. It does not explain the number of factorisations, even for the mirror quintic. Consider the numbers of factorisations for the mirror quintic for $p_{200}{\;\leq\;}p{\;\leq\;}p_{502}$ as shown in \fref{fig:ReducedFacnumsQuintic302}. The single factorisations at the high $p$ end of the plot, if random, represent more than $9\s$ deviations.
\vskip20pt

\begin{figure}[H]
\begin{center}
\includegraphics[width=\linewidth]{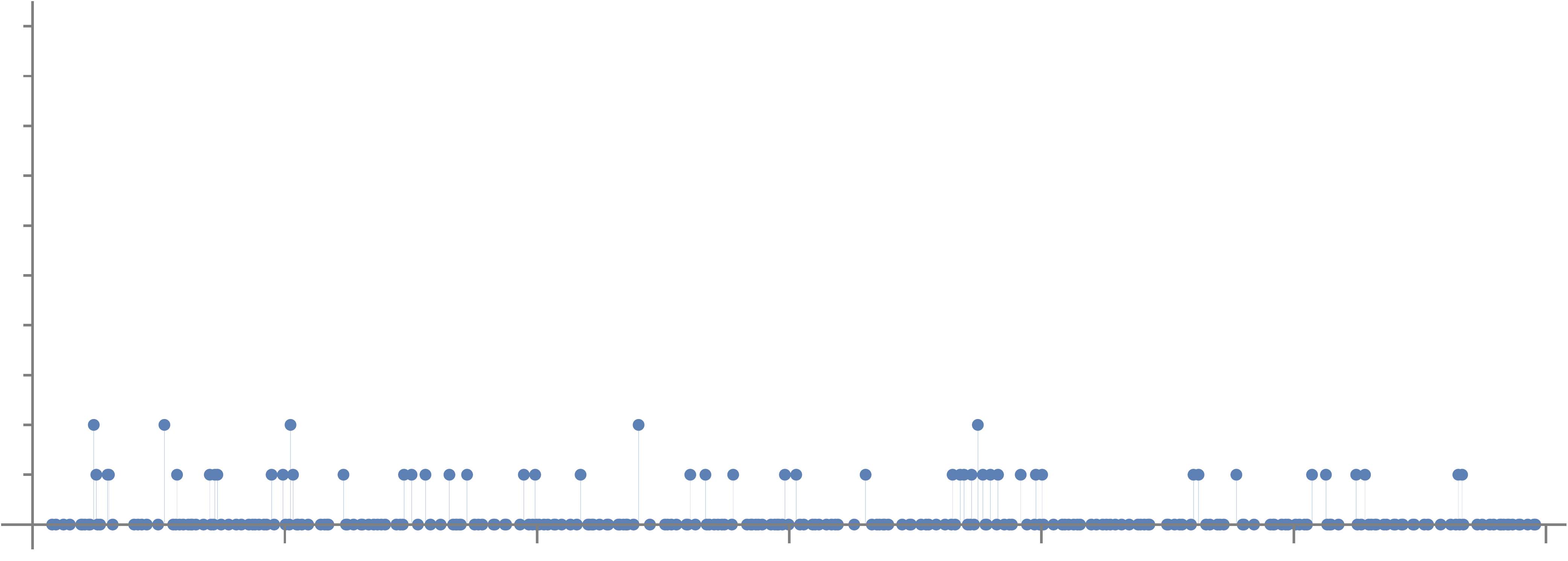}
\vskip0pt 
\place{-0.05}{0.54}{\scriptsize 1}
\place{-0.05}{0.75}{\scriptsize 2}
\place{-0.05}{0.96}{\scriptsize 3}
\place{-0.05}{1.16}{\scriptsize 4}
\place{-0.05}{1.37}{\scriptsize 5}
\place{-0.05}{1.57}{\scriptsize 6}
\place{-0.05}{1.78}{\scriptsize 7}
\place{-0.05}{1.99}{\scriptsize 8}
\place{-0.05}{2.20}{\scriptsize 9}
\place{-0.07}{2.41}{\scriptsize 10}
\place{0.00}{0.15}{\scriptsize 1200}
\place{1.05}{0.15}{\scriptsize 1600}
\place{2.10}{0.15}{\scriptsize 2000}
\place{3.15}{0.15}{\scriptsize 2400}
\place{4.20}{0.15}{\scriptsize 2800}
\place{5.25}{0.15}{\scriptsize 3200}
\place{6.29}{0.15}{\scriptsize 3600}
\place{6.6}{0.3}{$p$}
\capt{5.3in}{fig:ReducedFacnumsQuintic302}{The residual factorisations for the mirror quintic, after eliminating the primes up to $p_{200}$.}
\end{center}
\end{figure}
\newpage
\section{Review of Special Geometry}
\label{section: special geometry}
\vskip-10pt
We recall here the essential features of the Special Geometry of the moduli space of \cys given the prominent role it plays in this paper. A detailed account, in the spirit of the present discussion may be found in~\cite{candelas1991moduli}.

Denote by $\Omega$ and $\omega$ the holomorphic 3-form and K\"ahler form of $X$ respectively. There are natural \K geometries on the space of complex structures and space of \K forms. The \K potential for the space of complex structures is given by
\beq
K \;= -\text{log}\bigg({-}i\int\!\Omega\,\overline{\Omega}\bigg)
\notag\eeq

A fundamental observation is that $\Omega$ is defined only up to a parameter dependent scale transformation 
\beq
\Omega \rightarrow  f(\vph )\,\Omega
\notag\eeq
for any holomorphic $f$, so $\Omega$ should be understood as a section of a line bundle on the parameter space. Indeed, it is this observation that leads to the choice of $K$ as the natural choice of \K potential.

Although we are here concerned with one parameter spaces, let us allow for several complex structure parameters and denote these by $z^{\alpha}$. Consider also a quantity $\Psi^{(a,b)}$, which transforms under scale transformations with weight $(a,b)$, by which we mean
\beq
\Psi \rightarrow f^a \bar{f}^b\,\Psi~.
\notag\eeq
Thus $\O$ has weight $(1,0)$ and $\ee^{-K}$ has weight $(1,1)$.
We define a covariant derivative for this gauge transformation by
\beq\begin{split}
D_\a\Psi &\= \nabla_\a\Psi + a\, (\del_\a K)\,\Psi\\[3pt]
D_{\bar\b}\Psi &\= \nabla_{\bar\b}\Psi + b\, (\del_{\bar\b}K)\,\Psi
\end{split}\notag\eeq
where $\nabla_\a$ is the Levi-Civita connection. The virtue of this derivative is that $D_\a\Psi$ transforms in a manner parallel to $\Psi$
\beq
D_\a\Psi \rightarrow f^a \bar{f}^b\,D_\a\Psi~.
\notag\eeq
Note that $\ee^{\pm K}$ has weight $(\mp 1,\mp 1)$ so 
\beq
D_\a \ee^{\pm K}\= 0 ~~~\text{and}~~~D_{\bar\b} \ee^{\pm K}\= 0~.
\notag\eeq

Now $\O\in H^{3,0}$ and $\del_\a\O \in H^{3,0}\oplus H^{2,1}$, however the covariant derivatives 
$D_\a\O$ lie entirely in $H^{2,1}$ and form a basis for this cohomology group. In a similar way, the second and third covariant derivatives of $\O$ lie entirely in $H^{1,2}$ and $H^{0,3}$ respectively.

It is a standard exercise to derive the special geometry relations
\begin{align*}
D_\a\O &\;=\+ \chi_\a  & D_\a\O &\;=\+ \chi_\a \\[3pt]
D_\a\chi_\b\! &\;= -\ii\, y_{\a\b\g}\,\ee^K\bar{\chi}^\g & 
                             D_\a\chi_\b\! &\;= - y_{\a\b\g}\,\widetilde{\chi}^\g\\[3pt]
D_\a \bar{\chi}^\g\! &\;=\+ \d_\a{}^\g\,\overline{\Omega} &
                             D_\a \widetilde{\chi}^\g\! &\;=\+ \d_\a{}^\g\,\widetilde{\O}\\[3pt]
D_\a\overline{\O}\,&\;=\+ 0 & D_\a\widetilde{\O} \,&\;=\+ 0~,
\end{align*}
where, in these relations,
\beq
\widetilde{\chi}^\g \= \ii\,\ee^K g^{\g\bar\b}\chi_{\bar\b}~;~~~~\widetilde{\O} \= \ii\,\ee^K \overline{\O}
~;~~~~
y_{\a\b\g}\;= - \int \O\, \del_{\a\b\g} \O
\notag\eeq
and $g_{\a\bar\b}$ is the metric that derives from $K$.

The \K potential is simply written in terms of the integral periods, in virtue of \eqref{eq:IntegralPeriods} and
\eqref{eq:PiVector} we have

\beq
\ee^{-K}\;= -\ii\, \Pi^{\dagger}\Sigma\Pi
\label{eq:KahlerPotential}\eeq
where 
\beq
\Sigma\=
\begin{pmatrix}
\+0 & \mathbbl{1}\\
-\mathbbl{1} & 0
\end{pmatrix}~.
\notag\eeq
\vfill
\section*{Acknowledgements}
\vskip-8pt
It is a pleasure to acknowledge fruitful conversations and communications with Kilian B\"{o}sich, Noam Elkies, Minhyong Kim, Albrecht Klemm, Bruno Klingler and John Voight. The authors wish to thank the MITP, Mainz for hospitality and support during the 2018 Workshop on String Theory, Geometry and String Model Building. The authors also wish to thank KIAS for hospitality and support associated with the 2019 International Conference on Arithmetic Geometry and Quantum Field Theory, PC and XD are grateful also for support and hospitality of KIAS
     in 2018.
\newpage
\bibliographystyle{utcaps}
\bibliography{bibfilePC}

\providecommand{\href}[2]{#2}\begingroup\raggedright\begin{thebibliography}{10}

\bibitem{Ferrara:1995ih}
S.~Ferrara, R.~Kallosh, and A.~Strominger, ``{N=2 {E}xtremal {B}lack
  {H}oles},'' {\em Phys. Rev.} {\bf D52} (1995) R5412--R5416,
\href{http://arXiv.org/abs/hep-th/9508072}{{\tt hep-th/9508072}}.

\bibitem{Pioline:2006ni}
B.~Pioline, ``{Lectures on black holes, topological strings and quantum
  attractors},'' {\em Class. Quant. Grav.} {\bf 23} (2006) S981,
\href{http://arXiv.org/abs/hep-th/0607227}{{\tt hep-th/0607227}}.

\bibitem{Sen:2007qy}
A.~Sen, ``{Black Hole Entropy Function, Attractors and Precision Counting of
  Microstates},'' {\em Gen. Rel. Grav.} {\bf 40} (2008) 2249--2431,
\href{http://arXiv.org/abs/arXiv 0708.1270}{{\tt arXiv 0708.1270}}.

\bibitem{Moore:1998pn}
G.~W. Moore, ``{Arithmetic and {A}ttractors},''
\href{http://arXiv.org/abs/hep-th/9807087}{{\tt hep-th/9807087}}.

\bibitem{verrill1996}
H.~A. Verrill, ``Root lattices and pencils of varieties,'' {\em J. Math. Kyoto
  Univ.} {\bf 36} (1996), no.~2, 423--446.

\bibitem{verrill2004sums}
H.~A. Verrill, ``{Sums of squares of binomial coefficients, with applications
  to Picard-Fuchs equations},''
  \href{http://arXiv.org/abs/math.CO/0407327}{{\tt math.CO/0407327}}.

\bibitem{hulek_verrill_2005}
K.~Hulek and H.~Verrill, ``On modularity of rigid and nonrigid Calabi-Yau
  varieties associated to the Root Lattice $A_4$,'' {\em Nagoya Mathematical
  Journal} {\bf 179} (2005) 103--146.

\bibitem{lmfdb}
{The LMFDB Collaboration}, ``The L-functions and Modular Forms Database.''
  \url{http://www.lmfdb.org}.

\bibitem{Vanhove:2018mto}
P.~Vanhove, ``{Feynman integrals, toric geometry and mirror symmetry},'' in
  {\em {Proceedings, KMPB Conference: Elliptic Integrals, Elliptic Functions
  and Modular Forms in Quantum Field Theory: Zeuthen, Germany, October 23-26,
  2017}}, pp.~415--458.
\newblock 2019.
\newblock
\href{http://arXiv.org/abs/1807.11466}{{\tt 1807.11466}}.
\newblock

\bibitem{Bailey:2008ib}
D.~H. Bailey, J.~M. Borwein, D.~Broadhurst, and M.~L. Glasser, ``{Elliptic
  integral evaluations of Bessel moments},'' {\em J. Phys.} {\bf A41} (2008)
  205203,
\href{http://arXiv.org/abs/0801.0891}{{\tt 0801.0891}}.

\bibitem{Denef_2000}
F.~Denef, ``Supergravity flows and D-brane stability,'' {\em Journal of High
  Energy Physics} {\bf 2000} (aug, 2000) 050.

\bibitem{Denef_2001}
F.~Denef, B.~Greene, and M.~Raugas, ``Split attractor flows and the spectrum of
  {BPS} D-branes on the quintic,'' {\em Journal of High Energy Physics} {\bf
  2001} (may, 2001) 012.

\bibitem{Dwork1960Rationality}
B.~Dwork, ``On the Rationality of the Zeta Function of an Algebraic Variety,''
  {\em American Journal of Mathematics} {\bf 82} (1960), no.~3, 631--648.

\bibitem{KoblitzPadicNumbers}
N.~Koblitz, {\em p-adic Numbers, p-adic Analysis, and Zeta Functions}.
\newblock Graduate Texts in Mathematics. Springer, 1984.

\bibitem{Candelas:2007mb}
P.~Candelas and X.~de~la Ossa, ``{The Zeta-Function of a p-Adic Manifold, Dwork
  Theory for Physicists},'' {\em Commun. Num. Theor. Phys.} {\bf 1} (2007)
  479--512,
\href{http://arXiv.org/abs/0705.2056}{{\tt 0705.2056}}.

\bibitem{AST_1975__24-25__109_0}
J.-P. Serre, ``Valeurs propres des op\'erateurs de Hecke modulo $\ell$,'' in
  {\em Journ\'ees arithm\'etiques de Bordeaux}, no.~24-25 in Ast\'erisque,
  pp.~109--117.
\newblock Soci\'et\'e math\'ematique de France, 1975.

\bibitem{serre1987}
J.-P. Serre, ``Sur les représentations modulaires de degré $2$ de
  $\mathrm{Gal}(\overline{\mathbf{Q}}/\mathbf{Q})$,'' {\em Duke Math. J.} {\bf
  54} (1987), no.~1, 179--230.

\bibitem{10.2307/2118559}
A.~Wiles, ``Modular Elliptic Curves and Fermat's Last Theorem,'' {\em Annals of
  Mathematics} {\bf 141} (1995), no.~3, 443--551.

\bibitem{10.2307/2118560}
R.~Taylor and A.~Wiles, ``Ring-Theoretic Properties of Certain Hecke
  Algebras,'' {\em Annals of Mathematics} {\bf 141} (1995), no.~3, 553--572.

\bibitem{10.2307/827119}
C.~Breuil, B.~Conrad, F.~Diamond, and R.~Taylor, ``On the Modularity of
  Elliptic Curves over Q: Wild 3-Adic Exercises,'' {\em Journal of the American
  Mathematical Society} {\bf 14} (2001), no.~4, 843--939.

\bibitem{dieulefait2009modularity}
L.~Dieulefait, ``On the modularity of rigid Calabi-Yau threefolds: Epilogue,''
  \href{http://arXiv.org/abs/math/0908.1210}{{\tt math/0908.1210}}.

\bibitem{Khare2009article1}
C.~Khare and J.-P. Wintenberger, ``Serre's modularity conjecture (I),'' {\em
  Inventiones mathematicae} {\bf 178} (Jul, 2009) 485.

\bibitem{Khare2009article2}
C.~Khare and J.-P. Wintenberger, ``Serre's modularity conjecture (II),'' {\em
  Inventiones mathematicae} {\bf 178} (Jul, 2009) 505.

\bibitem{kisin2007}
M.~Kisin, ``Modularity of 2-dimensional Galois representations,'' {\em Current
  Developments in Mathematics} {\bf 2005} (2007) 191--230.

\bibitem{GuveaYui}
F.~Q. Gouvêa and N.~Yui, ``Rigid Calabi–Yau threefolds over Q are modular,''
  {\em Expositiones Mathematicae} {\bf 29} (2011), no.~1, 142 -- 149.

\bibitem{Stevenhagen1996}
P.~Stevenhagen and H.~W. Lenstra, ``Chebotar{\"e}v and his density theorem,''
  {\em The Mathematical Intelligencer} {\bf 18} (Mar, 1996) 26--37.

\bibitem{lauder2004}
A.~G.~B. Lauder, ``Deformation theory and the computation of zeta functions,''
  {\em Proceedings of the London Mathematical Society} {\bf 88} (2004), no.~3,
  565–602.

\bibitem{LocalZetaFunctionsI}
P.~Candelas, X.~de~la Ossa, and D.~Van~Straten, ``Local Zeta Functions I.'' To
  appear.

\bibitem{almkvist2005tables}
G.~Almkvist, C.~van Enckevort, D.~van Straten, and W.~Zudilin, ``Tables of
  Calabi--Yau equations,'' \href{http://arXiv.org/abs/math/0507430}{{\tt
  math/0507430}}.

\bibitem{Moore:2004fg}
G.~W. Moore, ``{{S}trings and {A}rithmetic},'' in {\em {Proceedings, Les
  Houches School of Physics: Frontiers in Number Theory, Physics and Geometry
  II: On Conformal Field Theories, Discrete Groups and Renormalization: Les
  Houches, France, March 9-21, 2003}}, pp.~303--359.
\newblock 2007.
\newblock
\href{http://arXiv.org/abs/hep-th/0401049}{{\tt hep-th/0401049}}.
\newblock

\bibitem{cattanidelignekaplan}
E.~{Cattani}, P.~{Deligne}, and A.~{Kaplan}, ``{On the locus of Hodge
  classes.},'' {\em {J. Am. Math. Soc.}} {\bf 8} (1995), no.~2, 483--506.

\bibitem{candelas1991moduli}
P.~Candelas and X.~de~la Ossa, ``{Moduli Space of {Calabi-Yau} Manifolds},''
  {\em Nucl. Phys.} {\bf B355} (1991) 455--481.

\bibitem{candelas1991pair}
P.~Candelas, X.~de~la Ossa, P.~S. Green, and L.~Parkes, ``{A Pair of Calabi-Yau
  manifolds as an exactly soluble superconformal theory},'' {\em Nucl. Phys.}
  {\bf B359} (1991) 21--74.
[AMS/IP Stud. Adv. Math.9,31(1998)].

\bibitem{Hosono:1994ax}
S.~Hosono, A.~Klemm, S.~Theisen, and S.-T. Yau, ``{Mirror symmetry, mirror map
  and applications to complete intersection Calabi-Yau spaces},'' {\em Nucl.
  Phys.} {\bf B433} (1995) 501--554,
  \href{http://arXiv.org/abs/hep-th/9406055}{{\tt hep-th/9406055}}.
[AMS/IP Stud. Adv. Math.1,545(1996)].

\bibitem{Halverson:2013qca}
J.~Halverson, H.~Jockers, J.~M. Lapan, and D.~R. Morrison, ``{Perturbative
  Corrections to Kaehler Moduli Spaces},'' {\em Commun. Math. Phys.} {\bf 333}
  (2015), no.~3, 1563--1584,
\href{http://arXiv.org/abs/1308.2157}{{\tt 1308.2157}}.

\bibitem{Deligne79Lvalues}
P.~Deligne, ``Valeurs de fonctions L et p\'eriodes d'integrales,'' vol.~33 of
  {\em Proc. Symp. Pure Math.}, pp.~313--346.
\newblock 1979.

\bibitem{2017arXiv170909751C}
S.~{Cynk} and D.~{van Straten}, ``{Periods of double octic Calabi--Yau
  manifolds},'' {\em arXiv e-prints} (Sept., 2017)
  \href{http://arXiv.org/abs/1709.09751}{{\tt 1709.09751}}.

\bibitem{BonischAndKlemm}
{K.~B\"{o}nisch} and {A.~Klemm}. Private communication.

\bibitem{Bender:1978aa}
C.~M. Bender and S.~A. Orszag, {\em Advanced mathematical methods for
  scientists and engineers}.
\newblock International series in pure and applied mathematics. McGraw-Hill,
  1978.

\bibitem{Bershadsky:1993ta}
M.~Bershadsky, S.~Cecotti, H.~Ooguri, and C.~Vafa, ``{Holomorphic anomalies in
  topological field theories},'' {\em Nucl. Phys.} {\bf B405} (1993) 279--304,
  \href{http://arXiv.org/abs/hep-th/9302103}{{\tt hep-th/9302103}}.
[AMS/IP Stud. Adv. Math.1,655(1996)].

\bibitem{Bershadsky:1993cx}
M.~Bershadsky, S.~Cecotti, H.~Ooguri, and C.~Vafa, ``{Kodaira-Spencer theory of
  gravity and exact results for quantum string amplitudes},'' {\em Commun.
  Math. Phys.} {\bf 165} (1994) 311--428,
\href{http://arXiv.org/abs/hep-th/9309140}{{\tt hep-th/9309140}}.

\bibitem{Gopakumar:1997dv}
R.~Gopakumar and C.~Vafa, ``{Branes and fundamental groups},'' {\em Adv. Theor.
  Math. Phys.} {\bf 2} (1998) 399--411,
\href{http://arXiv.org/abs/hep-th/9712048}{{\tt hep-th/9712048}}.

\bibitem{van38monodromy}
C.~van Enckevort and D.~van Straten, ``Monodromy calculations of fourth order
  equations of Calabi--Yau type, Mirror symmetry. V, 539--559,'' {\em AMS/IP
  Stud. Adv. Math} {\bf 38}.

\bibitem{Cynk2019}
S.~Cynk, M.~Schütt, and D.~van Straten, ``Hilbert modularity of some double
  octic Calabi–Yau threefolds,'' {\em Journal of Number Theory} (2019).

\bibitem{DissertationMeyer}
C.~Meyer, {\em A dictionary of modular threefolds}.
\newblock PhD thesis, Johannes Gutenberg Universit{\"a}t Mainz, 2005.

\bibitem{Katz:1996ht}
S.~H. Katz, D.~R. Morrison, and M.~R. Plesser, ``{Enhanced gauge symmetry in
  type II string theory},'' {\em Nucl. Phys.} {\bf B477} (1996) 105--140,
\href{http://arXiv.org/abs/hep-th/9601108}{{\tt hep-th/9601108}}.

\bibitem{hulek2005modularity}
K.~Hulek and H.~Verrill, ``On the modularity of Calabi-Yau threefolds
  containing elliptic ruled surfaces,''
  \href{http://arXiv.org/abs/math/0502158}{{\tt math/0502158}}.

\bibitem{cynkfreitagsalvati-manni}
S.~{Cynk}, E.~{Freitag}, and R.~{Salvati Manni}, ``{The geometry and arithmetic
  of a Calabi-Yau Siegel threefold.},'' {\em {Int. J. Math.}} {\bf 22} (2011),
  no.~11, 1585--1602.

\bibitem{vanGeemen1995}
B.~{van Geemen} and N.~Nygaard, ``On the Geometry and Arithmetic of Some Siegel
  Modular Threefolds,'' {\em Journal of Number Theory - J NUMBER THEOR} {\bf
  53} (07, 1995) 45--87.

\bibitem{vangeemenvanstraten}
B.~{van Geemen} and D.~{van Straten}, ``{The cusp forms of weight 3 on
  \(\Gamma_ 2(2,4,8)\).},'' {\em {Math. Comput.}} {\bf 61} (1993), no.~204,
  849--872.

\bibitem{Ooguri:2004zv}
H.~Ooguri, A.~Strominger, and C.~Vafa, ``{Black hole attractors and the
  topological string},'' {\em Phys. Rev.} {\bf D70} (2004) 106007,
\href{http://arXiv.org/abs/hep-th/0405146}{{\tt hep-th/0405146}}.

\bibitem{Vafa:1995ta}
C.~Vafa, ``{A stringy test of the fate of the conifold},'' {\em Nucl. Phys.}
  {\bf B447} (1995) 252--260,
\href{http://arXiv.org/abs/hep-th/9505023}{{\tt hep-th/9505023}}.

\bibitem{Klemm:1996kv}
A.~Klemm and P.~Mayr, ``{Strong coupling singularities and nonAbelian gauge
  symmetries in N=2 string theory},'' {\em Nucl. Phys.} {\bf B469} (1996)
  37--50,
\href{http://arXiv.org/abs/hep-th/9601014}{{\tt hep-th/9601014}}.

\bibitem{DissertationSamol}
K.~Samol, {\em Frobenius Polynomials for Calabi-Yau Equations}.
\newblock PhD thesis, Johannes Gutenberg Universit{\"a}t Mainz, January, 2010.

\bibitem{kedlaya2008}
K.~Kedlaya and A.~Sutherland, ``Hyperelliptic Curves, L-Polynomials, and Random
  Matrices,'' {\em Arithmetic, Geometry, Cryptography and Coding Theory,
  Contemporary Mathematics} {\bf 487} (2009) 119--162.

\end{thebibliography}\endgroup
\end{document}